\documentclass[floats,floatfix,amssymb,prd,twocolumn,superscriptaddress,nofootinbib]{revtex4-1}

\usepackage{subcaption}
\usepackage{ragged2e}
\DeclareCaptionJustification{justified}{\justifying}
\makeatletter
\newcommand{\subsetsim}{\mathrel{\mathpalette\subset@sim\relax}}
\newcommand{\subset@sim}[2]{%
  \vtop{\offinterlineskip\m@th
    \ialign{\hfil##\cr
      $#1\subset$\cr\noalign{\kern0.5pt}\scalebox{0.9}{$#1\sim$}\cr
    }%
  }%
}
\makeatother

\usepackage{amssymb,amsmath,verbatim,mathtools,needspace,enumitem,etoolbox,graphicx,physics,microtype,afterpage,bm}
\usepackage[dvipsnames, usenames]{xcolor}
\definecolor{linkcolor}{rgb}{0.0,0.3,0.5}
\usepackage{booktabs}

\definecolor{oucrimsonred}{rgb}{0.6, 0.0, 0.0}
\definecolor{persianblue}{rgb}{0.11, 0.22, 0.73}
\definecolor{forestgreen}{rgb}{0.13,0.35,0.13}
\usepackage[unicode, 
colorlinks=true, 
linkcolor=persianblue, 
citecolor=forestgreen, 
filecolor=persianblue,
urlcolor=oucrimsonred, 
pdfusetitle]{hyperref}

\usepackage[all]{hypcap}
\usepackage[T1]{fontenc}
\usepackage[utf8]{inputenc}
\usepackage{tabularx}
\usepackage{adjustbox}
\usepackage{float}
\usepackage{ulem}
\usepackage{xfrac}

\renewcommand{\arraystretch}{1.4}

\definecolor{azure}{rgb}{0.0, 0.5, 1.0}
\definecolor{VioletRed4}{rgb}{0.55, 0.13, .32}
\newcommand{\auu}[1]{{\color{black}{#1}}}

\usepackage{multirow}
\usepackage{pifont}
\usepackage{fontawesome}
\usepackage{lmodern}

\usepackage{multirow}

\allowdisplaybreaks
\usepackage{tikz}
\usepackage{color}
\usepackage{framed}

\definecolor{rossos}{cmyk}{0,1,1,0.55}
\definecolor{bluscuro}{rgb}{0.15, 0.2, .85}
\definecolor{bluchiaro}{cmyk}{1,.3,0.,0.1}
\definecolor{ForestGreen}{rgb}{0.13, 0.55, 0.13}

\def\nn{\nonumber}

\def\bea{\begin{eqnarray}}
\def\eea{\end{eqnarray}}

\def\d{{\mathrm{d}}}
\def\PBH{\text{\tiny{PBH}}}
\def\ABH{\text{\tiny{ABH}}}
\def\NS{\text{\tiny{NS}}}

\newcommand{\bs}{\begin{subequations}}
\newcommand{\es}{\end{subequations}}

\newcommand{\be}{\begin{equation}}
\newcommand{\ee}{\end{equation}}
\renewcommand{\d}{{\rm d}}

\newcommand{\llp}{\left [}
\newcommand{\rrp}{\right ]}
\newcommand{\lp}{\left (}
\newcommand{\rp}{\right )}

\def\lsim{\mathrel{\rlap{\lower4pt\hbox{\hskip0.5pt$\sim$}}
    \raise1pt\hbox{$<$}}}         
\def\gsim{\mathrel{\rlap{\lower4pt\hbox{\hskip0.5pt$\sim$}}
    \raise1pt\hbox{$>$}}}         

\newcommand{\ms}{M_\text{\tiny S}}
\newcommand{\ml}{M_\text{\tiny L}}

\makeatletter
\def\l@subsubsection#1#2{}
\makeatother

\newcommand{\sapienza}{Dipartimento di Fisica, Sapienza Università 
	di Roma, Piazzale Aldo Moro 5, 00185, Roma, Italy}
\newcommand{\infn}{INFN, Sezione di Roma, Piazzale Aldo Moro 2, 00185, Roma, Italy}

\begin{document}

\title{
From inflation to black hole mergers and back again:\\ Gravitational-wave data-driven constraints on inflationary scenarios\\ with a first-principle model of primordial black holes across the QCD epoch
\\
}

\author{Gabriele Franciolini}
\affiliation{\sapienza}
\affiliation{\infn}

\author{Ilia Musco}
\affiliation{\infn}

\author{Paolo Pani}
\affiliation{\sapienza}
\affiliation{\infn}

\author{Alfredo Urbano}
\affiliation{\sapienza}
\affiliation{\infn}


\begin{abstract}
Recent population studies have searched for a subpopulation of primordial black holes (PBHs) in the gravitational-wave (GW) events so far detected by LIGO/Virgo/KAGRA (LVK), in most cases adopting a phenomenological PBH mass distribution. When deriving such population from first principles in the standard scenario, however, the equation of state of the Universe at the time of PBH formation may strongly affect the PBH abundance and mass distribution, which ultimately depend on the power spectrum of cosmological perturbations. Here we improve on previous population studies on several aspects: (i) we adopt state-of-the-art PBH formation models describing the collapse of cosmological perturbations across the QCD epoch; (ii) we perform the first Bayesian multi-population inference on GW data including PBHs and directly using power spectrum parameters instead of phenomenological distributions; (iii) we critically confront the PBH scenario with LVK phenomenological models describing the GWTC-3 catalog both in the neutron-star and in the BH mass ranges, also considering PBHs as subpopulation of the total events. Our results confirm that LVK observations prevent the majority of the dark matter to be in the form of stellar mass PBHs. We find that the best fit PBH model can comprise a small fraction of the total events, in particular it can naturally explain events in the mass gaps. If the lower mass-gap event GW190814 is interpreted as a PBH binary, we predict that LVK should detect up to a few subsolar mergers and one to $\approx 30$ lower mass gap events during the upcoming O4 and O5 runs. Finally, mapping back the best-fit power spectrum into an ultra slow-roll inflationary scenario, we show that the latter predicts detectable PBH mergers in the LVK band, a stochastic GW background detectable by current and future instruments, and may include the entirety of dark matter in asteroid-mass PBHs.
\end{abstract}

\maketitle

{
  \hypersetup{linkcolor=black}
  \tableofcontents
}

\normalem

\section{Introduction}

Primordial black holes~(PBHs)~\cite{Zeldovich:1967lct,Hawking:1974rv,Chapline:1975ojl,Carr:1975qj} might have formed in the early universe after inflation from the collapse of large amplitude cosmological  perturbations~\cite{Ivanov:1994pa,GarciaBellido:1996qt,Ivanov:1997ia,Blinnikov:2016bxu} or by other mechanisms.
In the standard formation scenario, their characteristic mass depends mostly on the time these inhomogeneities re-enter the cosmological horizon, whereas their abundance and mass distribution depend strongly on the equation of state~(EoS) of the Universe at that epoch~\cite{Carr:1975qj,Shibata:1999zs,Niemeyer:1997mt,Jedamzik:1999am,Musco:2004ak,Musco:2008hv,Musco:2012au,Byrnes:2018clq,Muscoinprep}, and it is ultimately controlled by the power spectrum of cosmological curvature perturbations. In particular, the mass of PBHs can span several orders of magnitude and is not bounded from below ($M\gtrsim 2M_\odot$) as in the case of stellar-origin BHs, providing one of the key distinctive features~\cite{Franciolini:2021xbq} of this scenario. 

Besides being unique messengers of the early-time cosmology and inflationary models, in certain mass ranges PBHs could comprise the entirety of the dark matter, and could seed supermassive BHs at high redshift~\cite{Volonteri:2010wz,Clesse:2015wea,Serpico:2020ehh}.
These tantalizing possibilities have motivated the recent growing interest in searching for PBHs (see~\cite{Carr:2020gox} for a recent review), especially using gravitational-wave~(GW) data.

PBHs could contribute to at least a fraction of the BH merger events detected by the LIGO-Virgo-KAGRA~(LVK) Collaboration~\cite{LIGOScientific:2018mvr, LIGOScientific:2020ibl,LIGOScientific:2021djp} so far~\cite{Bird:2016dcv,Sasaki:2016jop,Eroshenko:2016hmn, Wang:2016ana, Ali-Haimoud:2017rtz, Chen:2018czv,Raidal:2018bbj,Liu:2019rnx, Hutsi:2019hlw, Vaskonen:2019jpv, Gow:2019pok,Wu:2020drm,DeLuca:2020qqa, Hall:2020daa,Wong:2020yig,Hutsi:2020sol,Kritos:2020wcl,DeLuca:2021wjr,Deng:2021ezy,Kimura:2021sqz,Franciolini:2021tla,Bavera:2021wmw,Liu:2021jnw}, and to those that will be detected by future GW instruments~\cite{DeLuca:2021wjr,DeLuca:2021hde,Pujolas:2021yaw,Ng:2021sqn,Franciolini:2021xbq,Ng:2022agi,Martinelli:2022elq,Cole:2022ucw} (see Refs.~\cite{Sasaki:2018dmp,Green:2020jor,Franciolini:2021nvv} for reviews on PBHs as GW sources).
In addition to outstanding events such as GW190425 (with a total mass that exceeds that one of known galactic neutron star~(NS) binaries) and the mass-gap events (such as GW190814~\cite{Clesse:2020ghq}, GW190521~\cite{DeLuca:2020sae}, and GW190426\_190642) which do not fit naturally in the standard astrophysical scenarios and might have a different origin, a subpopulation of PBHs may be competitive with certain astrophysical population models for explaining a fraction of events~\cite{Franciolini:2021tla}.
Population studies~\cite{2021arXiv211103634T}  will inevitably become very relevant as the number of detections increases, both during future LVK runs and especially in the era of next-generation detectors~\cite{Maggiore:2019uih,Kalogera:2021bya}.

So far population studies aimed at identifying a (sub)population of PBHs in LVK data (e.g.,~\cite{Hall:2020daa,Wong:2020yig,Hutsi:2020sol,DeLuca:2021wjr,Franciolini:2021tla}) have adopted phenomenological PBH mass distributions, such as a lognormal or a power-law function, that should approximately capture different underlying formation mechanisms.
However, in a realistic setting starting from first principles, the computation of the mass distribution should take into account several aspects: a given underlying early Universe model directly determines the power spectrum of primordial curvature perturbations, which in turns affects the collapse and eventually the PBH mass function. The latter might show several features which are not necessarily captured by simple parameterizations.

For example, the QCD phase transition of the early Universe, when free quarks are confined within hadrons, strongly affects the EoS of the cosmological fluid at energy scales corresponding to the formation of solar-mass PBHs~\cite{Jedamzik:1998hc,Byrnes:2018clq,Carr:2019hud,Carr:2019kxo,Jedamzik:2020omx}. 
As a rule of thumb, any drop of the EoS parameter $w=p/\rho$ (being $p$ and $\rho$ the pressure and energy density of the cosmological fluid, respectively) relative to the radiation-dominated case ($w=1/3$) is associated with an enhancement of PBH production, since the pressure contribution to balance gravity is weaker.
Thus, as a consequence of the QCD phase transition at few hundred MeV, one would generically expect a peak of the PBH mass function in the solar mass range, provided the 
power spectrum is sufficiently large at those specific scales.
However, being the gravitational collapse a non linear process, several details of the initial power spectrum might affect the final PBH mass function, also providing characteristic tails and subtle correlations between different mass scales that, as we shall discuss, should be taken into account.

In this paper we go beyond phenomenological models and build a framework to link the formation of PBH binaries and their GW signatures from first principles. Our final goal is to use GW data to constrain ab-initio models and inform inflationary dynamics. This allows us to build a self-consistent scenario which, on the one hand, is compatible with current constraints and, on the other hand, makes concrete predictions across a wide range of PBH masses. Indeed, owning to the specific shape of the mass distribution arising from an ab-initio model, constraints on a given mass range can percolate on different mass scales, making ab-initio models much more predictive (and hence falsifiable) than generic parameterizations.

One of the key novel ingredients of our framework is 
the inclusion of state-of-the-art PBH formation models describing
the collapse of radiation across the QCD epoch, incorporating the effect of critical collapse in shaping the QCD enhancement~\cite{Muscoinprep}.
A scenario in which the QCD era was deemed responsible for shaping the mass distribution of PBHs in the solar mass range was devised in Refs.~\cite{Carr:2019kxo,Jedamzik:2020omx} (see also \cite{Clesse:2020ghq,Bagui:2021dqi,Braglia:2021wwa,Braglia:2022icu}), 
where the power spectrum of curvature perturbations was specifically tuned to be nearly, but not exactly, scale invariant, which enhances the relevance of the QCD peak around the solar mass scale.
However, the physics of the collapse across the QCD epoch alone does not determine the entire PBH mass function, which chiefly depends also on the shape of the curvature perturbation spectrum.
It follows that the ratio between the abundance of PBHs at
${\cal O}(M_\odot)$ and ${\cal O}(30 M_\odot)$ (relevant for LVK detections), cannot by predicted by the QCD effect alone, unless strong assumptions on the spectral amplitudes at those two scales are made. 
Ref.~\cite{Juan:2022mir} specifically analysed such scenario, and concluded that the GW bound in the subsolar mass range (from the absence of subsolar events during O1/O2/O3 LVK runs) sets the most important constraint.
However, bounds on subsolar PBHs rely on assuming a specific PBH mass distribution~\cite{Nitz:2022ltl}, which is not necessarily the one assumed to come from the QCD phase transition in previous works and by the ab-initio model considered here.
Furthermore, Ref.~\cite{Juan:2022mir} concluded that PBH mergers shaped by the QCD EoS may not contribute to current LVK observations, unless an ad-hoc mass evolution for the PBH mass function and a cut-off in the power spectrum very close to the QCD scale are artificially introduced by hand.

We will extend the scope of these analyses, by exploring the role of the spectral tilt, which was previously fixed to a specific value
(and is not a priori related to the one constrained by CMB observations at much larger scales).
As we shall later discuss in details, 
we will leave the tilt as a free parameter of the model, which is eventually inferred from the data.
Due to the exponential dependence of the PBH abundance on the density variance, small modifications to the tilt (around 10\%) greatly reduce the QCD solar mass peak and render the scenario insensitive to the high-scale (i.e. low-mass) spectrum cut-off.

We revisit previous constraints by performing the first Bayesian population inference on GW data including a subpopulation of PBHs and directly using ab-initio power spectrum parameters (including the tilt and the effect of the QCD phase) instead of phenomenological distributions, and confronting the PBH scenario with the most recent GWTC-3 dataset~\cite{LIGOScientific:2021djp,2021arXiv211103634T}.
We allow the PBH model to produce subsolar merger events, and the constraint deriving from the absence of such binaries in LVK data is consistently included in our analysis by construction.
This constraint was not included in Ref.~\cite{Chen:2021nxo}, where the fit was arbitrarily cut at $\approx M_\odot$ and no constraining power from the absence of subsolar mergers is included in the inference. 
Another important addition of our analysis relative to~\cite{Chen:2021nxo} is the inclusion of a phenomenological fit describing the NS population~\cite{2021arXiv211103634T}, which is crucial to assess the nature of events in the solar-mass range.

Finally, employing the reverse engineering approach devised in Ref.~\cite{Franciolini:2022pav}, we show how the GW data-driven power spectrum can be naturally accommodated into an ultra slow-roll~(USR) inflationary scenario~\cite{Inomata:2016rbd,Garcia-Bellido:2017mdw,Ballesteros:2017fsr,Hertzberg:2017dkh,Kannike:2017bxn,Dalianis:2018frf,Inomata:2018cht,Cheong:2019vzl,Ballesteros:2020qam,Iacconi:2021ltm,Kawai:2021edk}. Remarkably, a single USR model informed by current observational constraints may explain the entirety of the dark matter in asteroid-mass PBHs while also allowing~\cite{Franciolini:2022pav} for detectable PBH mergers in the LVK band due to the enhancement of the PBH distribution around the solar-mass range induced by both the QCD phase transition and spectral features, and for detectable stochastic GW background~(SGWB) signals from the nano-Hertz to the kilo-Hertz band.

Throughout all this paper we assume geometrical units with $c=G=1$.

\section{PBH formation across the QCD epoch}\label{sec:PBHMFQCD}

Within the standard PBH formation scenario, which assumes PBHs form out of the collapse of large amplitude cosmological perturbations in the radiation dominated early Universe, a crucial role is played by the power spectrum of primordial curvature perturbations, $P_\zeta(k)$, and the corresponding value of the threshold $\delta_c$ for PBH formation. In this section, we summarize the results of~\cite{Muscoinprep}, where a state-of-the-art derivation of the threshold for the formation of PBHs during the QCD epoch is given.
This is obtained using detailed general relativistic numerical simulations, assuming spherical symmetry. 

We start with a brief introduction of the QCD phase transition, followed by a quick review of the mathematical formalism one needs to describe consistently the initial condition for PBH formation, clarifying the key ingredients used in the computation of the threshold. This allow us to discuss the impact on the formation of PBHs characterized by a solar-mass range of scale, and how we can include the effects of this phase within the computation of the PBH mass distribution.

\subsection{The QCD phase transition}

During the confinement of quarks into hadrons the particle degrees of freedom are varying with the temperature $T$. This results in a ratio between the pressure $p$ and the total energy density $\rho$ of the medium being not constant --~as in the case of a gas of ultrarelativistic particles~-- but varying with time according to
\begin{equation}
w(T) \equiv \frac{p}{\rho} = \frac{4g_{*,s}(T)}{3g_{*}(T)} - 1 \,.
\end{equation}
The functions $g_{*}(T)$ and $g_{*,s}(T)$ denote the two relevant measures of the effective number of relativistic degrees of freedom, defined as 
\begin{equation}
   g_{*}(T) = \frac{30\rho}{\pi^2T^4} 
    \quad \textrm{and} \quad 
   g_{*,s}(T) = \frac{45s}{2\pi^2T^3}\,,
\end{equation}
where $s$ is the entropy density of the medium and the pressure $p$ is given by
\begin{equation}
    p=sT-\rho=w(T)\rho\,.
\end{equation}

In the top panel of Fig.~\ref{fig:EoS/Threshold} we show the behavior of $w$ and the sound speed squared $c_s^2 \equiv \partial p/\partial \rho$ during the QCD phase transition, obtained from lattice QCD simulations~\cite{Hindmarsh:2005ix,Borsanyi:2013bia}, using the cosmological horizon mass $M_H$ as a measure of the fluid temperature\footnote{The cosmological horizon $R_H$ is a marginally trapped surface~\cite{Helou:2016xyu}, as the apparent horizon of a black hole, with $R_H=2M_H$.}. As we will discuss later, the non-negligible change of these two quantities during the QCD epoch, with respect the constant value ($w=c_s^2=1/3$) they have when the Universe is radiation dominated, plays a crucial role during the collapse of cosmological perturbations, and gives rise to a reduction of the threshold for the formation of PBHs (see the bottom panel of Fig.~\ref{fig:EoS/Threshold}).  

\subsection{Gradient expansion}\label{sec on Phi}

The threshold $\delta_c$ for PBH formation is defined as the critical value of the cosmological perturbation amplitude $\delta$ such that, for $\delta>\delta_c$ an apparent horizon appears during the collapse and a PBH is formed, while for $\delta<\delta_c$ the collapse bounces and the cosmological perturbation is dispersed into the surrounding medium.

To compute the value of the threshold one needs to specify initial conditions of the numerical simulations on super-horizon scale, when the asymptotic form of the space-time metric is given by
\begin{equation}
    \d s^2 = - \d t^2+a^2(t)e^{2\zeta(r)}  \left[ \d r^2 + r^2\d\Omega^2 \right] 
\end{equation}
where $a(t)$ is the scale factor, while $\zeta(r)$ is the conserved comoving curvature perturbations defined on a super-Hubble scale, converging to zero at infinity where the Universe is taken to be unperturbed and spatially flat. 

In this regime, using the so called gradient expansion or long wavelength approximation~\cite{Salopek:1990jq,Polnarev:2006aa,Harada:2015yda},
the energy density contrast $\delta\rho/\rho_b$ for adiabatic perturbations (the ones generated by a curvature profile $\zeta(r)$) can be written as~\cite{Yoo:2020dkz}
\begin{equation} 
\frac{\delta\rho}{\rho_b}(r,t) = - 
\frac{4}{3} \Phi
\left(\frac{1}{aH}\right)^2 e^{-5\zeta(r)/2} \nabla^2 e^{\zeta(r)/2},
\label{eqn:non-linear}
\end{equation}
where $H \equiv \dot{a}/{a}$ is the Hubble parameter, while the function $\Phi(t)$ depends on the equation of state of the Universe and is obtained by solving the following equation~\cite{Polnarev:2006aa}
\begin{equation}\label{eq:Phievo}
    \frac{1}{H}\frac{\d \Phi(t) }{\d t} + \frac{5+3 w(t)}{2} \Phi(t)-\frac{3}{2} (1+ w(t)) = 0
\end{equation}
integrated from past infinity to the time when the amplitude of the perturbation is computed. In standard models of the very early Universe (i.e. just after inflation) this is assumed to be dominated by a radiation dominated medium, with EoS $p=w\rho$ and $w=1/3$.

\begin{figure}[t!]
	\centering
    \includegraphics[width=0.49\textwidth]{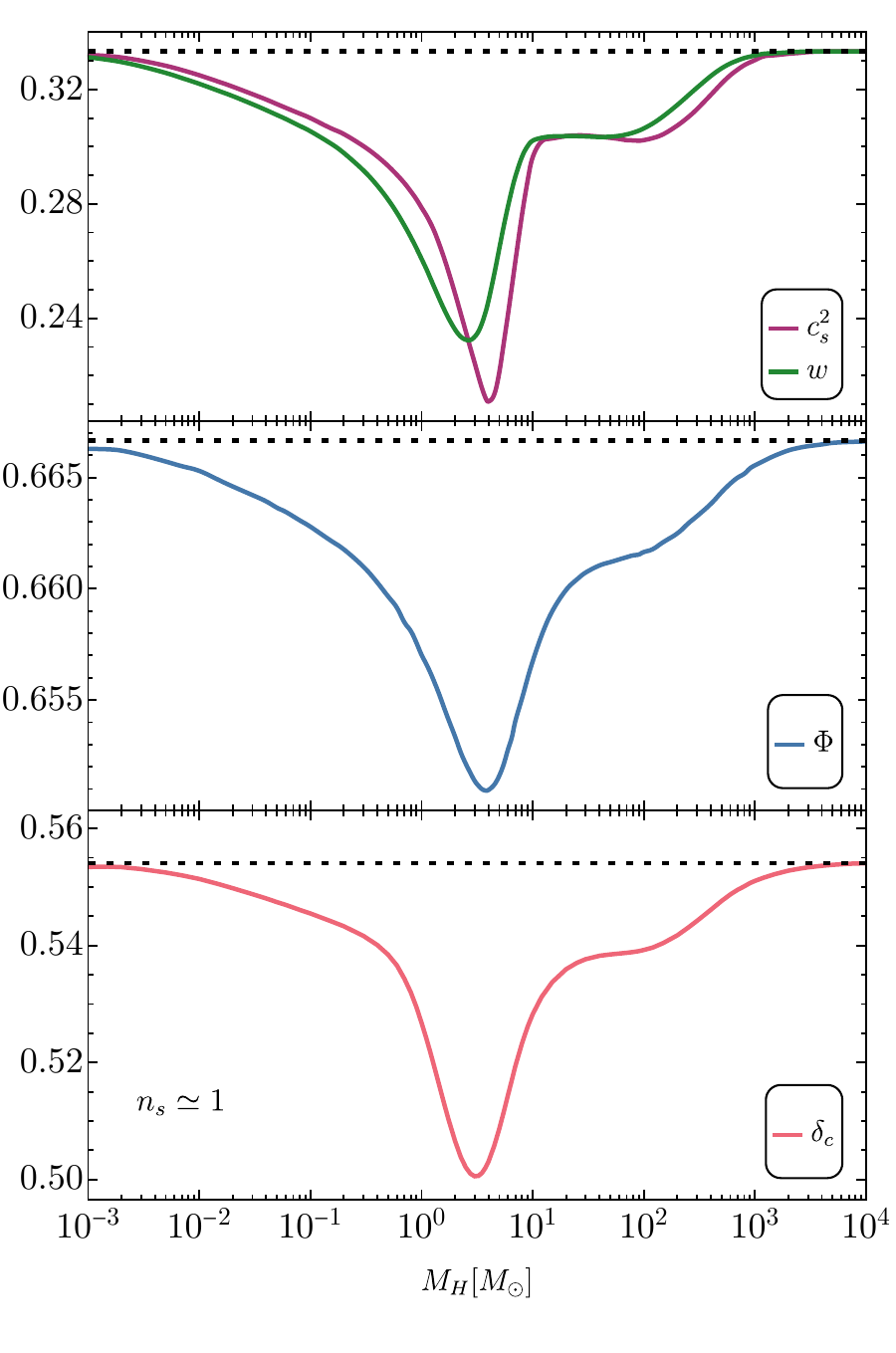}
	\caption{
	{\bf Top panel:} the EoS parameter $w=p/\rho$ (red) and squared speed of sound (blue) as functions of the cosmological horizon mass $M_H$.
	{\bf Central panel:} 
	Evolution of the EoS dependent parameter $\Phi$, relating the density contrast to the curvature perturbation as functions of the cosmological horizon mass $M_H$.
	{\bf Bottom panel:} 
	Same as above but showing the threshold for PBH formation.
	The dashed horizontal lines refer to the values obtained in the perfect radiation-fluid case.
	}
\label{fig:EoS/Threshold}
\end{figure}

When a constant $w(t) = \bar w$ characterises the fluid dominating the energy budget of the Universe, we have 
${\d \Phi(t) }/{\d t} = 0$ and one obtains 
\begin{equation}\label{solPhi}
\bar \Phi = \frac{3(1+ \bar w)}{(5+3 \bar w)},
\end{equation}
yielding $\bar \Phi=2/3$ for a radiation fluid with $\bar w = 1/3$. Equation~\eqref{solPhi} is an attractor solution of Eq.~\eqref{eq:Phievo}, i.e. if $w(t)$ slowly varies in time, ${\d \Phi(t) }/{\d t} \simeq 0$ and the evolution of $\Phi$ approaches the value given by Eq.~\eqref{solPhi}. The behavior of $\Phi$ across the QCD phase transition, obtained by solving Eq.~\eqref{eq:Phievo}, differs from the average $\bar \Phi$, particularly in the region where $w$ and $c_s^2$ are quickly varying with respect $M_H$. This is shown in the middle panel of Fig.~\ref{fig:EoS/Threshold}.

It was shown that a consistent way to define the threshold for PBH formation is in terms of the smoothed density contrast $\delta_m$ computed at horizon crossing time, i.e. $aH=1/r_m$. Using a top-hat window function with areal radius $R=a(t)\exp[\zeta(r_m)]r_m$, where $r_m$ indicates the location of the maximum of the mass excess, also called compaction function, the amplitude of spherically symmetric peaks in the smoothed density field is related to the curvature perturbation as~\cite{Musco:2018rwt}
\begin{equation}
\delta_m = - \Phi \, r_m \zeta'(r_m)\left[ 2 + r_m\zeta'(r_m) \right].
\label{eqn:smoothNonLinear}
\end{equation} 

Although strictly speaking the gradient expansion approach is valid only on super horizon scales, to compute the perturbation amplitude $\delta_m$ it is useful to extend this approach up to the cosmological horizon crossing time. Since then the region involved in the formation of a PBH becomes causally connected, and the collapse starts shortly afterwards. This gives a well defined criterion to quantify the amplitude of cosmological perturbations, comparing different initial configuration collapsing at different epochs.

\begin{figure*}[t!]
	\centering
	\includegraphics[width=0.49\textwidth]{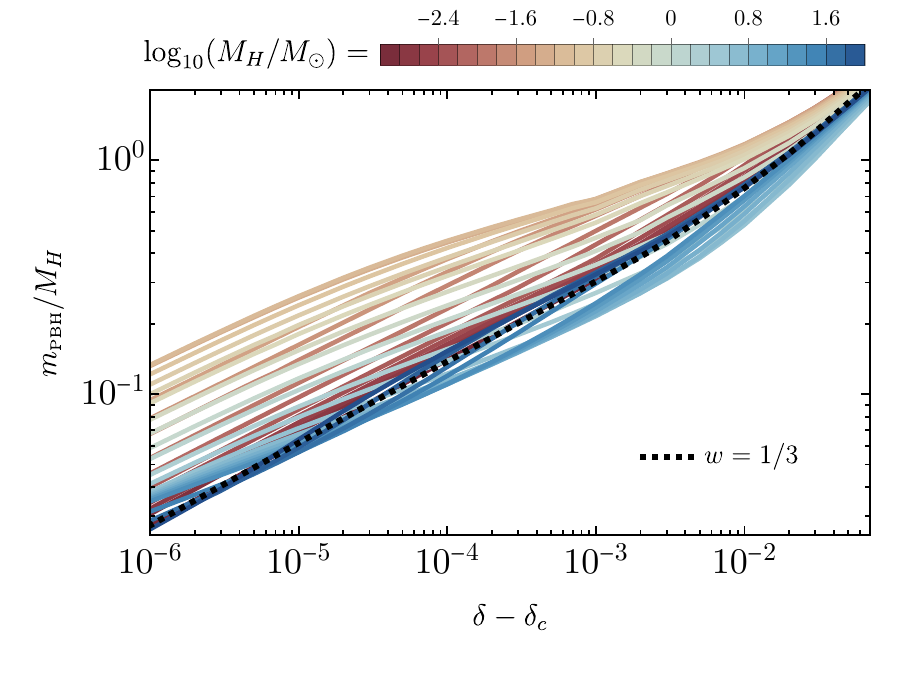}
	\includegraphics[width=0.49\textwidth]{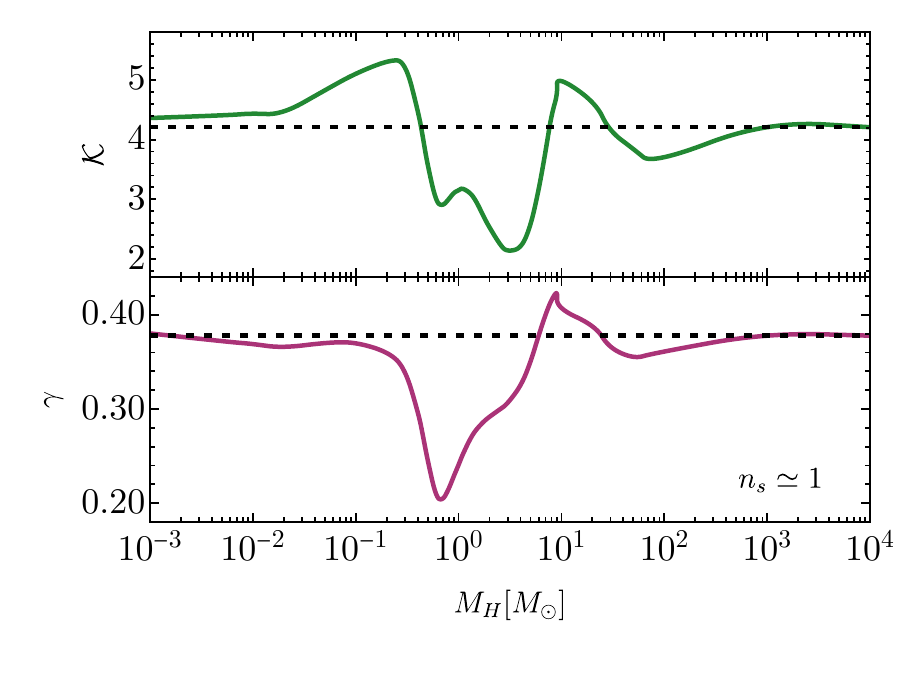}
\caption{{\bf Left panel:} PBH mass $m_\PBH$ plotted as a function of $\delta-\delta_c$ computed at the cosmological horizon crossing (see Ref.~\cite{Muscoinprep} for more details). The behavior for a radiation dominated medium is plotted with a black dashed line.\\
{\bf Right panel:} The values of the power law coefficients in Eq.~\eqref{critical collapse rel} found by fitting the results of numerical simulations shown in the left panel.}
\label{fig:MF_critical_collapse_1}
\end{figure*}

\subsection{The threshold for PBHs}\label{sec on critical}
The spherically symmetric numerical simulations used to compute the threshold $\delta_c$, and the mass distribution discussed in the next section, have been performed with a numerical code developed in~\cite{Musco:2004ak} that has been widely used and tested~\cite{Musco:2008hv,Musco:2012au}, including also an adaptive mesh refinement (AMR) scheme, which makes the code very flexible. This allows to compute the threshold with very high accuracy, a crucial point for calculating the mass spectrum discussed in Sec.~\ref{sec on mPBH}. 

The behavior of the threshold $\delta_c$ during the QCD phase transition, computed when a nearly scale invariant power spectrum (see Sec.~\ref{sec:CurvaturePS}) is assumed, is shown in the bottom panel of Fig.~\ref{fig:EoS/Threshold}. Looking at the top panel we can appreciate the varying EoS during this epoch: the value of threshold $\delta_c$ is affected by the change of both $w$ and $c_s^2$, with a minimum value $\delta_c\simeq0.5$ reached at $M_H=3M_\odot$ (between the minimum of these two quantities), about $10\%$ less than the value $\delta_c\simeq0.55$ one has during the standard radiation scenario (dashed line). The effect of $\Phi$ shown in the central panel is to give and additional lowering, accounting up to $25\%$ of the total decrement of the threshold, smoothing the whole behaviour of $\delta_c$, monotonically decreasing for $M_H \leq 3M_\odot$, and monotonically increasing afterwards, for $M_H \geq 3M_\odot$  .

This is quite different from the behavior of $\delta_c$ obtained in~\cite{Byrnes:2018clq} where the variation of the threshold during the QCD epoch was obtained simply from a fit of the numerical results given in~\cite{Musco:2012au} where only $w$ is varying\footnote{In~\cite{Musco:2012au} the threshold was not defined at the maximum of the compaction function but at the edge of the overdensity, as it was used to be done in past works~\cite{Jedamzik:1999am,Musco:2004ak,Musco:2008hv}. This does not allow to make a direct comparison with the new numerical results without a proper rescaling (see~\cite{Muscoinprep} for more details).}. This neglects completely the effects of $c_s^2$ during the dynamics of the collapse, and a correct computation of $\Phi$ entering in the definition of $\delta_m$ given in \eqref{eqn:smoothNonLinear}.  

More recently an attempt to improve the calculation, including also the effects of the sound speed, has been investigated in~\cite{Papanikolaou:2022cvo}, computing an analytic estimation of the threshold based on the three zone model used in~\cite{Harada:2013epa}. 
This however has the well known drawback of not being able to include the effects of the pressure gradients during the collapse, corresponding to an underestimation of the threshold which is strongly shape dependent~\cite{Musco:2018rwt}, varying with the initial curvature power spectrum of cosmological perturbation~\cite{Musco:2020jjb}. 

Even looking at the qualitative behavior of~\cite{Papanikolaou:2022cvo}, one can see a non monotonic behaviour in $\delta_c$ in the two key regions ($M_H$ smaller/larger than $3M_H$) which do not appear in the full numerical results shown in the bottom panel of Fig.~\ref{fig:EoS/Threshold}. This is a clear evidence of the intrinsic limit of making an analytic approximation of a non linear collapse process, as in~\cite{Papanikolaou:2022cvo} it was also pointed out, which is not able to include properly all the combined non linear effects related to the behavior of $w$ and $c_s^2$ when a cosmological perturbation is collapsing during the QCD phase. In general a proper computation of the threshold $\delta_c$, to be used in precise estimation of the abundance of PBHs, requires necessarily fully relativistic numerical simulations as the ones used here (see~\cite{Muscoinprep} for more details). 

\subsection{The mass spectrum for PBHs}\label{sec on mPBH}
In Fig.~\ref{fig:MF_critical_collapse_1} we show the resulting mass spectrum of PBHs obtained from the numerical simulaions of~\cite{Muscoinprep} obtained after the computation of the threshold, plotting $m_\PBH/M_H$ against $(\delta-\delta_c)$ during the QCD phase transition. As it is well known, in the standard scenario of a radiation dominated medium a critical collapse arises~\cite{Choptuik:1992jv,Gundlach:1999cu} and the mass spectrum of PBHs is characterized by a scaling law~\cite{Niemeyer:1997mt,Green:1999xm,Jedamzik:1999am,Musco:2004ak,Musco:2008hv,Musco:2012au} given by 
\begin{equation}\label{critical collapse rel}
m_\PBH(\delta) = {\cal K} M_H (\delta-\delta_c)^\gamma\,,
\end{equation}
where for $\delta-\delta_c \lesssim 10^{-2}$ the critical exponent $\gamma$ depends only on the parameter of the equation of state, i.e the value of $w$, completely independent on the initial configuration of the initial conditions, given by the initial profile of $\zeta(r)$, which affect instead the value of $\mathcal{K}$. This is shown on both plots of Fig.~\ref{fig:MF_critical_collapse_1} with a dashed line when $w=1/3$, which gives $\gamma \simeq 0.36$ and $\mathcal{K} \simeq 4$ for a nearly scale invariant curvature power spectrum, as the one considered here.

The QCD phase transition introduce an additional degree of freedom into the problem, which is the characteristic scale of the horizon crossing of the cosmological perturbation. This makes $\delta_c$, $\gamma$ and $\mathcal{K}$ to depend also on $M_H$, i.e. when the perturbation is crossing the cosmological horizon. The different lines shown in Fig.~\ref{fig:MF_critical_collapse_1} with a color varying between red, for smaller values of $M_H$, and blue for larger values, shows how the scaling law is modified by the characteristic scale of the problem. 

For the calculation of the mass distribution (see Section~\ref{sec PBH MF}) it is important to include these effects due to the variation of $\gamma$ and $\mathcal{K}$ in terms of $M_H$: an exact power-law critical behaviour is only obtained close enough to the density threshold $(\delta-\delta_c \lesssim 10^{-5})$, where the PBH masses are significantly smaller than the cosmological horizon mass, not able to affect significantly the collapse, while for larger values the EoS during the QCD epoch induces further modifications. We fit the relation between the PBH and horizon mass using the power-law template~\eqref{critical collapse rel} in the range of $\delta$ which most contributes to the abundance, i.e. $(\delta - \delta_c)\in [10^{-5},2\times 10^{-2}]$, and find that deviations from the functional form used in Eq.~\eqref{critical collapse rel} would only induce a small correction which we can neglect. 

The resulting values of ${\cal K}(M_H)$ and ${\gamma}(M_H)$ used here are shown in the right plot of Fig.~\ref{fig:MF_critical_collapse_1}: one could appreciate the significant variation of these quantities when $\delta_c$ is also significantly varying with respect $M_H$, compared to the constant values of the critical collapse during the radiation dominated epoch of the early Universe, indicated here with a black dashed line.
A general trend is observed: for $M_H\lesssim3 M_\odot$, there is a tendency to generate heavier PBHs, while the opposite is found when $M_H    \gtrsim 3 M_\odot$. This can be seen in the left panel of Fig.~\ref{fig:MF_critical_collapse_1}, where orange (light blue) lines fall above (below) the dashed black line indicating the result for a radiation perfect fluid. The fitted values of ${\cal K}(M_H)$ and ${\gamma}(M_H)$ shown in the right panel of Fig.~\ref{fig:MF_critical_collapse_1} aim to describe with enough accuracy this trend.

\subsection{Curvature power spectrum}\label{sec:CurvaturePS}

Our model is based on a parametrization of the curvature power spectrum, which we assume to have a nearly scale invariant shape of the form
\begin{equation}\label{PS_zeta}
  \mathcal{P}_\zeta (k)
    =A 
    \left ( \frac{k}{k_\text{\tiny min}}\right )^{n_s-1}
    \Theta(k-k_\text{\tiny min})
    \Theta(k_\text{\tiny max}-k),
\end{equation}
where $A$ defines the characteristic amplitude, $n_s$ is the spectral tilt\footnote{We warn the reader that the spectral tilt $n_s$ defined Eq.~\eqref{PS_zeta} specifically refers to the small (PBH) scales. The tilt observed at large (CMB) scales will be referred to as $n_s(k_*)$, where $k_*$ is the CMB pivot scale, see discussion in Sec.~\ref{sec:USR}. We stress that, as we shall discuss, they are not a priori related to each other.}, and $k_\text{\tiny min}$ and $k_\text{\tiny max}$ are the cut-off scales in momentum ($k$) space. 
This functional form generally describes broad spectra~\cite{MoradinezhadDizgah:2019wjf,DeLuca:2020ioi}, whose consequent PBH mass distribution may be modulated by the QCD epoch. 
Notice that, due to the exponential dependence of the PBH abundance to the spectral amplitude, even mildly tilted spectra with $n_s\neq 1$ generate {\it narrow mass distributions} strongly peaked towards small (when blue with $n_s>1$) or large (when red with $n_s<1$) masses.  See Sec.~\ref{sec PBH MF} for more details.

In reality, sharp cut-offs in momentum space do not appear in physically motivated curvature power spectra, which are also typically constrained to obey maximum growth or decay rates as a function of the wavenumber~\cite{Byrnes:2018txb,Kalaja:2019uju,Cole:2022xqc}. 
However, due to the exponential dependence of the PBH abundance to the variance of the density contrast, we do not expect such a simplification to affect our result.
Indeed, in Sec.~\ref{sec:USR} we shall show how the features of the parametrization~\eqref{PS_zeta} are naturally reproduced in a USR inflationary model.

Even though we restrict our parameter space to nearly scale invariant spectra, the variations of the spectral tilt would require considering potentially different shapes of collapsing overdensities (see e.g. \cite{Musco:2020jjb}).
Capturing this effect on the threshold and the other parameters of collapse, would necessitate numerically simulating the PBH formation across the QCD epoch over a fine grid of variations beyond the scale invariant spectrum, which is computationally very demanding, and is left to future extensions of this work.

We also assume the absence of primordial non-Gaussianities of the curvature perturbations (see e.g. Ref.~\cite{Franciolini:2018vbk,Atal:2019cdz,DeLuca:2021hcf,Taoso:2021uvl}) while we fully account for the
unavoidable intrinsic non-Gaussianities induced by the non-linear relation between the curvature perturbation and the energy density contrast~\cite{DeLuca:2019qsy,Young:2019yug}.

Notice that, for fixed spectrum shape parameters $[n_s, k_\text{\tiny min},k_\text{\tiny max}]$, 
the overall PBH abundance $f_\PBH$ (to be defined later on) is degenerate with the amplitude $A$.
Also, the minimum and maximum scales at which the power spectrum is cut correspond to characteristic horizon mass scales $\ms\equiv M_H(k_\text{\tiny max})$ and $\ml\equiv M_H(k_\text{\tiny min})$.
In other words, $\ms$ and $\ml$ are respectively the smallest and largest horizon masses bracketing the PBH formation epoch. 
Therefore, we equivalently choose to adopt the following hyperparameters describing the PBH model as
\begin{equation}
    {\bm \lambda}_\PBH 
    = 
    [\log_{10}f_\PBH,
    n_s,
    \log_{10}\ms,
    \log_{10}\ml
    ],
\end{equation}
where, if not explicitly indicated, the mass scales $\ms$ and $\ml$ are intended as expressed in units of the solar mass $M_\odot$. In Table~\ref{TbPBH:parameters}, we summarise the choice of priors of the PBH model later adopted in the GWTC-3 Bayesian inference analyses.

{
\renewcommand{\arraystretch}{1.4}
\setlength{\tabcolsep}{4pt}
\begin{table}
\caption{Hyperparameters of the PBH model and their prior ranges adopted in the inference analysis.
The mass scales $\ms$ and $\ml$ are intended as 
expressed in units of the solar mass $M_\odot$. }
\begin{tabularx}{1 \columnwidth}{|X|c|c|c|c|}
\hline
\hline
 Model &\multicolumn{4}{c|}{PBH}   
\\
\hline
${\bm \lambda}$ & $ \log_{10}f_\PBH$ & $n_s$ &  $\log_{10}\ms$ & $\log_{10}\ml$ 
\\
\hline
 Prior & $[-6,0]$ & $[0,1.5]$ & $[-2.5,\log_{10}\ml]$ & $[\log_{10}\ms, 4]$   \\
\hline
\hline
\end{tabularx}
\label{TbPBH:parameters}
\end{table}
}

\subsection{Computation of the mass distribution}\label{sec PBH MF}

In this section we report the computation of the PBH mass distribution starting from the primordial power spectrum defined in Eq.~\eqref{PS_zeta}.
We shall follow the derivation reported in Ref.~\cite{Young:2019yug}, to which we refer for more details. 

Looking at \eqref{eqn:smoothNonLinear} it has already been observed that  this equation can be written in terms of a  Gaussian component linearly related to the curvature perturbation $\delta_l \equiv  - 2 \Phi r_m\zeta'(r_m)$ as
\begin{equation}
\delta_m =
\lp  \delta_l - \frac{1}{4 \Phi} \delta_l^2\rp .
\label{eqn:NL}
\end{equation}
The probability density function of the linear component of the smoothed energy density contrast $\delta_l$ is Gaussian, and thus can be written as 
\begin{equation}
P(\delta_l) = \frac{1}{\sqrt{2\pi\sigma_0^2 }}\exp \left( -\frac{\delta_l^2}{2\sigma_0^2} \right).
\label{eqn:gaussianPDF}
\end{equation}
The variance $\sigma_0^2$ and the first momentum of the distribution $\sigma_1^2$ are
\begin{align}
&\sigma^2_i (r_m)
= \frac{4}{9}\Phi^2
\int\limits_0^\infty \frac{\mathrm{d}k}{k}(k r_m)^4 
\tilde{W}^2(k, r_m)
T^2 (k, r_m) 
k^{2i}
\mathcal{P}_\zeta(k),
\label{eqn:variance}
\end{align}
where $i=0,1$; $\tilde{W}(k,r_m)$ is the Fourier transform of the top-hat smoothing function,
\begin{equation}
\tilde{W}(k,r_m) = 3 
\llp \frac{\mathrm{sin}(k r_m)- k r_m \mathrm{cos}(k r_m)}{(k r_m)^3}\rrp ,
\label{eqn:window}
\end{equation}
and $T(k,r_m)$ is the linear transfer function
 \begin{equation}
T(k,r_m) = 3 
\llp \frac{\mathrm{sin}({k r_m}/{\sqrt{3}}) - {k r_m}\mathrm{cos}({k r_m}/{\sqrt{3}})
/{\sqrt{3}}
}{({k r_m}/{\sqrt{3}})^3}\rrp .
\label{eq:T}
\end{equation}
In the following, we are going to identify the smoothing scale $r_m$ with the corresponding horizon mass $M_H$ (fixed by the horizon crossing condition $a H r_m =1$~\cite{Musco:2018rwt}) using the relation with power spectral modes 
\begin{equation}
    r_m k  \equiv \kappa = 4.49\,,
\end{equation}
found for a broad (and nearly scale invariant) power spectrum~\cite{Musco:2020jjb}. 
This relation is strictly valid for a shape parameter  $\alpha = 3$ \cite{Musco:2020jjb}, consistently with the approximations described above.

By consequence one finds that the horizon mass $M_H$ is related to power spectral modes through
\begin{equation}
    M_H \simeq  
 17 M_\odot\left(\frac{g_*}{10.75}\right)^{-1/6}
 \left(\frac{k/\kappa}{{\rm pc}^{-1}}\right)^{-2}, 
 \label{M-k}
\end{equation}
where $g_*$ is the number of degrees of freedom of relativistic particles. 
We reiterate here for clarity that Eq.~\eqref{M-k} relates the horizon mass $M_H$ to the epoch of horizon crossing of the peak of the compaction function (of size $r_m$) produced by the single mode $k$.
This differs from the horizon mass corresponding to the crossing time of modes $k$ themselves and we point the attention of the reader to the relating coefficient $\kappa$ that has been frequently (but incorrectly) omitted in the past.

In principle, the transfer function defined in Eq.~\eqref{eq:T} is derived using linear perturbation theory in a radiation dominated Universe ($w = 1/3$). While $T(k,r_m)$ is modified by varying the EoS, and this would lead to a modified evolution of subhorizon modes, the presence of a window function already efficiently smooths curvature perturbations with $k r_m \gg 1$ and
the impact of a softer EoS should be small. 
As discussed in Sec.~\ref{sec semianalytical MF}, we will capture both $T(k,r_m)$ and $\tilde W(k,r_m)$ with an effective smoothing function, neglecting further modifications of $T(k,r_m)$ from a time-dependent $w$ around the QCD epoch. 

The threshold for PBH formation can be translated into a critical amplitude of the linear component $\delta_{c,l\pm}$ by inverting Eq.~\eqref{eqn:NL} as
\begin{equation}
\delta_{c,l\pm} = 2 \Phi \left( 1 \pm \sqrt{1-\frac{\delta_c}{\Phi}} \right) .
\label{eq:delta-pm}
\end{equation}
In the computation of the mass distribution we only include values of $\delta_l$ falling in the range
\begin{equation}\label{eq.deltalp}
\delta_{c,l-} < \delta_l <  2 \Phi 
\equiv  \delta_{l}^+,
\end{equation}
corresponding to type-I PBH formation~\cite{Musco:2018rwt}, and neglect the contribution from PBHs formed in the second branch whose contribution is exponentially suppressed.

The number density of sufficiently high peaks can be computed adopting the theory of  random Gaussian fields~\cite{Bardeen:1985tr}, which gives
\begin{equation}
\mathcal{N} = \frac{\sigma_1^3}{4 \pi^2 \sigma_0^3} \nu^3 \exp \left( -\frac{\nu^2}{2} \right),
\label{eqn:peakdensity}
\end{equation}
where we introduced the rescaled peak height $\nu\equiv\delta_l/\sigma_0$.
The mass fraction for each peak of given height $\nu$ which collapses to form a PBH can be expressed by evaluating
\begin{equation}
\beta_\nu = \frac{
m_\PBH
(\nu)}{M_{H}}\mathcal{N}(\nu)\theta(\nu - \nu_c),
\label{beta-first-time}
\end{equation}
where the Heaviside step 
function $\theta$ implements the threshold for collapse. 

The total energy fraction of the Universe composed by PBHs formed at a given time (equivalently identified with a single horizon mass $M_H$) is given by integrating the relevant range of $\nu$ between $\nu_{c-}\equiv\delta_{c,l-}/\sigma_0$ 
and $\nu_+ = \delta_{l}^+/\sigma_0$ 
(using Eqs.~\eqref{eq:delta-pm} and \eqref{eq.deltalp}), which can be written as
\begin{align}
\beta (M_H) = 
\int\limits_{\nu_{c-}}^{\nu_+} \mathrm{d}\nu 
\frac{\mathcal{K}}{3\pi} 
\left( \nu\sigma_0-\frac{1}{4 \Phi}(\nu\sigma_0)^2 - \delta_c \right)^\gamma \nonumber \\
\times \left( \frac{\sigma_1}{aH \sigma_0} \right)^3 \nu^3 \exp \left( -\frac{\nu^2}{2} \right)\,.
\label{eqn:beta}
\end{align}
The term $1/a H$ is fixed by the horizon crossing condition $a H  = 1 /  r_m = k_H / \kappa $~\cite{Musco:2018rwt}.
Finally, the entire energy fraction composed by PBHs after  formation is found by integrating over all relevant epochs (corresponding to the time span when modes within $k_\text{\tiny min}< k < k_\text{\tiny max}$ cross the Hubble horizon) as
\begin{equation}
\Omega_{\text{\tiny PBH}} = 
\int\limits_{\ms}^{\ml}
\mathrm{d} \ln M_{H} \left( \frac{M_{\rm eq}}{M_{H}} \right)^{1/2}\beta(M_{H}),
\label{eqn:omega}
\end{equation}
where $M_{{\rm eq}}=2.8\times10^{17}M_\odot$ is the horizon mass at the time of matter-radiation equality~\cite{Nakama:2016gzw}. 
The corresponding total PBH abundance is then simply defined as
\begin{equation}
    f_\PBH \equiv \frac{\Omega_{\text{\tiny PBH}}}{\Omega_{\text{\tiny DM}}},
\end{equation}
where $\Omega_{\text{\tiny DM}} = 0.265$.

The mass function $\psi(m_\PBH)$ is defined as the fraction of PBHs with mass in the infinitesimal interval 
$(m_\PBH,m_\PBH +\d m_\PBH)$. This can be obtained by differentiating $\Omega_{\text{\tiny PBH}}$ with respect to the PBH mass as
\begin{equation}
\psi(m_\PBH) 
= 
\frac{1}{\Omega_\PBH} 
\frac{\mathrm{d}\Omega_\PBH}
{\mathrm{d} m_\PBH}.
\label{eqn:massFunction}
\end{equation}
Our definition of the mass distribution implies unit normalisation under integration as
\begin{equation}
   \int {\rm d} m_\PBH \psi(m_\PBH)  =1,
\end{equation}
so that $\psi(m_\PBH)$ has the dimensions of [1/mass].
Notice that an alternative definition of the mass distribution may be given in terms of logarithmic mass intervals.
This is found by computing
\begin{equation}
f(m_\PBH) 
\equiv
\frac{1}{\Omega_\text{\tiny DM}} 
\frac{\mathrm{d}\Omega_\PBH}
{\mathrm{d}\ln m_\PBH}
= m_\PBH f_\PBH \psi(m_\PBH),
\label{eqn:massFunctionlog}
\end{equation}
yielding a dimensionless function.
This alternative quantity will be useful when comparing the mass distribution resulting from our analysis with PBH constraints~\cite{Carr:2020gox}, see Sec.~\ref{implications}.

In order to compute the full mass distribution, it is convenient to invert the relation between horizon and PBH mass through the critical collapse relation~\eqref{critical collapse rel}, focusing only on the type-I branch,  as 
\begin{equation}\label{change_var}
     \delta_l = 
2 \Phi 
\left(1 - 
\sqrt{\Lambda}
     \right),
\end{equation}
where
\begin{equation}
    \Lambda = 
    1-\frac{\delta_c}{\Phi}
     -\frac{1}{\Phi} \left(
     \frac{m_\PBH}{{\cal K}M_H}
     \right)^{1/\gamma }.
\end{equation}
At this point, using Eq.~\eqref{change_var}, we can change variable of integration in Eq.~\eqref{eqn:beta} and write
\begin{align} \label{mfder}
&\psi (m_\PBH) =
\frac{8}{3 \pi\, \Omega_\PBH m_\PBH}
\int\limits_{\ms}^{\ml}
\frac{\mathrm{d} M_{H}}{M_H}
\left( \frac{M_{\rm eq}}{M_{H}} \right)^{1/2}
\lp \frac{\sigma_1}{a H \sigma_0 } \rp  ^3
\nonumber \\
&
\times 
\frac{\Phi ^3 {\cal K}}{\gamma \sigma_0 ^4}
\left(\frac{m_\PBH}{{\cal K} M_H}\right)^{\frac{1+\gamma}{\gamma}}
\frac{\lp 1- \sqrt{\Lambda}\rp ^3}{\Lambda^{1/2}}
\exp\llp
- \frac{2 \Phi^2 }{\sigma_0^2} 
\lp 1-\sqrt{\Lambda}
\rp^2
\rrp ,
\end{align}
and the integration range of $M_H$ is subject to the condition  $\Lambda>0$ (because we require $\delta>\delta_c$). 
The quantities ${\cal K}(M_H)$, $\gamma(M_H)$,
 $\Phi(M_H)$, $\delta_c(M_H)$, and $\sigma_i(M_H)$ are left within the integration over the horizon mass scale, as they all explicitly depend on $M_H$ when thermal effects are included. 
In the low mass limit, i.e. $m_\PBH \ll \ms$, one can find that the mass distribution~\eqref{mfder} scales as
\begin{equation} \label{tail}
    \psi(m_\PBH) \propto
    \lp 
    \frac{m_\PBH}{{\cal K} \ms}
    \rp^{1/\gamma},
\end{equation}
which gives the characteristic tail $\psi(m_\PBH) \propto m_\PBH^{2.8}$ if one assumes the energy density of the Universe behaving as a relativistic fluid with $w=1/3$, which gives $\gamma \approx 0.36$~\cite{Niemeyer:1997mt}.

\begin{figure}[!t]
	\centering
	\includegraphics[width=0.49\textwidth]{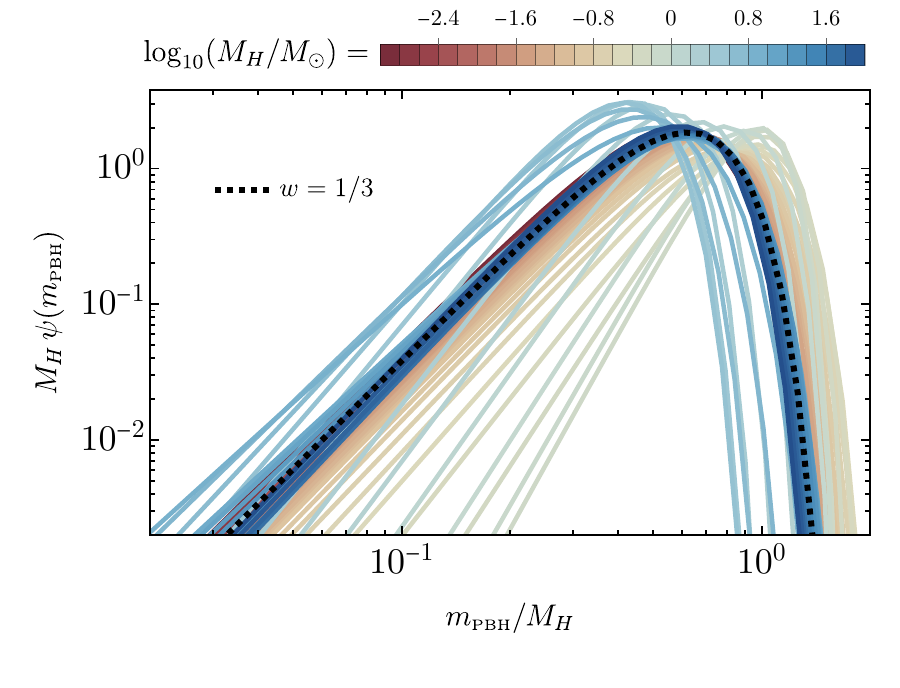}
\caption{
	Mass distribution resulting from the collapse of a single spectral scale $k_H$ crossing the horizon at various $M_H$.
The low mass tail is dictated by the critical collapse scaling $\gamma$, see Eq.~\eqref{tail}.
	The dashed black line denotes the result assuming a Universe with the radiation EoS $w = 1/3$.
 In that case, the peak of the mass function for given $M_H$ contribution would sit at $m_\PBH/M_H =0.602$.
}
\label{fig:MF_critical_collapse}
\end{figure}

In Fig.~\ref{fig:MF_critical_collapse} we show the mass distribution generated by the collapse of a single mode $k_H$. 
Depending on the exact moment of the cosmological horizon crossing, which fix exactly at which epoch across the QCD era the collapse takes place, the consequent mass distribution deviates from the one obtained when the Universe is radiation dominated. 
In particular, we observe differences in the low mass tail and in the location of the peak of $\psi(m_\PBH)$. Modes collapsing before (after) $M_H\approx M_\odot$ tends to generate a mass distribution peaked at larger (smaller) values compared to the reference result of a radiation dominated medium. This generates a pile-up effect around the solar mass, which additionally contributes to enhance the QCD peak induced by the reduced value of the threshold around $M_H \approx M_\odot$. 

To summarize, with ${\cal K}(M_H)$, $\gamma(M_H)$, $\Phi(M_H)$, and $\delta_c(M_H)$ computed from the simulations in Ref.~\cite{Muscoinprep}
and shown in Figs.~\ref{fig:EoS/Threshold} and~\ref{fig:MF_critical_collapse_1}, alongside $\sigma_i(M_H)$ from Eq.~\eqref{eqn:variance}, the algorithm presented above can be applied to compute the PBH mass function and the corresponding total abundance in terms of the parameters of the cosmological power spectrum.

\begin{figure}[!h]
	\centering
	\includegraphics[width=0.495\textwidth]{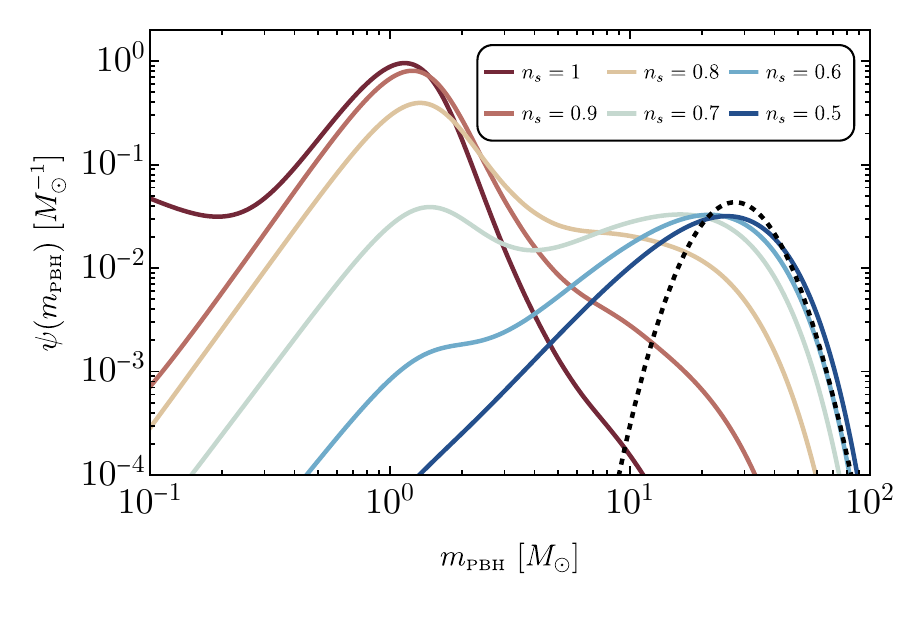}
	\caption{
	The mass function obtained with a few choices of the curvature power spectrum compatible with the
	posterior distribution inferred by the analysis presented in Sec.~\ref{sec:mixedpopinference} (see Table~\ref{TbPBH:posparameters}). This plot assumes $f_\PBH = 10^{-3}$, the minimum horizon mass to be smaller than $\ms \lesssim 10^{-2.5} M_\odot$, the largest mass in the spectrum $\ml = 10^{2.8} M_\odot$ and a variable tilt $n_s$.
	(Only for an exactly scale invariant spectrum $n_s=1$, does the mass function depend on $\ms$, in which case we fix $\ms = 10^{-2.5}M_\odot$.) The black dashed line reports the lognormal mass distribution found as the best fit in the analysis of Ref.~\cite{Franciolini:2021tla}.
	Overall, the ab-initio distribution shaped by the QCD phase transition has larger support for PBHs with $m_\PBH\lesssim 10 M_\odot$ compared to the lognormal parameterization.
	}
\label{fig:MF_full}
\end{figure}

In Fig.~\ref{fig:MF_full}, we show few representative examples of such a mass distribution, obtained by fixing the hyperparameters ${\bm \lambda}_\PBH$ of the curvature power spectrum. 
In particular, we focus the attention on the role of the tilt $n_s$. 
In case $n_s = 1$, the spectrum is sizeable at modes collapsing during the QCD epoch and a bump around the solar mass is obtained~\cite{Byrnes:2018clq}, on top of what is expected from a scale invariant spectrum $\psi (m_\PBH) \approx m_\PBH^{-3/2}$~\cite{DeLuca:2020ioi}. 
On the other hand, already for slightly red spectra (with $n_s<1$), the mass distribution becomes independent of the UV spectrum cut-off $k_\text{\tiny max}$ (i.e., of $\ms$) and  increasingly tilted towards larger masses, up to the point where the QCD enhancement becomes irrelevant, due to the slightly reduced power at the QCD scale, compensating the reduced value of threshold $\delta_c$ with respect $w=1/3$. For intermediate values of $n_s$ (e.g. $n_s\approx 0.7$), a doubly peaked mass distribution can be realised, where the location of the light peak is fixed by the QCD epoch and the heavy one is instead controlled by $\ml$.

We can compare these examples with the best-fit lognormal mass distribution obtained in the analysis of Ref.~\cite{Franciolini:2021tla} (black dashed line in Fig.~\ref{fig:MF_full}). As one can see, red tilted spectra may produce similar mass distributions peaked at around $m_\PBH \approx 30 M_\odot$, for which the QCD softening of the EoS plays no role. The critical collapse, however, generates an asymmetry in the mass distributions that can only be taken into account by introducing additional parameters controlling the skewness of the distribution, as pointed out in Ref.~\cite{Gow:2020cou}. 

\subsection{The semi-analytical mass distribution}\label{sec semianalytical MF}

The computation of the integral~\eqref{mfder}, 
which should be performed on a sufficiently dense grid of values of $m_\PBH$ for each choice of the PBH hyperparameters ${\bm \lambda}_\PBH$, may be rather time consuming, because it requires computing numerically the integrals~\eqref{eqn:variance} at each $M_H$.

In order to simplify the description of the PBH abundance and speed up the hierarchical Bayesian analysis, we absorb the effect of both the window function and linear transfer function, 
which are cutting subhorizon curvature modes, in a single Gaussian window function of the form 
\begin{equation}
    \hat  W(k, \hat R) = \exp\llp-(k \hat  R)^2/4\rrp,
\end{equation}
where the smoothing scale $\hat R$ is fitted appropriately. In particular, $\hat R$ have been adjusted to match the average smoothing between $r_m$ and $r_m/\sqrt{3}$ through the factor $\hat R = sr_m$ with 
\begin{equation}
    s ={( 1+ 1/\sqrt{3})\over 2} \simeq 0.78.
\end{equation}
We checked that this approximation, solely intended to speed up the computation of the mass distribution when running the Monte Carlo Markov Chain analysis, does not introduce any appreciable modification to the mass distribution.

Within this simplifying assumption, one can solve  Eq.~\eqref{eqn:variance} analytically, 
\begin{align}
  &\sigma_0 ^2
  = 
  \frac{4}{9}\Phi^2
  2^{{({n_s}+1)}/{2}}
  A  
  s^{-4}
    (k_\text{\tiny min} {r_m} s )^{1-{n_s}}
    \nonumber \\
  & \times \left[
  \Gamma
  \left(\frac{{n_s}+3}{2},
  \frac{(k_\text{\tiny min} {r_m} s)^2}{2}\right)
  -
  \Gamma \left(\frac{{n_s}+3}{2},
  \frac{(k_\text{\tiny max} {r_m} s)^2}{2}\right)
  \right],
  \label{eq:sigma_analytical}
\end{align}
while $\sigma_1 = k_\text{\tiny min} \sigma_0(n_s\rightarrow n_s+2)$, 
and $\Gamma (a,z) = \int_z^\infty 
    t^{a-1} e^{-t}
    {\rm d} t$ is the incomplete Gamma function.
The variance can be expressed in terms of the model hyperparameters by setting 
\begin{align}
  k_\text{\tiny min} r_m 
  & = \kappa \sqrt{M_H/\ml}\,,
  \nonumber \\
  k_\text{\tiny max} r_m 
  & = \kappa \sqrt{M_H/\ms},
\label{eq:scalingkM}
\end{align}
while the amplitude $A$ is fixed with a bisection method to select the desired PBH abundance $f_\PBH$ (within subpercent accuracy on the latter).

\begin{figure}[t!]
	\centering
	\includegraphics[width=0.49\textwidth]{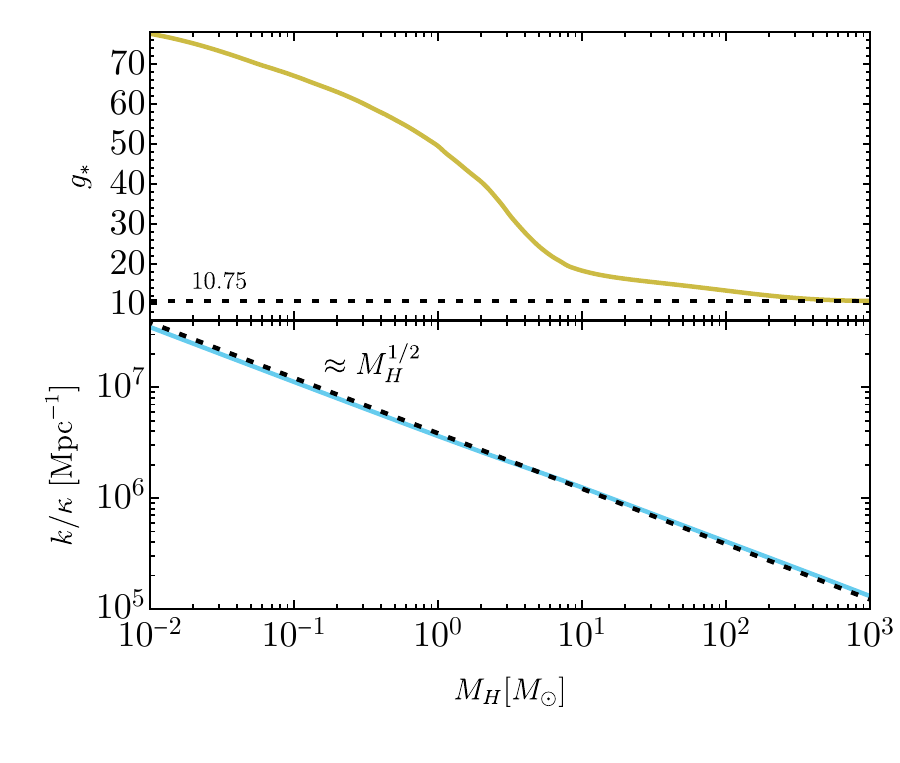}
\caption{
The top plot shows how the effective degrees of freedom $g_*$ varies as a function of $M_H$. 
The bottom plot shows the impact $g_*$ has on the relation between spectral modes and the horizon mass at horizon crossing time. 
}
\label{fig:gstarkH}
\end{figure}

In Eq.~\eqref{eq:scalingkM} we implicitly fixed the number of the effective degrees of freedom $g_*$ appearing in Eq.~\eqref{M-k}, and in Fig.~\ref{fig:gstarkH} (top panel) we show how $g_*$ varies as a function of the temperature of the Universe, that is tracked by $M_H$ in our formalism. This induces a small deviation from the scaling reported in Eq.~\eqref{eq:scalingkM} (see bottom panel of Fig.~\ref{fig:gstarkH}), which we neglect to make the variance fully analytical with the aim of speeding up the computations. 
Therefore, in the following we will fix $g_* = 25$, i.e. the value at the central region of our interest. By fixing $g_*$, we neglect a small running of $n_s$ effectively induced by the change of degrees of freedom when computing the mass distribution.

One final simplification one may attempt is to neglect the 
critical collapse and remove the integration over the horizon mass scale $M_H$. 
This, however, cannot be done consistently as the width of the QCD 
modulation around the solar mass is {\it narrower}
than the one induced by the critical collapse. As it can be seen in Fig.~\ref{fig:MF_full}, the critical mass distribution has a crucial role in shaping the peak of the mass distribution around the solar mass. 
This can be deduced by realising that the mass distributions obtained with different $n_s$ have the same scaling $\approx m_\PBH^{1/\gamma}$ below the QCD peak, induced by the critical collapse of the mode $k$ corresponding to the minimum of the threshold $\delta_c$, while they are basically insensitive to the variance $\sigma_0$ at $M_H<M_\odot$.

\section{The PBH merger rate}\label{sec:PBH merger rate}

The standard PBH formation mechanism we consider assumes PBHs are generated from the collapse of sizable Gaussian cosmological pertubations in the radiation dominated epoch of the early Universe~\citep{Ivanov:1994pa,GarciaBellido:1996qt,Ivanov:1997ia,Blinnikov:2016bxu}. 
In this scenario, PBHs are predicted to be characterised by small natal spins~\citep{DeLuca:2019buf, Mirbabayi:2019uph}, and are not clustered at high redshift~\citep{Ali-Haimoud:2018dau,Desjacques:2018wuu,Ballesteros:2018swv,MoradinezhadDizgah:2019wjf,Inman:2019wvr,DeLuca:2020jug}.
Furthermore, the PBH merger rate at low redshift is dominated by binaries that gravitationally decouple from the Hubble flow before the matter-radiation equality~\cite{Nakamura:1997sm,Ioka:1998nz}.
We compute the differential volumetric PBH merger rate density following~Refs.~\cite{Raidal:2018bbj, Vaskonen:2019jpv,DeLuca:2020jug,DeLuca:2020qqa} as
\begin{align}
\label{eq:diffaccrate}
 \frac{\d  {\cal R}_\PBH}{\d m_1 \d m_2}
& = 
\frac{1.6 \times 10^6}{{\rm Gpc^3 \, yr}} 
f_\PBH^{\frac{53}{37}} 
\lp \frac{t(z)}{t_0} \rp^{-\frac{34}{37}}  
\eta^{-\frac{34}{37}}
\lp \frac{M}{M_\odot} \rp^{-\frac{32}{37}}  
 \nonumber \\
& \times
S(M, f_\PBH, \psi, z)
\psi(m_1) \psi (m_2)
\end{align}
where 
 $M = m_1+m_2$,  $\eta = m_1 m_2/M^2$, and $t_0$ is the current age of the Universe. 
 
 The suppression factor $S<1$  accounts for environmental effects in both the early- and late-time Universe. We can separately define each contribution as
\begin{equation}
S \equiv S_\text{\tiny early}(M, f_\PBH, \psi) \times S_\text{\tiny late}(f_\PBH,z).    
\end{equation}
An analytic expression for $S$ can be found in Ref.~\cite{Hutsi:2020sol}, which we report here for completeness. 
In the early Universe, suppression results as a consequence of interactions between PBH binaries and both the surrounding dark matter inhomogeneities, as well as neighboring PBHs at high redshift~\cite{Eroshenko:2016hmn,Ali-Haimoud:2017rtz,Raidal:2018bbj,Liu:2018ess}. 
This factor takes the form\footnote{
The suppression factor in Eq.~\eqref{S1} was tested against N-body simulations in Ref.~\cite{Raidal:2018bbj},
also assuming a wide (but lognormal) mass distribution. 
While, in this work, we adopt a different mass distribution, derived from first principles, its width in the stellar mass range is compatible with the one tested in Ref.~\cite{Raidal:2018bbj}, supporting our adoption of Eq.~\eqref{S1}.
}
\begin{align}
S_\text{\tiny early}& 
	\thickapprox 
	1.42 \llp \frac{\langle m^2 \rangle/\langle m\rangle^2}{\bar N(y) +C} + \frac{\sigma ^2_\text{\tiny M}}{f^2_\PBH}\rrp ^{-21/74} 
	\exp \llp -  \bar N(y ) \rrp, 
    \label{S1}
\end{align}
with
\begin{align}
\bar N(y) 
\equiv 
\frac{M}{\langle m \rangle }
\lp \frac{f_\PBH}{f_\PBH+ \sigma_\text{\tiny M}} \rp ,
\end{align}
and the rescaled variance of matter density perturbations takes the value $\sigma_\text{\tiny M}\simeq 0.004$.
In Eq.~\eqref{S1}, the constant factor $C$ is defined as (see Eq.~(A.5) of Ref.~\cite{Hutsi:2020sol})
\begin{align}
    C 
     &=  \frac{ f_\PBH^{2}}{\sigma_\text{\tiny M}^{2}} \frac{\langle m^2\rangle}{\langle m\rangle^2}
    \nonumber \\
    & \times \left\{ \left[ \frac{\Gamma(29/37)} {\sqrt{\pi}} U\left(\frac{21}{74},\frac{1}{2},\frac{5f_\PBH^{2}}{6 \sigma_\text{\tiny M}^{2}}\right) \right]^{-\frac{74}{21}} - 1 \right\}^{-1},
\end{align}
where $\Gamma(x)$ is the Euler Gamma function and $U(a,b,z)$ denotes the confluent hypergeometric function.
We warn the reader that we are adopting a different notation for the mass distribution compared to the one used in Ref.~\cite{Hutsi:2020sol}, 
which here is normalised such that $\int \d m \psi (m) = 1$.
 With this choice, the mass average reads
\begin{equation}
\langle m^{n} \rangle = \int m^{n} \psi(m) \d  m.
\end{equation}

In the late Universe, multiple encounters with other PBHs that 
 populate small clusters formed from the initial Poisson conditions lead to a thermalisation of the eccentricity distribution, which enhances the merger time and effectively reduces the late-time universe merger rate~\cite{Jedamzik:2020ypm,Young:2020scc,Jedamzik:2020omx,Trashorras:2020mwn,Tkachev:2020uin}. By accounting for the fraction of binaries which avoids dense enough clusters and are not disrupted, one can write down this additional suppression factor as~\cite{Vaskonen:2019jpv,DeLuca:2020jug,Hutsi:2020sol,link}
\begin{align}
S_\text{\tiny late} (x) & \thickapprox \text{min} \llp 1, 9.6 \cdot 10^{-3} x ^{-0.65} \exp \lp 0.03 \ln^2 x \rp  \rrp,
\end{align}
where we introduced the variable $x \equiv (t(z)/t_0)^{0.44} f_\PBH$.
Notice also that, for $f_\PBH \lesssim 0.003$, one always finds \mbox{$S_\text{\tiny late}\simeq 1$}, i.e. the suppression of the merger rate due to disruption inside PBH clusters is negligible. This is also supported by the results obtained through cosmological N-body simulations finding that PBHs are essentially isolated when their abundance is small enough~\cite{Inman:2019wvr}.

It is important to mention that the late-time suppression factor was only computed for a sufficiently narrow mass distribution~\cite{Vaskonen:2019jpv,DeLuca:2020jug,Hutsi:2020sol}. 
So far, a full computation considering wide distributions was not performed in the literature. However, we do not expect this extension to modify significantly the formulation used here as $S_\text{\tiny late}$ is found to be only mildly dependent on the mass scale (see e.g.~\cite{Franciolini:2022htd}). 
Furthermore, we generically expect $S_\text{\tiny late}\approx1$ for the PBH abundance inferred a posteriori by our analysis.
In the computation of the merger rate, we are also neglecting the contribution from binaries that can form dynamically within PBH clusters from either capture or three-body interactions.
This is justified because, in this scenario and for the small values of  $f_\PBH$ we obtain, the contribution of those channels to the total merger rate is subdominant relative to the early universe binaries~\cite{Franciolini:2022ewd}.

Finally, the natal distribution of PBH masses and spins (the latter being initially negligible~\cite{DeLuca:2019buf, Mirbabayi:2019uph}, see also~\cite{Koga:2022bij}) can be modified if PBHs undergo an efficient accretion phase during cosmic history~\cite{DeLuca:2020bjf,DeLuca:2020fpg,DeLuca:2020qqa}.
For a given accretion model, the peculiar accretion-driven and redshift-dependent mass-spin distribution can be used to add extra information in the inference~\cite{Franciolini:2021tla,Franciolini:2021xbq,Franciolini:2022iaa} and also impact the merger rate~\cite{DeLuca:2020bjf,DeLuca:2020fpg,DeLuca:2020qqa}.
However, while certain features of PBH accretion are robust and should be model-independent, there remain large uncertainties in the mass (and, especially, spin) accretion. Thus, in order to remain agnostic and conservative, here we neglect PBH accretion and do not include spin information in the merger events. In practice, in the inference we shall only use the dependence of the merger rate on the individual masses and redshift, conservatively limiting the information that can be inferred from single merger events~\cite{Franciolini:2021xbq,Franciolini:2022iaa}.

Let us conclude by stressing that, while in this work we compute the mass distribution from first principles across the QCD era as described in Sec.~\ref{sec:PBHMFQCD}, thus going beyond the  parametrization often used in the literature, certain characteristics of the PBH model are general and arise from the form of the PBH merger rate in Eq.~\eqref{eq:diffaccrate}. 
These features are the monotonic merger rate evolution with redshift, ${\cal R}_\PBH \approx (t/t_0)^{-34/37}$, 
a scaling of the merger rate with the PBH abundance,
\begin{align}
	{\cal R}_\PBH
	\propto
	\begin{cases}
f_\PBH^{2/3} &\text{for}\quad f_\PBH\gtrsim 10^{-3},
		\\
f_\PBH^{2} &\text{for}\quad f_\PBH \lesssim 10^{-3},
	\end{cases}
\end{align}
 a lack of preference towards symmetric mass ratios enforced by the term $\eta^{-34/47}$, and an exponential suppression of heavy (i.e. $M \gg \langle m \rangle$) mergers due to the suppression factor~\eqref{S1}.

\section{LVK phenomenological models}

In the following section, we will compare and mix our ab-initio PBH channel with phenomenological models used by the LVK Collaboration to fit the 
BH and NS binary events in the GWTC-3 catalog. This approach is very conservative, because we choose to confront the PBH scenario with the best working model specifically tailored to describe the coarse-grained properties of the observed merger population. 
As such, this approach is not meant to be used to \emph{search for} a subpopulation of PBHs in the data, but rather to place an \emph{upper bound} on the PBH abundance compatible with the data and to assess whether certain events are more likely ascribed to a putative primordial channel.
As we will discuss in the conclusions, one natural extension of this analysis would entail considering ab-initio astrophysical models, as attempted in Ref.~\cite{Franciolini:2021tla}. 

According to the LVK prescription, compact objects with masses below $3M_\odot$ are labelled as NSs, whereas heavier objects are labelled as BHs. Two different mass distributions are used to describe mergers of these families, as discussed below.
However, at variance with the LVK analysis, we shall adopt a more 
agnostic approach and allow for the light events to be BH binaries (of primordial origin), with the exception of GW170817~\cite{LIGOScientific:2017vwq} for which 
sufficient evidence for the interpretation as a NS binary was gathered with the observation of an electromagnetic counterpart compatible with a NS merger~\cite{LIGOScientific:2017ync}.

In the standard scenario, PBH mergers at low redshift are due to binaries that had gravitationally decoupled from the Hubble flow before the matter-radiation equality~\cite{Nakamura:1997sm,Ioka:1998nz}, i.e. much before the first stars were born.
Thus, ``mixed'' binaries formed by an isolated PBH and either an astrophysical-origin BH or a NS can be assembled only through dynamical capture, e.g. in dense clusters. The probability of forming these binaries is very low~\cite{Kritos:2020wcl,Sasaki:2021iuc} and we shall neglect such possibility. In other words, we shall assume that all primordial binaries are formed by two PBHs and that all astrophysical-origin binaries are formed by astrophysical BHs and/or NSs.

{
\renewcommand{\arraystretch}{1.4}
\setlength{\tabcolsep}{4pt}
\begin{table*}[t!]
\caption{Population hyperparameters ${\bm \lambda}$ 
for the ABH and NS models considered in this work, along with their prior distributions. We refer to a uniform distribution between two values $\theta_\text{\tiny min }$ and $\theta_\text{\tiny max}$ as $[\theta_\text{\tiny min },\theta_\text{\tiny max}]$.
Rates $({\cal R}^0)$ are reported in units of $[{\rm yr^{-1} Gpc^{-3}}]$ while ($m,\mu,\sigma$) are written in units of $[M_\odot]$.} 
\begin{tabularx}{2.05 \columnwidth}{|X|c|c|c|c|c|c|c|c|c|c|c|c|c|}
\hline
  \hline
  Model &
  \multicolumn{8}{c|}{ABH}   &
  \multicolumn{5}{c|}{NS}   
  \\
  \hline
${\bm \lambda}$ &
$\log_{10}{\cal R}^0_\ABH$ &
$\alpha$             &
$\beta$              &
$m_\text{\tiny min}$ &
$m_\text{\tiny max}$ &
$\lambda_\text{\tiny peak}$ &
$\mu_m$  &
$\sigma_m$ &
$\log_{10}{\cal R}^0_\NS$ &
$m_\text{\tiny min}^\NS$ &
$m_\text{\tiny max}^\NS$ &
$\mu_m^\NS$ &
$\sigma_m^\NS$
  \\
  \hline
  Prior &
  [-3,3]&
  [0,5] &
  [0,7] &
  [3,10]&
  [30,100]&
  [0,1]&
  [20,50]&
  [1,10] &
  [-1,5]&
  [1,1.5] &
  [1.5,3] &
  [1,3] &
  [0.01,2] \\
 \hline
  \hline
\end{tabularx}
\label{tab:priors_ABH_NS}
\end{table*}
}

\subsection{Astrophysical BH binaries}
We describe the merger rate of astrophysical BH binaries with the reference population model called \textsc{Power Law + Peak}~\cite{Talbot:2018cva} adopted by the recent LVK population analyses (see e.g. Ref.~\cite{2021arXiv211103634T}). 
Henceforth we shall refer to this as the ``astrophysical'' BH (ABH) population, although it should be kept in mind that the model is phenomenological and not based on ab-initio astrophysical simulations.
The ABH model assumes that the distribution of primary binary BH mass $m_1$ is described by a mixture of a power law model,
 \begin{equation}
     P_\ABH(m_1|\lambda,m _\text{\tiny min},m_\text{\tiny max}) \propto 
     m_1^{-\alpha}
 \end{equation} 
 and a Gaussian peak,
 \begin{equation}
 N_\ABH(m_1|\mu_m,\sigma_m,m _\text{\tiny min},m_\text{\tiny max}) \propto \exp\left[-\frac{(m_1-\mu_m)^2}{2\sigma_m^2}\right],    
 \end{equation}
 normalized to unity across the range $ m _\text{\tiny min} \leq m_1 \leq m_\text{\tiny max}$. 
 The mixing fraction between the two components is dictated by $\lambda_\text{\tiny peak}$ as
\begin{align}
    p^{m_1}_\ABH (m_1) 
    &=
    (1-\lambda_\text{\tiny peak})
    \,P_\ABH(m_1) 
    + \lambda_\text{\tiny peak} N_\ABH(m_1).
    \label{eq:pm}
\end{align}
We describe the distribution of mass ratio via a power law as
    \begin{equation}
    p^{m_2}_\ABH (q|m_1,\beta) \propto q^{\beta},
    \label{eq:pq}
    \end{equation}
constrained within the range $m_\text{\tiny min}/m_1 \leq q \leq 1$.
For simplicity, we do not introduce the term $S_\ABH(m|\delta_m)$ adopted in LVK analyses to smooth the sharp cutoff below $m_\text{\tiny min}$.
Finally, the evolution of the merger rate at high redshift follows
\begin{equation}
    p^z_\ABH(z|\kappa) 
    \propto 
 \left(1+z\right)^\kappa.
    \label{eq:pz}
\end{equation}
The number of events can be found by integrating the merger rate density with the additional factor of ${dV_c}/{dz}/(1+z)$. 
Since observations are limited to small redshift, the merger rate evolution 
is still rather poorly constrained. In order to simplify the analysis, we fix the power law evolution of the astrophysical phenomenological model to the best-fit value of $\kappa = 2.9$~\cite{2021arXiv211103634T}.
To summarise, we write the differential merger rate density of ABH as
\begin{equation}
\frac{\d  {\cal R}_\ABH}{\d m_1 \d m_2}
=
{\cal R}_\ABH^0 p^z_\ABH(z)
p^{m_1}_\ABH (m_1)
p^{m_2}_\ABH (m_2|m_1),
\end{equation}
and the hyperparameters of the ABH model are
\begin{equation}
     {\bm \lambda}_\ABH 
     =
     [\log_{10}{\cal R}^0_\ABH,
    \lambda_\text{\tiny peak},
     \alpha,\beta,
     m_\text{\tiny min},
     m_\text{\tiny max},
     \mu_m,\sigma_m
     ],
\end{equation}
where we introduced the quantity ${\cal R}^0_\ABH \equiv {\cal R}_\ABH (z=0)$ controlling the present-day ABH merger rate density. 
In Table~\ref{tab:priors_ABH_NS} we report the prior ranges for the ABH model parameters adopted in the following Bayesian analysis.

While evidence of additional features on top of the \textsc{Power Law + Peak} coarse grained-model was found by the LVK Collaboration~\cite{2021arXiv211103634T} (see also Refs.~\cite{Callister:2021fpo,Tiwari:2020otp,Edelman:2021zkw,Tiwari:2021yvr,Li:2022jge,Franciolini:2022iaa,Biscoveanu:2022qac}) we do not expect our results --~especially the upper bound on $f_\PBH$~-- to be affected by potential systematic effects in our choice of benchmark mass model.

\subsection{Binaries involving NSs}
Following the LVK population analysis, we model the distribution of NSs as an underlying Gaussian mass distribution that is common to all NSs, with random pairing into compact binaries. 
For mixed NSBH mergers, the BH mass distribution is fixed to be uniform between $[3\divisionsymbol 60 ]M_\odot$. 
 The joint mass distribution takes the form
 \begin{equation}
p_\NS(m_1,m_2) =
\begin{cases}
N_\NS(m_1) N_\NS(m_2), \\
U(m_1,[3 M_\odot , 60 M_\odot]) N_\NS(m_2),
\end{cases}
\end{equation}
for NS and mixed NSBH binaries, respectively, 
where the Gaussian peak is defined as
\begin{equation}
N_\NS(m|\mu_m^\NS,\sigma_m^\NS,m _\text{\tiny min}^\NS,m_\text{\tiny max}^\NS) \propto \exp\left[-\frac{(m-\mu_m^\NS)^2}{2(\sigma_m^\NS)^2}\right]
,
 \end{equation}
normalized to unity across the range 
$ m_\text{\tiny min}^\NS \leq m \leq m_\text{\tiny max}^\NS$.
We assume the redshift evolution of the merger rate for this channel follows the same behaviour of the ABH model, namely ${\cal R}_\NS(z) \approx {\cal R}_\NS^0 (1+z)^{2.9}$. 
This evolution is, however, practically  irrelevant, as light mergers are currently observable only at $z \approx 0$.
Finally, we can write the differential merger rate as
\begin{equation}
\frac{\d  {\cal R}_\NS}{\d m_1 \d m_2}
=
{\cal R}_\NS^0 p^z_\NS(z)
p_\NS(m_1,m_2),
\end{equation}
and the hyperparameters of the NS model are (see also Table~\ref{tab:priors_ABH_NS})
\begin{equation}\label{parns}
     {\bm \lambda}_\NS 
     =
     [\log_{10}{\cal R}^0_\NS,
     m_\text{\tiny min}^\NS,
     m_\text{\tiny max}^\NS,
     \mu_m^\NS,\sigma_m^\NS].
\end{equation}
One may also consider splitting the rate of BNS and NSBH binaries, thus introducing an additional parameter in Eq.~\eqref{parns}.
However, due to the small number of detections with at least one component lighter than $3 M_\odot$,
merger rate densities remain affected by large uncertainties in the light sector of the catalog \cite{2021arXiv211103634T}, and 
both contributions are broadly compatible with each other.



\section{Analysis setup}
In this section we summarise the statistical framework we use to perform the analysis and model comparison (see e.g.~\cite{Mandel:2018mve,Vitale:2020aaz}), alongside our event selection within the GWTC-3 dataset~\cite{LIGOScientific:2021djp}. 

\subsection{Hierarchical Bayesian inference}

The aim of the hierarchical Bayesian inference is to produce posterior distributions for the hyperparameters of a model ${\cal M}$ which is assumed to explain the GW dataset, alongside the corresponding evidence $Z_{\cal M}$ allowing for statistical model comparisons. 
The LVK Collaboration's Gravitational Wave Open Science Center~\cite{Vallisneri:2014vxa,GWOSCref} releases the output of the parameter estimation performed on each GW signal as a collection of posterior distributions for the parameters describing the properties of each individual merger.
We denote this as {\it event posteriors} 
$p({\bm \theta}|{\bm d}_i)$, 
where ${\bm \theta}$ indicates the binary {\it event parameters}. The index $i$ runs over all the detected GW events while, in our analysis, we restrict the set of intrinsic binary parameters to  
${\bm \theta} = (m_1,m_2,z)$.

We compute the number of GW events produced in a given model within the observation time as 
\begin{equation}
    N(\bm \lambda)\equiv 
    \int 
    \d {\bm \theta} 
     N_\text{\tiny pop}( \bm{\theta}| \bm \lambda)
=
    T_\text{\tiny obs} R({\bm \lambda})
    \int 
    \d {\bm \theta} 
    p_\text{\tiny pop}(\bm{\theta}|\bm{\lambda})
    ,
\end{equation} 
where $R({\bm \lambda})$ is the intrinsic merger rate,  $p_\text{\tiny pop}(\bm{\theta}|\bm{\lambda})$ is the \textit{population likelihood}, corresponding to the distribution of event parameters for the model ${\cal M}$ characterised by hyperparameters $\bm{\lambda}$, and $T_\text{\tiny obs}$ is the duration of the various LVK observing runs.

One can account for the selection effects induced by the finite sensitivity of the detectors 
by introducing the {\it observable} number of events
\begin{align}
N_\text{\tiny det}(\bm \lambda)\equiv \alpha(\bm \lambda) 
{{N(\bm \lambda)}{}
}\,,
\label{eq:selectionFunction}
\end{align}
where the selection bias parametrized by $\alpha(\bm \lambda) \leq 1 $ will be discussed in the next subsection.

Given a vector of hyperparameters $\bm{\lambda}$ (or {\it population parameters}) describing the model ${\cal M}$, 
the  posterior distribution inferred from the data is
\begin{align}
\frac{p({\bm \lambda}|{\bm d})}{\pi({\bm \lambda}) }
\propto
e^{- N_\text{\tiny det} ({\bm \lambda})} 
N({\bm \lambda})^{N_\text{\tiny obs}}
\prod_{i=1}^{N_\text{\tiny obs}} 
\int {\rm d}{\bm \theta_i} 
\frac{p({\bm \theta_i}|{\bm d})  p_\text{\tiny pop}({\bm \theta_i}|{\bm \lambda})}{\pi({\bm \theta_i})} \,,
\label{eq:posterior}
\end{align}
where the prefactor introduces the standard terms describing the statistics of an inhomogeneous Poisson process (see e.g. Refs.~\citep{2004AIPC..735..195L,2018PhRvD..98h3017T,Mandel:2018mve,Thrane:2018qnx} for detailed derivations),
$\pi({\bm \lambda})$ is the prior distribution assumed for the model hyperparameters, and  $\pi ({\bm \theta}_i)$ is the prior distribution over the intrinsic parameters adopted by the LVK Collaboration when performing the parameter estimation for each individual event.
The factor $\pi ({\bm \theta}_i)$ in the denominator removes the dependence of the analysis on the priors adopted by LVK Collaboration to perform parameter estimation, which was shown to potentially affect the interpretation of individual events \cite{Vitale:2017cfs,Zevin:2020gxf} (see also 
Ref.~\cite{Bhagwat:2020bzh} for this analysis with PBH informed priors).

In order to speed up the evaluation of Eq.~\eqref{eq:posterior}, the integral is performed  using importance sampling, i.e. by computing the expectation value of the prior-reweighted population likelihood as a discrete sum over the samples of the event posteriors. 
In practice, this can be equivalently written as
\begin{align}
\frac{p({\bm \lambda}|{\bm d})}{\pi({\bm \lambda}) }
 \propto 
e^{- N({\bm \lambda}) \alpha({\bm \lambda})}
\prod_{i=1}^{N_\text{\tiny obs}}\frac{1}{{\cal S}_i}\sum_{j=1}^{{\cal S}_i} 
\frac{N_\text{\tiny pop}(^j\bm{\theta}_i|\bm{\lambda})}{\pi(^j\bm{\theta}_i)},
\label{eq:populationPosterior_discrete}
\end{align}
where $j$ labels the $j$-th sample of the $i$-th event, and ${\cal S}_i$ identifies the length of the $i$-th posterior.
We sample Eq.~\eqref{eq:populationPosterior_discrete} using the MCMC package \texttt{emcee}~\cite{Foreman-Mackey:2012any}.

Given a model ${\cal M}$, the evidence $Z_{\cal M} $ is defined as the marginal population likelihood. This is found by performing the integral of the population posterior
\begin{equation}
Z_{\cal M} \equiv \int \d {\bm \lambda} \, p(\bm{\lambda}|\bm{d}).
\end{equation}
We compute the evidence for each model from the posterior data following Ref.~\cite{NR1994}.
One can then compare different models by computing the so-called Bayes factors, defined as 
\begin{equation}
{\cal B}^{{\cal M}_1}_{{\cal M}_2} \equiv \frac{ Z_{{\cal M}_1}}{Z_{{\cal M}_2}}.
\end{equation}
According to Jeffreys' scale criterion~\cite{Jeffreys}, a Bayes factor larger than $(10,10^{1.5},10^2)$ would imply a strong, very strong, or decisive evidence in favour of model ${\cal M}_1$ with respect to model ${\cal M}_2$ given the available dataset.

\subsection{Selection bias}\label{sec:selectionbias}

One of the most time consuming tasks when evaluating the likelihood function in Eq.~\eqref{eq:populationPosterior_discrete} is the computation of the selection bias $\alpha({\bm \lambda})$, quantifying the fraction of observable events in model ${\cal M}$ characterised by the hyperparameters  ${\bm \lambda}$.
Following recent work (see e.g. Ref.~\cite{Zevin:2020gbd}), we estimate the selection bias by computing the SNR for LIGO Hanford, LIGO Livingston and Virgo operating at \texttt{midhighlatelow} sensitivity~\cite{Aasi:2013wya} while adopting the \texttt{IMRPhenomPv2} waveform approximant~\cite{Hannam:2013oca,Khan:2015jqa} built in the PyCBC package~\cite{alex_nitz_2019_3546372}.
The network SNR threshold for detection is set by requiring the quadrature sum of the SNRs from the three detectors to be above \mbox{$\rho_\text{\tiny th}$ = 10}, a value which is consistent with the false-alarm-rate threshold used as a detection criterion for events in LVK searches~\cite{Aasi:2013wya}. 

Analogously to what is done in the LVK analyses, we speed up the computation of the observable number of events [Eq.~\eqref{eq:selectionFunction}] by building an injection which covers all the parameter space reached by the models we consider (which is larger than the injection released by the LVK Collaboration). 
We select successfully found injections (i.e., SNR$>\rho_\text{\tiny th}$) 
and reweight to the population with hyperparameters ${\bm \lambda}$ as 
\begin{equation}
\alpha({\bm \lambda}) = 
\frac{1}{N_\text{\tiny inj}} 
\sum_{j=1} ^{N_\text{\tiny found}}
\frac{p_\text{\tiny pop}({{\bm \theta}_j}|{\bm \lambda})}
 {p_\text{\tiny inj}({{\bm \theta}_j})}.
    \label{eq:sel-effects}
\end{equation}
In the previous step, we introduced $N_\text{\tiny found}$ as the number of recovered events, $N_\text{\tiny inj}$ as the total number of injections (including those that are not observable with low SNR) and $p_\text{\tiny inj}(\theta)$ as the reference distribution from which injections were built.
In particular, the injected distribution of masses follows $p_\text{\tiny inj}(m_1) \propto m_1^{-2.35}$ for $0.1 \,M_\odot \leq m_1 \leq 500\,M_\odot$ and $p_\text{\tiny inj}(q|m_1) \propto q^2$,
$ p_\text{\tiny inj} (z) \propto \left(1+z\right)^{2-1} {dV_c}/{dz}$, and again we neglect the binary spins.
In order to efficiently cover the wide mass range, we split the injection in two parts with primary mass below and above $5 M_\odot$. 
We analyse events for the latter region injecting a population up to redshift $z\leq 2$.
The light events, given the much smaller detection horizon, are injected with a redshift distribution extending up to redshift $z\leq 0.2$.

When computing the expected number of events during the future O4/O5 observing runs, we adopt the same framework presented here but with updated LIGO and Virgo future sensitivity curves from Ref.~\cite{O4sens}.

\subsection{The GWTC-3 dataset}

Out of the $\approx 90$ GW detection candidates found by the first three LVK observing runs, here
we use the same subset of confident detections selected for the GWTC-3 population analysis in Ref.~\cite{2021arXiv211103634T}. 
Following this choice, the GWTC-3 dataset contains 69 binary BH events and 7 potential NS-involving binaries (which are characterised by at least one object with mass below $3 M_\odot$.

It is particularly important to include light events in our analysis due to the potential PBH contribution to light binary components, in particular in the solar-mass range mostly affected by the QCD phase. 
This implies, in particular, that we do include the light events GW170817,  GW190425,  GW190426\_152155,  GW190814, 
GW190917\_114630,  GW200105\_162426,  GW200115\_042309, out of which only the first one is confidently regarded as a NS binary due to the observation of the electromagnetic counterpart~\cite{LIGOScientific:2017ync}.
We do not consider the additional candidate events found by independent searches performed outside the LVK Collaboration (e.g.~\cite{Nitz:2021zwj,Olsen:2022pin}), and leave such a task for future work. 

We adopt the \texttt{Overall\_posterior} samples provided in Ref.~\cite{PEresult_GWTC1} for the $11$ considered events from the GWTC-1 catalog, the \texttt{PrecessingSpinIMRHM} posteriors provided in Refs.~\cite{PEresult_GWTC2} and~\cite{ligo_scientific_collaboration_and_virgo_2021_5117703} for events in the GWTC-2 and GWTC-2.1 catalogs, respectively, while we adopt the \texttt{C01:Mixed} samples for the O3b events reported in the GWTC-3 dataset~\cite{ligo_scientific_collaboration_and_virgo_2021_5546663}.

\section{Constraints on PBHs from GWTC-3}\label{sec:mixedpopinference}

In this section we report the results of the Bayesian inference analyses of GWTC-3 data, assuming either the astrophysical phenomenological models or PBHs (or a mixture of both) are generating mergers of binary BHs and NSs. While it was already shown that PBHs alone are not able to explain all the features observed in the recent GW catalogs~\cite{Hall:2020daa,Hutsi:2020sol,DeLuca:2021wjr,Franciolini:2021tla} assuming a lognormal mass distribution, repeating such simplified analysis is useful to confirm this conclusion remains valid also in a first-principle model including the effects of the QCD phase transition. 

\subsection{Single-population inference}\label{sec:singlepopinference}

We start by discussing the inference on each model separately, focusing on the two subsets of the total events divided by the condition $m_2 \lessgtr 3 M_\odot$ (which are defined by the LVK Collaboration as events containing NS components or not). 

\begin{figure}[t!]
	\centering
	\includegraphics[width=0.49\textwidth]{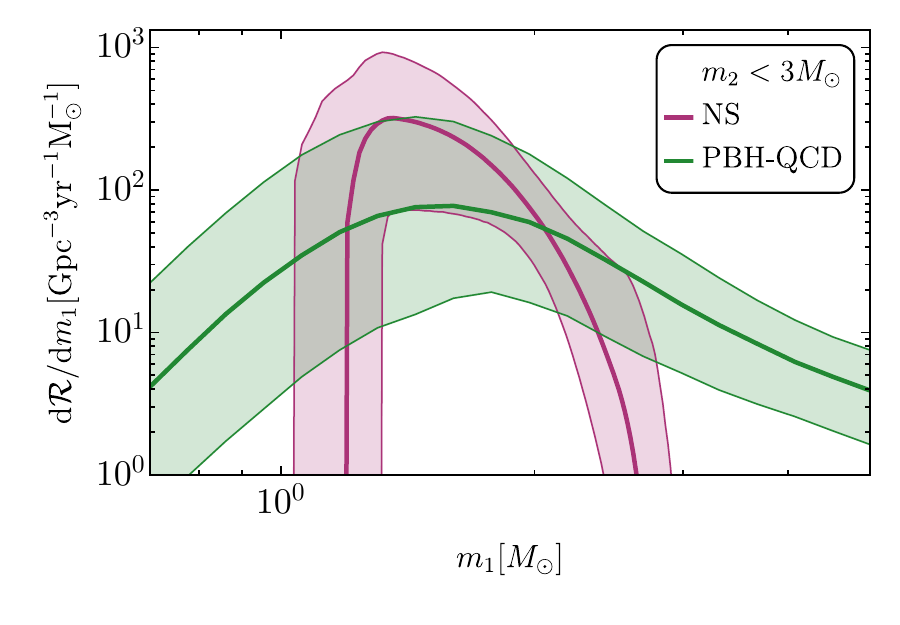}
	\caption{NS and PBH merger rate distributions as a function of primary mass as inferred from a subset of the GWTC-3 catalog ($m_2<3 M_\odot$) assuming either the NS phenomenological model or the ab-initio PBH model with QCD effects and including GW190814.}
\label{fig:NS pop}
\end{figure}

Focusing on the light set of GWTC-3 events, in Fig.~\ref{fig:NS pop} we show the inferred merger rate distribution, where at least one of the binary components has mass smaller than $3M_\odot$. We either assume the NS phenomenological model or the PBH model.  In the latter case, we do not include the binary NS event GW170817.
We report the corresponding posterior distributions in Appendix~\ref{app:posteriors}.

As previously discussed, the probability of a binary formed by only one PBH is very low, so we neglect this possibility. This implies that, for a highly asymmetric binary with $m_2<3M_\odot$ and $m_1\gg 3 M_\odot$ (like, e.g., GW190814~\cite{Abbott:2020khf}), if the secondary is identified as a PBH then also the primary should be. This is not the case for the astrophysical channels, where the secondary is naturally identified as a NS and the primary as an ABH. This difference explains why the best-fit PBH merger rate distribution in Fig.~\ref{fig:NS pop} has support at larger masses compared to the NS case, although they both peak when $1\lesssim m_1/M_\odot\lesssim 2$, driven by the events with $m_{1,2}<3 M_\odot$ commonly identified as NS binaries. 
 As a consequence, one falsifiable prediction that follows from the interpretation of GW190814 as a PBH binary is the generation of events filling the lower mass gap potentially existing in the ABH sector between the heaviest allowed NS mass (see e.g. Ref.~\cite{Chatziioannou:2020pqz} for a review) and the lightest BH observed  \cite{1998ApJ...499..367B,2010ApJ...725.1918O,2011ApJ...741..103F} (see also \cite{Farah:2021qom}). 
Furthermore, due the features of the ab-initio PBH mass distribution previously discussed, the PBH merger rate is broader than in the NS case and inevitably has a nonnegligible support also in the subsolar range induced by the critical collapse tail. As we shall see, this is a general feature of the model that allows making predictions on subsolar mergers in the PBH scenario.

\begin{figure*}[t!]
	\centering
	\includegraphics[width=0.49\textwidth]{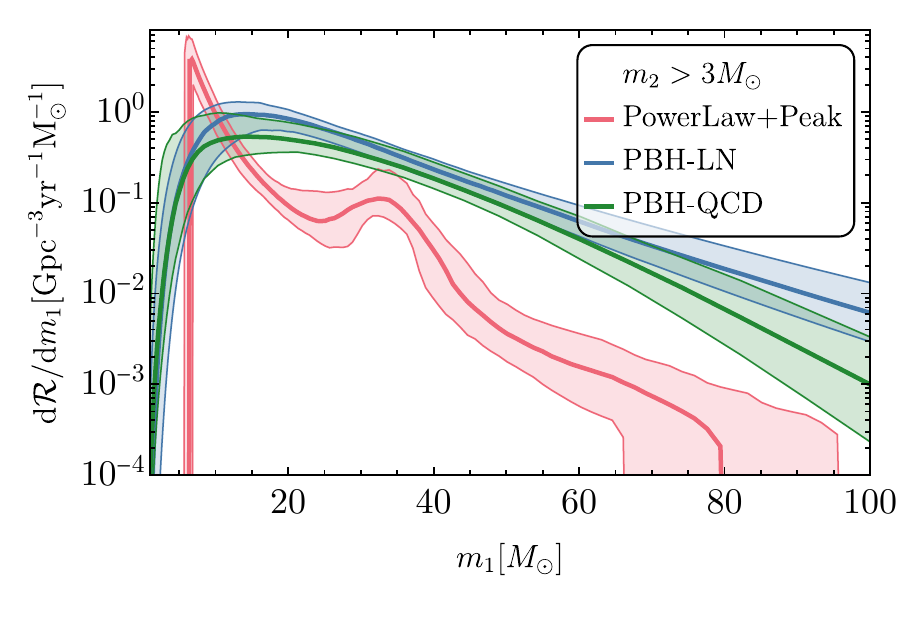}
	\includegraphics[width=0.49\textwidth]{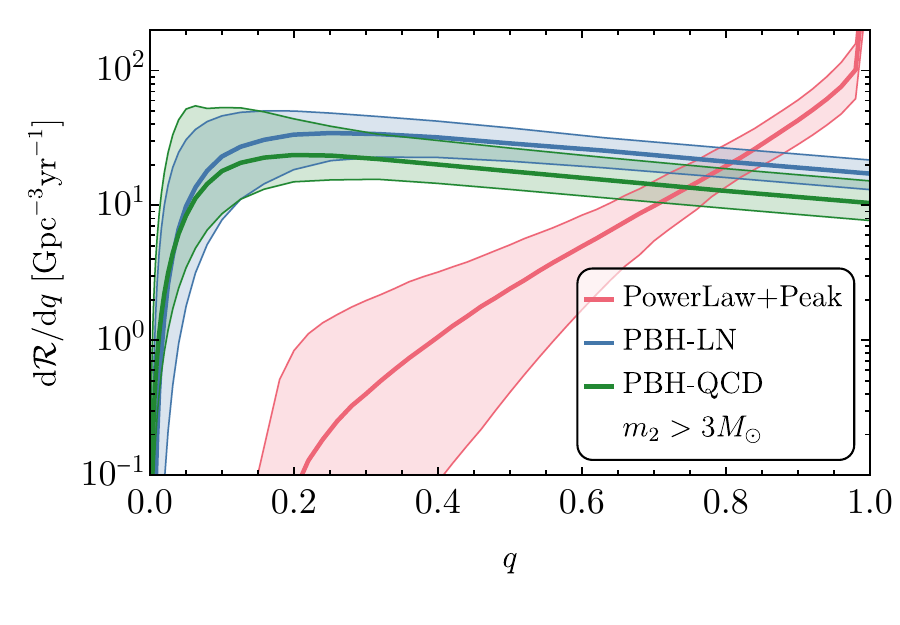}
	\caption{
	BH merger rates as a function of the primary mass (left panel) or of the mass ratio (right panel) as inferred from a subset of the GWTC-3 catalog ($m_2>3 M_\odot$) and assuming only a single binary BH population, either described by the ABH phenomenological model (red), the PBH model assuming a lognormal mass distribution (blue), or the ab-initio
	PBH model with a mass distribution fixed by the curvature spectrum in Eq.~\eqref{PS_zeta} and modulated by the QCD phase (green).
}
\label{fig:single pops}
\end{figure*}

In Fig.~\ref{fig:single pops}, we show the merger rate distribution inferred using only the heavy GWTC-3 events, where both binary components have mass larger than $3M_\odot$ and are therefore identified as BHs. We assume a single binary BH population, either described by the phenomenological ABH model (red) or by our ab-initio PBH model (green), or also by a phenomenological PBH model using a lognormal mass distribution (blue) often used in the literature (see, e.g.,~\cite{Hall:2020daa,Hutsi:2020sol,DeLuca:2021wjr,Franciolini:2021tla}) and shown here for comparison.
More details on these are given in Appendix~\ref{app:posteriors}.
Interestingly, in this case we observe that the two PBH models yield fairly similar distributions both for the primary mass (left panel) and mass ratio (right panel).
This is because the effects of the power spectrum and QCD phase are largely washed out by the absence of detections with masses below $\approx 6 M_\odot$ in this subset of events and some universal properties of the merger rate in Eq.~\eqref{eq:diffaccrate}, which make the final result largely independent of the details of the two specific parametrization of the PBH mass distribution.

Figure~\ref{fig:single pops} confirms previous results (e.g.,~\cite{Hall:2020daa,Hutsi:2020sol,DeLuca:2021wjr,Franciolini:2021tla}) finding that the PBH merger rate distribution is markedly different from the ABH one, in particular it lacks a double peak in the mass distribution, it predicts a larger merger rate at high masses, and (in the absence of accretion~\cite{Franciolini:2021tla,Franciolini:2021xbq,Franciolini:2022iaa,DeLuca:2020bjf,DeLuca:2020fpg,DeLuca:2020qqa}) it predicts a broader merger-rate distribution as a function of the mass ratio which does not favour $q = 1$.
The Bayes factors strongly disfavour the interpretation of the totality of the events as coming from the PBH channel alone. In particular, we find
$
\log_{10}{\cal B}^\ABH_\text{\tiny PBH,LN}
= 18.5
$ and 
$
\log_{10}{\cal B}^\ABH_\text{\tiny PBH,QCD}
= 17.0.
$
The value obtained for with a lognormal mass distribution is consistent with what estimated in previous analysis \cite{Hall:2020daa,DeLuca:2021wjr}, but scaled considering the larger statistical sample available with the newest GWTC-3 catalog.

\begin{figure*}[t!]
	\centering
	\includegraphics[width=0.99\textwidth]{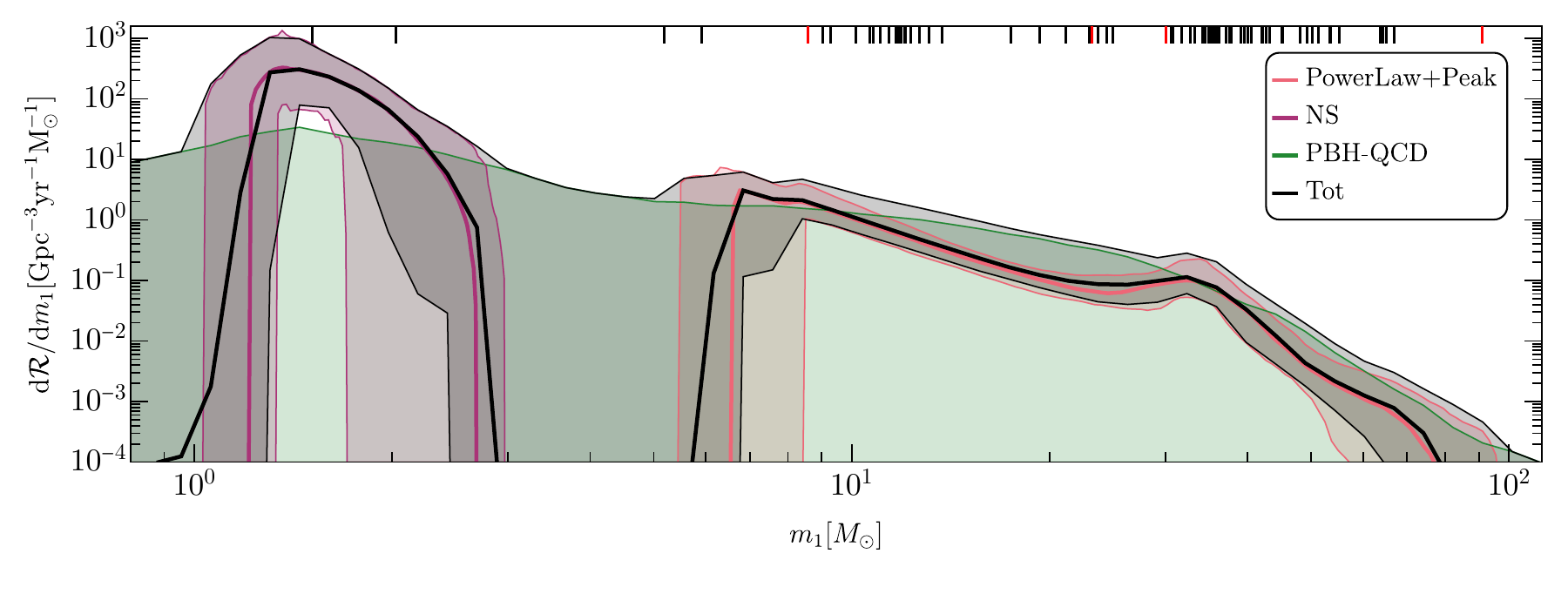}
	\caption{
	Merger rate distribution as a function of the primary mass for the GWTC-3 population inference and including contribution from three channels: ABH phenomenological model, NS phenomenological model, and ab-initio PBH model modulated by the QCD phase.
	Events in the lower mass gap (e.g., GW190814) are more naturally interpreted as PBHs rather than being included in the ABH or NS phenomenological channels. This is where the black line mostly deviates from the median distribution of the NS/BBH astrophysical channels.
	Each black ticks at the top of the frame indicate the median values for the primary mass of each GWTC-3 event. In red, we highlight those with nonnegligible probability (i.e. $>5\%$) of being of primordial origin in our analysis, see Table~\ref{Tb:parameters}. }
\label{fig:mixed NS BBH pops}
\end{figure*}

\subsection{Multi-population inference} \label{sec:multipop}
Let us now move to population inferences assuming multiple channels. The corresponding posterior corner plots are presented in Appendix~\ref{app:posteriors}.

In Fig.~\ref{fig:mixed NS BBH pops}, we show the (differential) merger rate distribution as a function of the primary mass for the entire GWTC-3 catalog allowing contributions from three channels: the LVK ABH and NS phenomenological models, and the ab-initio PBH model originating from the curvature power spectrum and modulated by the QCD phase transition.
The most striking feature of this plot is the fact that the PBH distribution can cover the entire mass range, from subsolar ($m_1\lesssim M_\odot$) to intermediate mass ($m_1\gtrsim 100 M_\odot$), with a large support also in the lower mass gap ($3\lesssim m_1/M_\odot\lesssim5$) which is instead avoided by the ABH and NS distributions. Note that this last property is nontrivial, since the ABH and NS models are phenomenological and not informed by astrophysical priors, so a priori there is no constraint preventing the best-fit ABH and NS distributions from having support in the lower mass gap.

The support of the PBH distribution at high masses is due both to heavy events potentially interpretable as PBHs and also (as in the previous case of single-population analyses) to the fact that if the secondary is interpreted as a PBH then automatically also the heavier primary is primordial, so the low-mass ($\leq 3 m_\odot$) and high-mass ranges ($\geq 3 m_\odot$) are intertwined. This is not the case for the ABH/NS models, since they allow for mixed BH-NS binaries following independent distributions.

Furthermore, the PBH merger rate distribution shown in Fig.~\ref{fig:mixed NS BBH pops} has significant support in the subsolar range. This contribution is only bounded from above by the non observation of subsolar events in the GWTC-3 catalog. As previously discussed, this interesting property is due to the inevitable broadness of the PBH mass function below the QCD peak induced by the critical collapse. 
Since the PBH distribution has support in the solar mass range, as it provides a competitive explanation for GW190814 and a marginal contribution to the otherwise NS binaries (although the PBH merger rate is 1-2 orders of magnitudes smaller than for the NS distribution), then in our ab-initio PBH model it is inevitable to have support for subsolar mergers. This is not the case for the NS phenomenological model, whose mass distribution abruptly drops in the subsolar range.

Finally, note that while the ABH and NS distributions shown in Fig.~\ref{fig:mixed NS BBH pops} have an upper and lower value given by their corresponding $90\%$ credible interval, the PBH distribution has no lower value since the posterior of $f_\PBH$ is also compatible with zero (see Appendix~\ref{app:posteriors}).
This property is natural in our analysis, since the ABH and NS distributions are phenomenological models built to reproduce most of the features of the data. 
\footnote{For example, the phenomenological ABH distribution can accommodate the upper mass-gap event GW190521~\cite{Abbott:2020tfl}, even if it is challenging to explain the latter in standard astrophysical scenarios due to the pulsational pair supernova instability preventing the formation of binaries with masses above the (uncertain) limit $\thickapprox 50 M_\odot$
\cite{Barkat:1967zz,Heger:2001cd,Woosley:2007qp,Belczynski:2016jno,Woosley:2016hmi,Stevenson:2019rcw,Farmer:2019jed,Renzo:2020rzx,Mapelli:2019ipt,Croon:2020oga,Marchant:2020haw,Ziegler:2020klg,Belczynski:2020bca}. One possibility widely investigated in the literature is the interpretation of such event as a second generation merger in globular clusters or galactic nuclei \cite{Fishbach:2017dwv,Gerosa:2019zmo,Rodriguez:2019huv,Baibhav:2020xdf,Kimball:2020opk,Samsing:2020qqd,Mapelli:2020xeq},
even though it may be challenging to explain the observed rate of this event (see also ~\cite{DeLuca:2020sae,Franciolini:2021tla}).} 
Therefore, as previously remarked, our analysis is not meant to search for a PBH subpopolation but rather to place an upper limit on the PBH abundance compatible with the data (see next section).

{
\renewcommand{\arraystretch}{1.4}
\setlength{\tabcolsep}{4pt}
\begin{table}[t!]
\caption{GWTC-3 events with highest PBH likelihood listed in chronological order. The two groups refer to $m_2>3 M_\odot$ (top) or $m_2<3 M_\odot$ (bottom). We also report the measured masses of each event.}
\begin{tabularx}{1 \columnwidth}{|X|c|c|c|}
\hline
\hline
GW event & PBH prob. [\%] & $m_1[M_\odot]$ & $m_2[M_\odot]$
\\
\hline 
\hline
GW151012 & 1.2 &
$23.2^{+14.9}_{-5.5}$ & $13.6^{+4.1}_{-4.8}$
\\
\hline
GW190412 & 25.4 &
 $30.1^{+4.7}_{-5.1}$ &  $8.3^{+1.6}_{-0.9}$
\\
\hline
  GW190512\_180714    & 1.6   &
   $23.3^{+5.3}_{-5.8}$ &  $12.6^{+3.6}_{-2.5}$
\\
\hline
  GW190519\_153544    &  1.5  &
   $66.0^{+10.7}_{-12.0}$ &      $40.5^{+11.0}_{-11.1}$
\\
\hline
GW190521 & 7.2 & 
$95.3^{+28.7}_{-18.9}$
&
$69.0^{+22.7}_{-23.1}$
\\
\hline
  GW190602\_175927    & 2.7   &
  $69.1^{+15.7}_{-13.0}$
  &
  $47.8^{+14.3}_{-17.4}$
  \\
\hline
  GW190701\_203306    &   1.4  &
  $53.9^{+11.8}_{-8.0}$
  &
  $40.8^{+8.7}_{-12.0}$
\\
\hline
GW190706\_222641      &  1.3  &
$67.0^{+14.6}_{-16.2}$
&
$38.2^{+14.6}_{-13.3}$
\\
\hline
 GW190828\_065509     & 2.8   &
 $24.1^{+7.0}_{-7.2}$
 &      
 $10.2^{+3.6}_{-2.1}$
\\
\hline
GW190924\_021846 & 40.3 &
$8.9^{+7.0}_{-2.0}$
&
$5.0^{+1.4}_{-1.9}$
\\
\hline 
GW191109\_010717 &2.9 &
$65^{+11}_{-11}$
&
$47^{+15}_{-13}$
\\
\hline
GW191129\_134029 & 1.2 &
$10.7^{+4.1}_{-2.1}$
&
$6.7^{+1.5}_{-1.7}$
\\
\hline 
\hline 
GW190425 &2.8&
$2.0^{+0.6}_{-0.3}$ &
$1.4^{+0.3}_{-0.3}$
\\
\hline 
GW190426\_152155 & 1.2 &
$5.7^{+3.9}_{-2.3}$
&
$1.5^{+0.8}_{-0.5}$
\\
\hline
GW190814 & 29.1 &
$23.2^{+1.1}_{-1.0}$
&
$2.59^{+0.08}_{-0.09}$
\\
\hline 
GW190917\_114630 &3.0 &
$9.3^{+3.4}_{-4.4}$
&
$2.1^{+1.5}_{-0.5}$
\\
\hline
GW200105\_162426 &3.6&
$8.9^{+1.2}_{-1.5}$
&
$1.9^{+0.3}_{-0.2}$
\\
\hline 
GW200115\_042309 & 1.2 &
$5.9^{+2.0}_{-2.5}$
&
$1.44^{+0.85}_{-0.29}$
\\
\hline
\hline
\end{tabularx}
\label{Tb:parameters}
\end{table}
}

Nonetheless, it is interesting that there exist events with a significant likelihood to be interpreted as PBH binaries by our inference, as shown in Table~\ref{Tb:parameters}. 
In general, the most interesting events are those being  either in the light or heavy portions of the catalog, 
close to either mass gaps, or being characterised by a small mass ratio.
While many events have ${\cal O}(\%)$ probability, for GW190924\_021846, GW190814, GW190412, and GW190521 the probability is approximately $40\%$, $29\%$, $25\%$, and $7\%$ respectively.
We stress that we are comparing an ab-initio PBH model with phenomenological LVK fits tailored to match current data without any astrophysical input. In particular, the LVK fits do not enforce any mass gap in the ABH/NS distribution, so it is possible that events like GW190814 and GW190924\_021846 (with masses $m_2\approx 2.7 M_\odot$ and $m_2\approx 5M_\odot$, which respectively lie squarely in the lower-mass gap and on its upper end) are well fitted by the ABH or NS phenomenological models. Thus, it is interesting and a priori not granted that precisely these events have a sizeable probability to be interpreted as primordial. 
This is due to the fact that they nevertheless lie in a relatively scarcely populated mass range, so the phenomenological distributions should stretch significantly to accommodate them, possibly reducing their ability to fit the many other heavier events in the catalog.
 Overall, these results may indicate that such events, regardless of their primordial interpretation, may not fit consistently within the population described by the LVK reference model and may belong to distinct populations of NS and BH binaries.

It is also interesting that the light events ($m_{1,2}\lesssim 3M_\odot$) that are interpreted as standard NS binaries by the LVK analysis (e.g., GW190425) have only ${\cal O}(\%)$ likelihood to be interpreted as PBHs. This is due to the fact that, even if the PBH distribution modulated by the QCD phase peaks at $m_1\approx M_\odot$, its magnitude is anyway much smaller than the inferred value of the NS distribution.
 This is most likely due to the combination of the critical collapse tail (which does not allow for a sharp drop of the mass function below the solar mass) and the constraint from the absence of sub-solar detections in GWTC-3. 

To conclude this section, we report the Bayes factors comparing the ABH+NS model to the one which includes a PBH subpopulation, found to be
\begin{align}
\log_{10}{\cal B}
_{\ABH+\NS}
^{\ABH+\NS+\text{\tiny PBH,QCD}}
= 0.9\,,
\end{align}
showing a marginal evidence in favour of a contribution from a PBH channel. This interpretation implicitly includes the downplaying effect of a larger set of parameters introduced in the model when a PBH subpopulation is allowed. 
Indeed, the ratio between the best-fit likelihood
of the two models is
\begin{align}
\log_{10}\lp 
\frac{L^*_{\ABH+\NS+\text{\tiny PBH,QCD}}}
{L^*_{\ABH+\NS}}
\rp
=1.4,
\end{align}
 Therefore, the PBH subpopulation improves the fit to the data but not to a sufficient level that would make their absence strongly disfavoured. 
 
 Overall, this analysis suggests the presence of more features in the GWTC-3 data than what is captured by the LVK NS and BBH phenomenological models. Even in the most conservative setting, we found that a PBH subpopulation may capture some of these features, even when the non-observation of a subsolar merger population is taken into account. We now proceed to discuss some interesting implications of our results for future detections and constraints on PBHs and early universe models of inflation.

\section{Implications for future GW experiments and PBH models}\label{implications}

In this section we discuss some implications of our results for the upcoming LVK observation runs and for the PBH scenario.

\subsection{Predicted rate of subsolar mergers and mass-gap events in future LVK searches}

As previously discussed, a general property of the ab-initio PBH model is to predict a significant merger rate in the subsolar range and in the lower mass gap, due to the broadness of the PBH mass function. Thus, once fixing the best-fit PBH abundance distribution through the Bayesian inference, it is possible to make \emph{falsifiable} predictions about the expected numbers of events in the subsolar mass range and in the lower mass gap, assuming some of the GWTC-3 events already detected is interpreted as a primordial binary.

In Table~\ref{TbPBH:O4O5}, we show these predictions, assuming GW190814 is primordial ($29\%$ likelihood in our analysis). Assuming a primordial origin for GW190924\_021846 ($40\%$ likelihood in our analysis) provides similar predictions\footnote{In the following we shall mostly assume that GW190814 is a primordial binary, even though the PBH likelihood of GW190924\_021846 is higher. Besides the fact that the two assumptions would provide similar results, GW190814 is more challenging to fit within standard astrophysical scenarios and the mass of its secondary \cite{LIGOScientific:2020zkf} motivates exploring other explanations for this event.}.
First of all, the first row in Table~\ref{TbPBH:O4O5} shows that the interpretation of at least GW190814 as a PBH binary implies 
the current catalog may include a fraction between $1\%$ and $29\%$ of PBH mergers. On the other hand, the number of expected subsolar mergers within the O1-O2-O3 observation runs is below unity, consistently with the absence of observations in that mass range.

Due to the much improved sensitivity of future observation runs, we notice that O4 and O5 would be bound to detect many PBH events, as expected. However, unless some of these events have smoking-gun features~\cite{Franciolini:2021xbq}, it would be hard to distinguish them from ordinary astrophysical channels.
Therefore, a more interesting prediction of Table~\ref{TbPBH:O4O5} is the number of subsolar and mass-gap events detectable in O4 and O5. In particular, in O5 there could be as many as $\approx 8$ subsolar events per year (but the $90\%$ confidence interval is also compatible with zero events). More interestingly, if GW190814 is assumed to be primordial then O5 should detect one to a few dozen events per year in the lower mass-gap (and up to $\approx 50$ upper mass-gap events), which might be more difficult to interpret in astrophysical scenarios.

While detecting a subsolar merger would be a unique smoking gun for PBHs (or would anyway call for new physics beyond the standard astrophysical formation scenario \cite{Shandera:2018xkn,Cardoso:2019rvt,Guo:2019sns,Bramante:2017ulk,Takhistov:2020vxs,Dasgupta:2020mqg,Giffin:2021kgb,Barsanti:2021ydd}), the lower mass gap~\cite{Gupta:2019nwj} could be populated also by second-generation mergers formed in dense stellar clusters, whose rates in this mass range are particularly uncertain. A way to distinguish second-generation BH mergers from PBH mergers is by measuring the binary spins, since in the former case the spin is expected to be nonnegligible \cite{Hofmann:2016yih,Gupta:2019nwj}, at variance with the latter case~\cite{Franciolini:2021xbq,Franciolini:2022iaa}.

We conclude this section by speculating that the existence of a lower mass gap population of PBHs may be compatible with the OGLE-2011-BLG-0462 low mass BH microlensing observation~\cite{2022arXiv220201903L,OGLE:2022gdj},
whose X-ray luminosity is consistent with the small radiative efficiency expected for a BH and disfavours a NS interpretation~\cite{Mereghetti:2022qgz}, see also Ref.~\cite{Abramowicz:2022mwb}.

{
\renewcommand{\arraystretch}{1.4}
\begin{table}[!t]
\caption{Assuming GW190814 had primordial origin, this table reports the 90\% C.I. for the number of detected PBH events within GWTC-3, 
and predicted events (per year) with O4 and O5 sensitivity.
We also indicate forecasted detections within the subsolar ($m_2<M_\odot$, SS), 
lower mass gap 
($m_1$ or $m_2\, \in [2.5,\, 5]\,M_{\odot}$, LMG), 
upper mass gap ($m_1>50 M_\odot$, UMG) ranges.}
\begin{tabularx}{1 \columnwidth}{|X|c|c|c|c|}
\hline
\hline
  & $N_\PBH^\text{\tiny det}$ &
 $N_\PBH^\text{\tiny det}$(SS) & 
 $N_\PBH^\text{\tiny det}$(LMG) &
 $N_\PBH^\text{\tiny det}$(UMG)
\\
\hline
 O1-O3 & 
$[0.8,22.4]$ &
$[0.0,0.6]$ & 
$[0.1,2.3]$ &
$[0.0,6.1]$
\\
\hline
O4 & 
$[1.9,43.7]$ &
$[0.0, 1.3]$ & 
$[0.3, 13.0]$ & 
$[0.0, 13.1]$ 
\\
\hline
O5 & 
$[10.3,216.7]$ &
$[0.0 ,8.6]$ & 
$[0.8, 25.2]$ & 
$[0.0, 47.3]$ 
\\
\hline
\hline
\end{tabularx}
\label{TbPBH:O4O5}
\end{table}
}

\subsection{PBH constraints}

The posterior distribution describing the parameters of the PBH population (see Appendix~\ref{app:posteriors}) can be used to set an upper bound on the PBH abundance in the solar mass range. 

\begin{figure}[!t]
	\centering
	\includegraphics[width=0.495\textwidth]{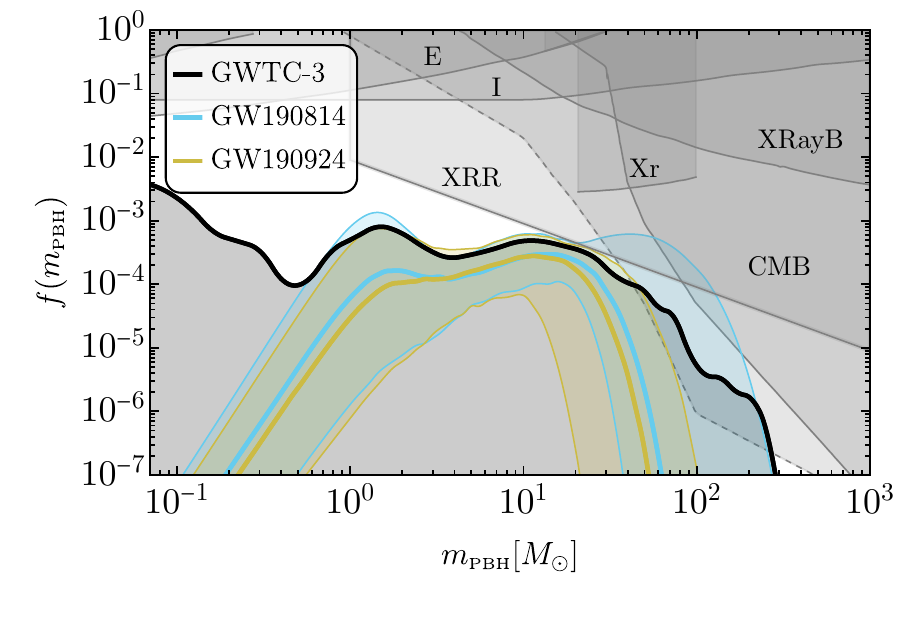}
	\caption{
	Constraints on the PBH mass distribution derived in this work and compared to existing ones~\cite{Carr:2020gox}.
	The black curve shows the upper bound (90\% C.I.) for the mass distribution obtained from the GWTC-3 inference. The cyan (yellow) band shows the posterior distribution assuming GW190814 (GW190924) is interpreted as a PBH binary. 
	}
\label{fig:PBH constraints}
\end{figure}

In Fig.~\ref{fig:PBH constraints} we show the posterior predictive distribution for the PBH mass function $f(m_\PBH)$ in a logarithmic scale obtained from the GWTC-3 inference, together with existing constraints
in this mass range (see, e.g.,~\cite{Carr:2020gox} for a recent review). 
In the mass range of interest for our discussion, the most relevant constraints come from CMB anisotropies produced by accreting PBHs in the early universe~\cite{Ali-Haimoud:2016mbv, Serpico:2020ehh}. 
Other constraints come from comparing the {\it late time} emission of
electromagnetic signals from interstellar gas accretion onto PBHs with observations of galactic radio and X-ray isolated sources~(XRay)~\cite{Gaggero:2016dpq,Manshanden:2018tze} and 
X-ray binaries~(XRayB)~\cite{Inoue:2017csr},
X-ray and radio backgrounds (XRR) \cite{Ziparo:2022fnc}, lensing searches of massive compact halo objects (MACHOs) towards the Large Magellanic Clouds (EROS,E)~\cite{Allsman:2000kg}, fast transient events near critical curves of massive galaxy clusters (ICARUS,I)~\cite{Oguri:2017ock}, and observations of stars in the Galactic bulge by the Optical Gravitational Lensing Experiment (OGLE,O)~\cite{Niikura:2019kqi}. 
The deflection of light by PBHs in the density spike likely existing around the M87 supermassive black hole combined with EHT measurements give rise to additional constraints~\cite{Silk:2022cck}, which are not shown as their conservative version would fall behind the region already excluded by CMB.
Consistently with the assumptions made in the previous sections, here we also do not account for the potential impact of PBH accretion that may shifting CMB constraint to higher masses \cite{DeLuca:2020fpg}.

The black area in Fig.~\ref{fig:PBH constraints} corresponds to the ($90\%$ confidence level) upper bound on the PBH mass distribution $f(m_\PBH)$ derived from the GWTC-3 multi-population inference and hence extends up to $f(m_\PBH)\to0$.
The cyan region is instead the posterior distribution assuming that the lower mass-gap event GW190814 is a primordial binary, which forces $f(m_\PBH)$ to be nonzero and therefore bounded from below. A similar bound can be obtained by assuming that GW190924 is primordial (yellow curves).
First of all, we note that the allowed region for the PBH model is not excluded by other constraints not based on GW events. Only a minor overlap between the cyan band and the CMB constraints is observed, which is not however sufficient to constrain the scenario. 
Finally, the mass distribution is allowed to gain a higher contribution going towards masses well below $\approx M_\odot$ due to the reduced sensitivity of LVK deep in the subsolar mass range.

\begin{figure}[!t]
	\centering
	\includegraphics[width=0.495\textwidth]{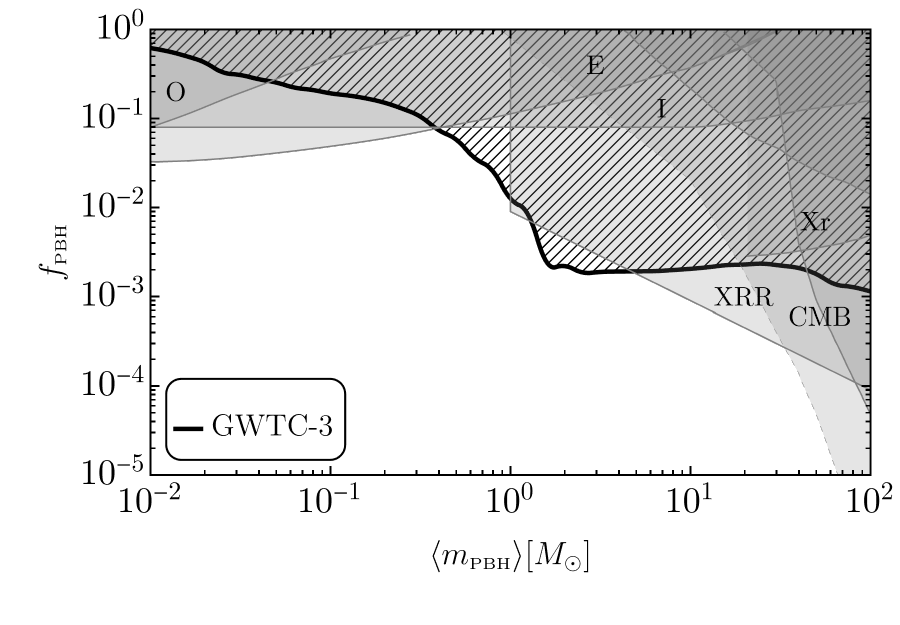}
	\caption{ Upper bound on the PBH abundance as a function of average mass $\langle m_\PBH \rangle$ derived at 90\% C.I. from the GWTC-3 dataset assuming a power-law parametrization of the primordial power spectrum and (conservatively) a dominant contribution of astrophysical mergers in the LVK band. 
	}
\label{fig:PBH constraints2}
\end{figure}

We can also translate the constraint on the mass distribution on the overall value of the abundance defined from Eq.~\eqref{eqn:massFunctionlog} as
\begin{equation}\label{fPBHintegral}
    f_\PBH \equiv  \int \d \ln m_\PBH f(m_\PBH)\,.
\end{equation}
While most of the posterior of $f_\PBH$ is constrained to be much smaller than unity, see Fig.~\ref{fig:posterior_mixedall}, there is also a small support for a tail reaching $f_\PBH=1$. This tail is correlated with blue spectra (i.e. large $n_s$) giving larger support to light masses, and small $M_\text{\tiny S}$.
This means that values of the PBH abundances of order unity can only be reached for light PBH populations where the LVK sensitivity sufficiently degrades. 
We can better visualise this result by computing the maximum $f_\PBH$ at 90\% C.I. as a function of the average PBH mass $\langle m_\PBH \rangle$.
This upper bound represents the maximum value of the fraction of the dark matter which can be explained by a PBH population derived assuming a power spectrum of the form \eqref{PS_zeta} and an average mass $\langle m_\PBH\rangle$, when also a ABH population of mergers is allowed to efficiently explaing the majority of mergers in the GWTC-3 dataset. 
This bound is shown in Fig.~\ref{fig:PBH constraints2}, showing a marked plateau around $f_\PBH\approx 2 \times 10^{-3}$, consistently with previous approximated studies \cite{Ali-Haimoud:2017rtz,Vaskonen:2019jpv,Wong:2020yig,Hutsi:2020sol,Franciolini:2021tla}, which drastically degrades at masses below $M_\odot$, eventually hitting other non-GW-based constraints.
This also confirms that LVK observations set the most stringent constraints in the mass range $\langle m_\PBH \rangle \in [0.3,50] M_\odot$.\footnote{While our constraint is derived assuming nearly-Gaussian perturbations and a consequent initial Poisson spatial distribution of PBHs,
it was recently shown that even assuming (more exotic) clustered initial condition does not allow to evade constraints preventing stellar mass PBHs from being a dominant component of the dark matter \cite{DeLuca:2022uvz}. }

Furthermore, as already shown in Table~\ref{Tb:parameters}, we see that the best fit PBH model allows for a certain number of events (GW190412, GW190924\_021846, GW190814, GW190521) to have a primordial origin with probability respectively about ($25\%$, $40\%$, $29\%$, $7\%$). This means, in contrast with the analysis of Ref.~\cite{Juan:2022mir}, that the absence of subsolar events during the past LVK runs (which is automatically included in our analysis) does not exclude the possibility that \emph{some} of the detected events have a primordial origin.
This is due to the fact that in our ab-initio model we allow the tilt $n_s$ to vary and its inferred value is given in Table~\ref{TbPBH:posparameters} (along with the posteriors of the other PBH population hyperparameters), while the same parameter was fixed to $n_s\approx 0.95$ (very close to its value at the much larger, and uncorrelated, CMB scales) in Ref.~\cite{Juan:2022mir} 
(following the choice made in Refs.~\cite{Carr:2019kxo,Jedamzik:2020omx}).

Although error bars on $n_s$ are large, the population inference systematically selects a \emph{redder} tilted curvature power spectrum which reduces the abundance in the (sub-) solar mass range and erases the dependence to the high scale $k_\text{\tiny max}$ (i.e. low mass $\ms$) cut-off, which is compatible with the left boundary of its prior range $\ms  = 10^{-2.5} M_\odot$.
This is needed in order to counteract the QCD enhancement at the solar mass and reduce the hierarchy in mass distribution between the solar mass and ${\cal O}$(tens) of solar masses (where GW190814 and other events gets support from). 
The PBH abundance is found to depend strongly on the tilt, so even a change by $\sim 10\%$ can change the abundance significantly. In particular, a smaller value of $n_s$ makes the QCD peak less pronounced and the slope at higher masses less steep, resulting in observable rates in the ${\cal O}(10M_\odot)$ range even in the absence of subsolar events.\footnote{We note that current constraints in the subsolar mass~\cite{Nitz:2022ltl} relies on assuming a given PBH mass distribution, which is not the one induced by the QCD phase transition considered here. We stress that our analysis automatically accounts for possible subsolar events and the absence thereof in GWTC-3.}
Finally, the contribution to the heavier portion of the catalog depends instead on  the scale where the power spectrum grows from the CMB values ($k_\text{\tiny min}$ or $M_\text{\tiny L}$). This is an inevitable ingredient in PBH models, as we shall discuss in the next section.

{
\renewcommand{\arraystretch}{1.4}
\setlength{\tabcolsep}{4pt}
\begin{table}[!t]
\caption{Posterior 90\% C.I. for PBH population parameters assuming GW190814 is primordial (similar results are found by assuming that GW190924 is primordial).}
\begin{tabularx}{1 \columnwidth}{|X|c|c|c|}
\hline
\hline
 Parameter & All & GW190814 & GW190924 
\\
\hline
\hline
$\log_{10}A$ & 
$-1.9^{+0.4}_{-0.6}$ & 
$-1.93^{+0.10}_{-0.05}$  & 
$-1.9^{+0.1}_{-0.1}$
\\
\hline
$n_s$ & 
$0.68^{+0.66}_{-0.61}$ & 
$0.68^{+0.18}_{-0.40}$ &
$0.64^{+0.29}_{-0.56}$
\\
\hline
$\log_{10}(k_\text{\tiny min}/{\rm Mpc^{-1}}) $ & 
$6.0^{+1.6}_{-0.6}$ & 
$5.9^{+0.2}_{-0.4}$  & 
$6.0^{+0.3}_{-0.2}$
\\
\hline
$\log_{10}(k_\text{\tiny max}/{\rm Mpc^{-1}}) $ &
$7.8^{+0.6}_{-0.9}$ & 
$8.1^{+0.3}_{-0.9}$ & 
$8.0^{+0.4}_{-1.2}$
\\
\hline
\hline
$\log_{10}f_\PBH$ & 
$-3.4^{+2.2}_{-2.3}$ & 
$-3.1^{+0.5}_{-0.4}$ & 
$-3.2^{+0.3}_{-0.5}$
\\
\hline
$\log_{10}(\ms/M_\odot)$ & 
$-1.2^{+1.8}_{-1.2}$ & 
$-1.6^{+1.7}_{-0.7}$ & 
$-1.6^{+2.5}_{-0.9}$
\\
\hline
$\log_{10}(\ml/M_\odot) $ & 
$2.4^{+1.3}_{-3.2}$ & 
$2.6^{+0.7}_{-0.3}$ & 
$2.5^{+0.5}_{-0.5}$
\\
\hline
\hline
\end{tabularx}
\label{TbPBH:posparameters}
\end{table}
}

\section{PBHs from inflationary dynamics: a data-driven model} \label{sec:USR}

We now come to the theoretical interpretation of our data-driven results. The goal we set in this section is simple but ambitious: we aim to construct a model of PBH formation that gives an abundance distribution compatible with the allowed region shown in Fig.~\ref{fig:PBH constraints}.
This question will be addressed in Sec.~\ref{sec:OnlySolar}.
Even more ambitiously, we may ask whether such PBH distribution could comprise the entirety of the dark matter observed in the universe. 
This question will be addressed in Sec.~\ref{sec:AlsoDarkMatter}. 

Before entering into the details, let us illustrate the general strategy.
At first sight, the answer to the last question is a resounding no -- after all the very same constraint extracted in Fig.~\ref{fig:PBH constraints2} limits the maximum abundance of PBHs to be far below order-one values. It is well-known, in fact, that the only mass range, consistent with observational bounds, in which dark matter could entirely consist of PBHs is $10^{-16} \lesssim m_\PBH/M_{\odot}\lesssim 10^{-12}$~\cite{Carr:2020gox} (dubbed the asteroid mass range in the following), that is for PBHs way lighter than the solar-mass range covered by LVK data.  
However, a first exception to this apparent incompatibility was pointed out in Ref.\,\cite{DeLuca:2020agl}, where it was shown that a broad curvature power spectrum in the form of a double-step Heaviside theta function could potentially give birth to a population of PBHs with a mass distribution covering vastly different scales (see also Ref.~\cite{Inomata:2016rbd}). 

In light of this result, one could be tempted to interpret a putative mass distribution compatible with the constraint in Fig.~\ref{fig:PBH constraints} as the proverbial tip of the iceberg, that is just the final part of a much wider mass distribution possibly reaching order-one abundances  at values of $m_\PBH$ compatible with the asteroid mass range.
The idea of Ref.~\cite{DeLuca:2020agl}---originally thought in reference to the stochastic signal of GWs generated, as a second-order effect, by the large scalar perturbations that form PBHs---was recently explored in much more detail in Ref.~\cite{Franciolini:2022pav}. Two of the results of this paper are worth emphasizing. First, it was shown how to engineer consistent (that is, compatible with CMB observations and the end of inflation) inflationary dynamics which give rise, starting from a handful of physically meaningful parameters, to a curvature power spectrum compatible with the toy model given in Ref.~\cite{DeLuca:2020agl}; 
second, and most importantly, it was shown that, once the relevant parameters that control the background dynamics have been identified, it is relatively simple to understand what are the conditions that are needed in order to generate a PBH mass distribution that gives an order-one abundance of dark matter in the asteroid mass range and, \emph{simultaneously}, a detectable fraction of solar-mass merger events.
In this section, we will add one more piece of information to this picture, quantitatively assessing and reinforcing its robust observational consequences in the solar-mass range.

As in Ref.~\cite{Franciolini:2022pav}, our starting point for the following discussion is the analytical {\it ansatz} 
\begin{align}
\eta(N) = & \frac{1}{2}\bigg\{\left[
\eta_{\rm I} - \eta_{\rm II} + (\eta_{\rm II}-\eta_{\rm I})\tanh\left(\frac{N - N_{\rm I}}{\delta N_{\rm I}}\right)
\right] + \nn \\
&
\left[
\eta_{\rm II} + \eta_{\rm III} + (\eta_{\rm III}-\eta_{\rm II})\tanh\left(\frac{N - N_{\rm II}}{\delta N_{\rm II}}\right)
\right] + \nn \\
&  \left[
\eta_{\rm IV}-\eta_{\rm III} + (\eta_{\rm IV}-\eta_{\rm III})\tanh\left(\frac{N - N_{\rm III}}{\delta N_{\rm III}}\right)
\right]\bigg\}\,.\label{eq:MainEqEta}
\end{align}
that describes the time evolution of the Hubble parameter $\eta \equiv -\ddot{H}/2H\dot{H}$,
where $\dot{H} = dH/dt$ is the 
cosmic-time derivative of the Hubble rate $H$ and $N$, defined by $dN/dt = H$, is the number of $e$-folds.

\begin{figure}[!t]
	\centering
	\includegraphics[width=0.495\textwidth]{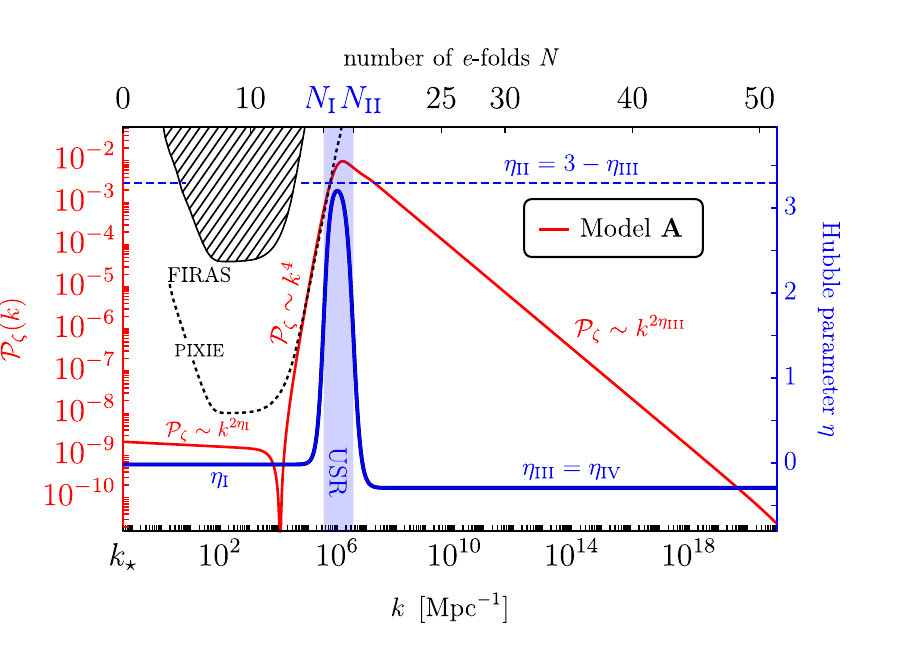}
	\caption{
	\auu{
	Curvature power spectrum (red, left-side $y$-axis) as a function of the comoving wavenumber $k$ (lower-side $x$-axis); on the upper-side $x$-axis, we indicate the number of $e$-folds $N$ according to the horizon-crossing condition $k=a(N)H(N)$, normalized at $N=0$ for the pivot scale $k_{\star} = 0.05$ Mpc$^{-1}$. 
	We superimpose (blue, right-side $y$-axis) the time evolution of the Hubble parameter $\eta$. 
	The figure refers to the explicit realization of our Model~\textbf{A} in Table~\ref{Tab:ModelTab}.
	The vertical region shaded in blue indicates the USR phase ($N_{\rm I}\leqslant N \leqslant N_{\rm II}$). 
	The meshed region shows the FIRAS constraint 
	from CMB spectral distortions computed for the steepest growth power spectrum, $\mathcal{P}_{\zeta} \sim k^4$, cf. Ref.~\cite{Byrnes:2018txb}.
	The dotted black line illustrates the 
	projected constraints from a future PIXIE-like spectral distortion experiment \cite{Chluba:2019nxa,Green:2020jor}.
	}
	}
\label{fig:DynSolarPeak}
\end{figure}

{\renewcommand{\arraystretch}{1.4}
\begin{table*}[t]
\begin{center}
\begin{adjustbox}{max width=1\textwidth}
\begin{tabular}{||c||c|c|c|c|c|c|c|c|c|c||c||}
\hline\hline
 \multirow{2}{*}{Model parameters in Eq.~\eqref{eq:MainEqEta}} & \multirow{2}{*}{$\epsilon_{\rm I}$, $\eta_{\rm I}$} &  \multirow{2}{*}{$N_{\rm ref}$} &  \multirow{2}{*}{$N_{\rm I}$} & 
 \multirow{2}{*}{$\eta_{\rm II}$, $\Delta N_{\rm USR}$} & \multirow{2}{*}{$\eta_{\rm III}$} & \multirow{2}{*}{$\Delta N_{\rm plateau}$} 
  & \multirow{2}{*}{$N_{\rm IV}$, $\eta_{\rm IV}$} & \multirow{2}{*}{$\delta N_{\rm I}$} & \multirow{2}{*}{$\delta N_{\rm II}$} 
  & \multirow{2}{*}{$\delta N_{\rm III}$} & \multirow{2}{*}{$f_\PBH = 1$} \\ 
 & & & & & & & & & & & \\ \hline\hline
\multirow{2}{*}{Model \textbf{A}} & $\epsilon_{\rm I} = 3.125\times10^{-4}$ & \multirow{2}{*}{$0$} & \multirow{2}{*}{$15.75$} & $\eta_{\rm II} = 3-\eta_{\rm III}$ & 
\multirow{2}{*}{$-0.292$} & \multirow{2}{*}{\ding{55}} & $N_{\rm IV} = 55$ & \multirow{2}{*}{$0.50$}  & \multirow{2}{*}{$0.59$} & \multirow{2}{*}{\ding{55}} & \multirow{2}{*}{{\color{red}{\ding{55}}}}  \\ 
  & $\eta_{\rm I} = -1.68\times 10^{-2}$ & & & $\Delta N_{\rm USR} = 2.342$ & & & $\eta_{\rm IV}=\eta_{\rm III}$ & & & & \\  \hline
  \multirow{2}{*}{Model \textbf{B}} & $\epsilon_{\rm I} = 3.125\times10^{-4}$ & \multirow{2}{*}{$0$} & \multirow{2}{*}{$15.5$} & $\eta_{\rm II} = 3.17$ & 
\multirow{2}{*}{$-0.294$} & \multirow{2}{*}{\ding{55}} & $N_{\rm IV} = 55$ & \multirow{2}{*}{$0.50$}  & \multirow{2}{*}{$0.59$} & \multirow{2}{*}{\ding{55}} & \multirow{2}{*}{{\color{red}{\ding{55}}}}  \\ 
  & $\eta_{\rm I} = -1.68\times 10^{-2}$ & & & $\Delta N_{\rm USR} = 2.47$ & & & $\eta_{\rm IV}=\eta_{\rm III}$ & & & & \\  \hline\hline
\multirow{2}{*}{Model \textbf{C}} & $\epsilon_{\rm I} = 3.125\times10^{-4}$ &  \multirow{2}{*}{$0$} & \multirow{2}{*}{$15.75$} & $\eta_{\rm II} = 3.197$ & 
\multirow{2}{*}{$0$} & \multirow{2}{*}{$17.56$} & $N_{\rm IV} = 55$ & \multirow{2}{*}{$0.50$} & \multirow{2}{*}{$0.68$} & \multirow{2}{*}{$0.50$} & \multirow{2}{*}{{\color{Green}{\ding{51}}}} \\ 
   & $\eta_{\rm I} = -1.68\times 10^{-2}$ & & & $\Delta N_{\rm USR} = 2.44$ & & & $\eta_{\rm IV} = -0.576$ & & & & \\ \hline
\multirow{2}{*}{Model \textbf{D}} & $\epsilon_{\rm I} = 3.125\times10^{-4}$ & \multirow{2}{*}{$0$} & \multirow{2}{*}{$15.5$} & $\eta_{\rm II} = 3.11$ & 
\multirow{2}{*}{$0.012$} & \multirow{2}{*}{$17.56$} & $N_{\rm IV} = 55$ & \multirow{2}{*}{$0.50$} & \multirow{2}{*}{$0.50$} & \multirow{2}{*}{$0.50$} & \multirow{2}{*}{{\color{Green}{\ding{51}}}}  \\ 
 & $\eta_{\rm I} = -1.68\times 10^{-2}$ & & & $\Delta N_{\rm USR} = 2.44$ & & & $\eta_{\rm IV}=-0.567$ & & & & \\  \hline\hline
\end{tabular}
\end{adjustbox}
\vspace{-0.05cm}
\caption{{
Free parameters of our models together with their numerical benchmark values. 
We define $\Delta N_{\rm USR}\equiv N_{\rm II} - N_{\rm I}$ and  
$\Delta N_{\rm plateau}\equiv N_{\rm III} - N_{\rm II}$. 
Consistently with Planck data~\cite{Planck:2015fie}, at the CMB pivot scale $k_{\star} = 0.05$ Mpc$^{-1}$, all models give 
$n_s (k_*) = 0.965$ and $A_s = 2.1\times 10^{-9}$ for, respectively, spectral index and amplitude of the curvature power spectrum, and a tensor-to-scalar ratio
$r = 0.005$.
Only Models~\textbf{C} and \textbf{D} produce PBHs that can account for the entirety of the dark matter without violating existing constraints  (see also Appendix~\ref{sec:Tach} for more details).
}}\label{Tab:ModelTab}
\end{center}
\end{table*}
}

\subsection{Solar-mass PBHs from inflationary dynamics}\label{sec:OnlySolar}

Consider first the limit $\eta_{\rm III} = \eta_{\rm IV}$ in Eq.~\eqref{eq:MainEqEta}. 
The last line vanishes, and we are left with the expression
\begin{align}
\eta(N) = & \frac{1}{2}\bigg\{\left[
\eta_{\rm I} - \eta_{\rm II} + (\eta_{\rm II}-\eta_{\rm I})\tanh\left(\frac{N - N_{\rm I}}{\delta N_{\rm I}}\right)
\right] + \nn \\
&
\left[
\eta_{\rm II} + \eta_{\rm III} + (\eta_{\rm III}-\eta_{\rm II})\tanh\left(\frac{N - N_{\rm II}}{\delta N_{\rm II}}\right)
\right]\bigg\}\,.\label{eq:MainEqEta2}
\end{align}
The meaning of the free parameters entering in 
Eq.~\eqref{eq:MainEqEta} becomes manifest by looking at Fig.~\ref{fig:DynSolarPeak}.
The right-side $y$-axis of this figure shows Eq.~\eqref{eq:MainEqEta2} as a function of the number of $e$-fold $N$ (upper-side $x$-axis) or, equivalently, the comoving wavenumber $k$ (lower-side $x$-axis); these two quantities are indeed related by 
the horizon-crossing condition $k=a(N)H(N)$ that we normalize in such a way that $N=0$ corresponds to the crossing time of the CMB pivot scale $k_{\star} = 0.05$ Mpc$^{-1}$. 
Once the time evolution of $\eta$ is given, it is immediately possible to obtain the time evolution of the Hubble parameter $\epsilon\equiv -\dot{H}/H^2$ by solving the differential equation $\eta = \epsilon - 1/2\,d\log\epsilon/dN$ with initial condition $\epsilon_{\rm I}$ at $N=0$; at this stage, therefore, the background inflationary dynamics is completely specified. The curvature power spectrum can be now obtained by solving numerically the Mukhanov-Sasaki equation (cf. Ref.~\cite{Franciolini:2022pav} for technical details).
In Fig.~\ref{fig:DynSolarPeak} we superimpose the curvature power spectrum $\mathcal{P}_\zeta (k)$ that corresponds to the 
time evolution of $\eta$ shown in the same figure.

The most important part of the dynamics is the presence of a phase of USR during which we have $\eta \gtrsim 3/2$ 
in our parametrization, inducing an exponential growth of a specific set of modes. Such USR phase takes place in the $e$-fold time interval $N_{\rm I}\lesssim N \lesssim N_{\rm II}$.

Curvature perturbations that cross the horizon well before the USR phase are not affected by the latter, and contribute to the power spectrum according to the usual slow-roll approximation. This part of the power spectrum follows the scaling $\mathcal{P}_{\zeta}\sim k^{2\eta_{\rm I}}$ (cf. Fig.~\ref{fig:DynSolarPeak}) and the numerical values of 
$\eta_{\rm I}$ and $\epsilon_{\rm I}$ are chosen in such a way to fit CMB data at the pivot scale. 

Curvature perturbations that cross the horizon right before the USR phase are those that are mostly affected by the latter. These modes are exponentially enhanced and give rise to a steep growth of the power spectrum that in our model follows the scaling 
$\mathcal{P}_{\zeta}\sim k^{4}$. Because of this growth, 
the curvature power spectrum experiences a parametric change (with respect to the preceding slow-roll value) of the order $\Delta\mathcal{P}_{\zeta} \sim e^{2\eta_{\rm II}\Delta N_{\rm USR}}$, with 
$\Delta N_{\rm USR} \equiv N_{\rm II}- N_{\rm I}$. To fix ideas, 
in order to get a seven orders-of-magnitude enhancement of the power spectrum (that would bring 
the typical slow-roll amplitude $\mathcal{P}_{\zeta} \sim 10^{-9}$ up to $\mathcal{P}_{\zeta} \sim 10^{-2}$) one needs $\eta_{\rm II}\Delta N_{\rm USR} \sim 8$. In other words, the combination of parameters $\eta_{\rm II}\Delta N_{\rm USR}$ 
controls the height that the curvature power spectrum reaches as a consequence of the USR phase.  

Curvature perturbations that cross the horizon well after the USR phase during the $e$-fold time interval $N \gtrsim N_{\rm II}$ give rise to the final part of the power spectrum with scaling 
$\mathcal{P}_{\zeta}\sim k^{2\eta_{\rm III}}$. 
In this part of the dynamics the slow-roll approximation is again applicable. 
The value of $\eta_{\rm III}$ is negative, and this is crucial for inflation to end. We fix the numerical value of $\eta_{\rm III}$ by imposing a total number of 55 inflationary $e$-folds; different choices of this benchmark value would slightly affect the parameters of our model while not significantly changing the dynamics.

Finally, the parameters $\delta N_{\rm I}$ and $\delta N_{\rm II}$ control the width of the tanh-transitions between different values of $\eta$ at, respectively, $N_{\rm I}$ and $N_{\rm II}$. The limit $\delta N_{\rm I,II} \to 0$ corresponds to a step transition.

The above discussion captures the most evident features of the curvature power spectrum and explains the formation of the peak shown in Fig.~\ref{fig:DynSolarPeak}. 
The precise form of the power spectrum at the tip of the peak is shaped by curvature modes that cross the horizon during and immediately after the USR phase.
This aspect is truly crucial for our analysis since this is the part of the power spectrum that should be compared with the model in Eq.~\eqref{PS_zeta}.
On the theory side, the key aspect is the possibility to establish the so-called Wands duality~\cite{Wands:1998yp} between the USR phase that takes place during the $e$-fold time interval $N_{\rm I} \leqslant N \leqslant N_{\rm II}$ and the subsequent phase $N>N_{\rm II}$. In short, the Wands duality is the statement that phases with $\eta$ and $3-\eta$ give rise to the same spectral slope in the curvature power spectrum (cf. also  Refs~\cite{Biagetti:2018pjj,Karam:2022nym}). 
In our model, this implies that if we set $\eta_{\rm II} = 3 - \eta_{\rm III}$ we expect that the form of the power spectrum right after the tip of the peak will take the same power-law form $\mathcal{P}_{\zeta}\sim k^{2\eta_{\rm III}}$ that, as discussed before, characterizes the last part of the dynamics. 
In Model~\textbf{A}, therefore, we enforce the condition $\eta_{\rm II} = 3 - \eta_{\rm III}$, see Table~\ref{Tab:ModelTab}.

\begin{figure*}[!t!]
\begin{center}
$$\includegraphics[width=.495\textwidth]{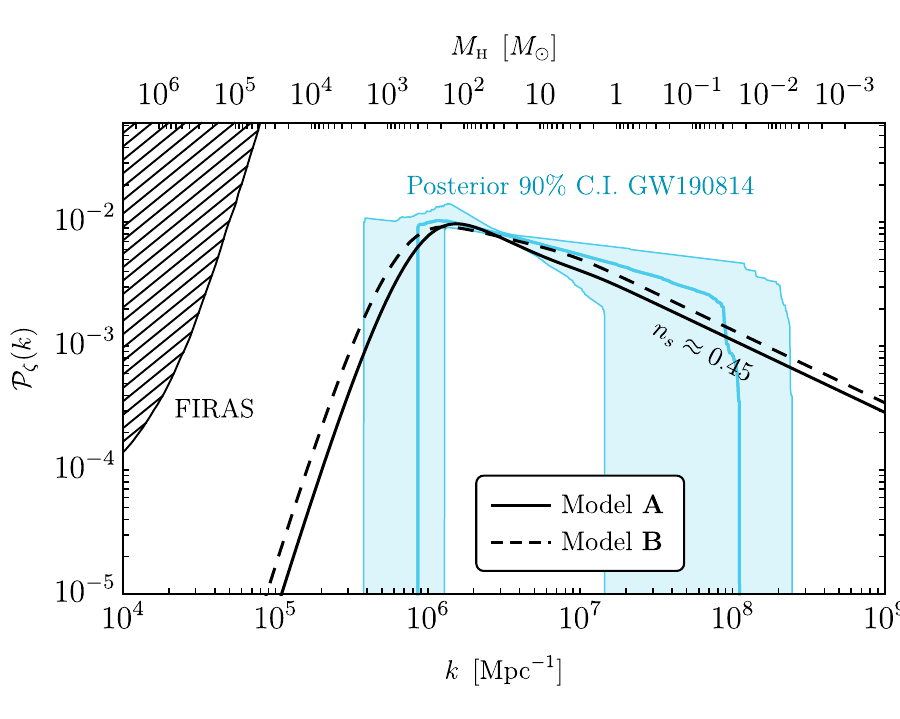}
\quad\includegraphics[width=.495\textwidth]{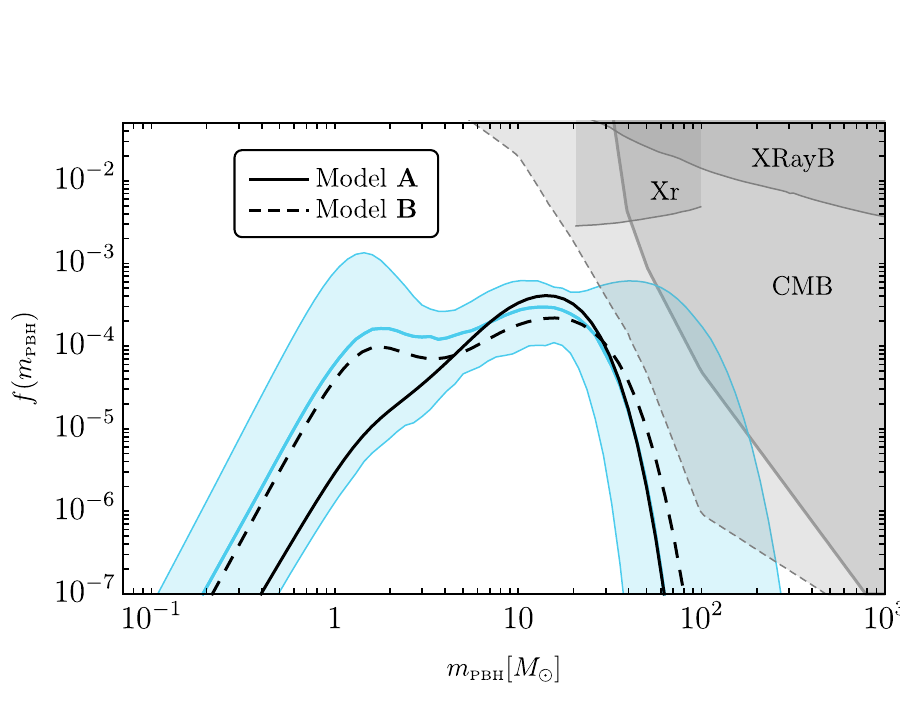}$$
\caption{
\auu{
{\bf Left panel:} zoom-in of Fig.~\ref{fig:DynSolarPeak} near the peak of the spectrum. 
On the top $x$-axis we consider, instead of comoving wavenumbers $k$, the horizon mass according to Eq.~\eqref{M-k}.
In addition to Model~\textbf{A} we also show the power spectrum that corresponds to Model~\textbf{B} in Table~\ref{Tab:ModelTab} (the latter is not shown in Fig.~\ref{fig:DynSolarPeak} since the small differences between Model~\textbf{A} and Model~\textbf{B} can be only appreciated near the peak). The cyan shaded region corresponds to the 90\% C.I. posterior distribution obtained in Sec.~\ref{sec:multipop}
assuming GW190814 is interpreted as a PBH
binary (cf. Table~\ref{TbPBH:posparameters}).
{\bf Right panel:} Mass distribution 
$f(m_\PBH)$ for both Model~\textbf{A} and Model~\textbf{B} compared with the 90\% C.I. posterior distribution
assuming GW190814 is interpreted as a PBH
binary (cf. Fig.~\ref{fig:PBH constraints}).
}
 }\label{fig:PS_fPBH_SolarPeak}  
\end{center}
\end{figure*}

In the left panel of Fig.~\ref{fig:PS_fPBH_SolarPeak} we zoom in on the peak of the power spectrum. 
The solid black line corresponds to Model~\textbf{A}. The numerical solution of the Mukhanov-Sasaki equation confirms our analytical intuition: right after the peak, the curvature power spectrum can be well approximated by a power-law with spectral index that, in the notation of Eq.~\eqref{PS_zeta}, takes the approximate value $n_s = 1 + 2\eta_{\rm III}$. 
Since $\eta_{\rm III}$ is negative and $|\eta_{\rm III}| < {\cal O}(1)$ (otherwise the inflaton will roll too fast towards the end of inflation) it is natural to expect a red tilted power spectrum, consistently with our previous analysis using GW data. In the explicit realization given by Model~\textbf{A}, we find $n_s \approx 0.45$, consistent with our population inference (see Table~\ref{Tb:parameters}).
In 
the left panel of Fig.~\ref{fig:PS_fPBH_SolarPeak} we also show (region shaded in cyan) the 90\% C.I. posterior distribution
assuming GW190814 is interpreted as a PBH
binary. This is the same region shown in Fig.~\ref{fig:PBH constraints} but re-computed in terms of the parameters of the power spectrum. 
The comparison shows that the USR dynamics in Model~\textbf{A} gives a good agreement with the data-driven results derived in Sec.~\ref{sec:multipop}. 
En route, we note that, after mapping the inference to the parameters of the power spectrum in Eq.~\eqref{PS_zeta}, the posterior distribution of $P_\zeta(k)$ is extremely well constrained at around $k = 3\times 10^6$ Mpc$^{-1}$, whereas the error bars become larger towards the two cut-off scales in
momentum. Intuitively, this is expected. 
We remind that this posterior is based on the assumption that 
GW190814 is a primordial binary, which forces $f(m_\PBH)$ to be nonzero (and pretty well determined in particular for $m_\PBH \simeq 20 M_\odot$ that is the primary mass of GW190814). Since the abundance has an exponential dependence on the amplitude $A$ of the power spectrum, the latter cannot change too much at around the corresponding wavenumbers. Additionally, the remaining parameters entering in the spectrum \eqref{PS_zeta} correlate in such a way to respect the stringent bound around $3\times 10^6$ Mpc$^{-1}$ while broadening the permitted regions at both sides.

In the right panel of Fig.~\ref{fig:PS_fPBH_SolarPeak} we show the mass distribution $f(m_\PBH)$ computed according to the formalism set\footnote{It should be noted that we now fully compute the variances in Eq.~\eqref{eqn:variance} numerically without relying on the analytical approximation in Eq.~\eqref{eq:sigma_analytical}. The agreement between both approaches confirms the validity of the approximations adopted to perform the MCMC Bayesian analysis. } in Sec.~\ref{sec:PBHMFQCD}. 
We compare the distribution given by Model~\textbf{A} 
with the 90\% C.I. posterior distribution
assuming GW190814 is interpreted as a PBH
binary. As expected, the model is consistent with the region bracketed by the confidence interval. 
Since we have $n_s \approx 0.45$, the model tends to under-produce PBHs in the subsolar mass range with respect to the median value.

It is important to stress that Model~\textbf{A} relies on the condition $\eta_{\rm II} = 3 - \eta_{\rm III}$.
If we break the Wands duality, it is no longer guaranteed that, after the tip of the peak, the curvature power spectrum will be described by a single power-law. On the contrary, we expect that curvature modes that cross the horizon during and immediately after the USR phase will give to the power spectrum a slightly different scaling compared to the one that characterizes the subsequent phase,   $\mathcal{P}_{\zeta}\sim k^{2\eta_{\rm III}}$. 
To better illustrate this point, we consider Model~\textbf{B} in Table~\ref{Tab:ModelTab}. In this model,  the Wands duality is broken, $\eta_{\rm II} \neq 3 - \eta_{\rm III}$. In Fig.~\ref{fig:PS_fPBH_SolarPeak} Model~\textbf{B} is represented by the black dashed line. As expected, we see that Model~\textbf{B} gives a curvature power spectrum that, right after the peak, is characterized by a broken power-law with two slightly different spectral indices. The second one is fixed by the last part of the dynamics and always given by $n_s = 1 + 2\eta_{\rm III}$. The first one, on the contrary, can be tuned to match more accurately the central value $n_s \approx 0.68$ given in Table\,\ref{TbPBH:posparameters} and, therefore, it would enhance the number of PBHs 
in the subsolar mass range.
As shown in the right panel of Fig.~\ref{fig:PS_fPBH_SolarPeak}, 
this is exactly what Model~\textbf{B} was designed for and it matches the data-driven distribution much more closely than Model~\textbf{A}.

\begin{figure*}[t!]
	\centering
$$\includegraphics[width=.495\textwidth]{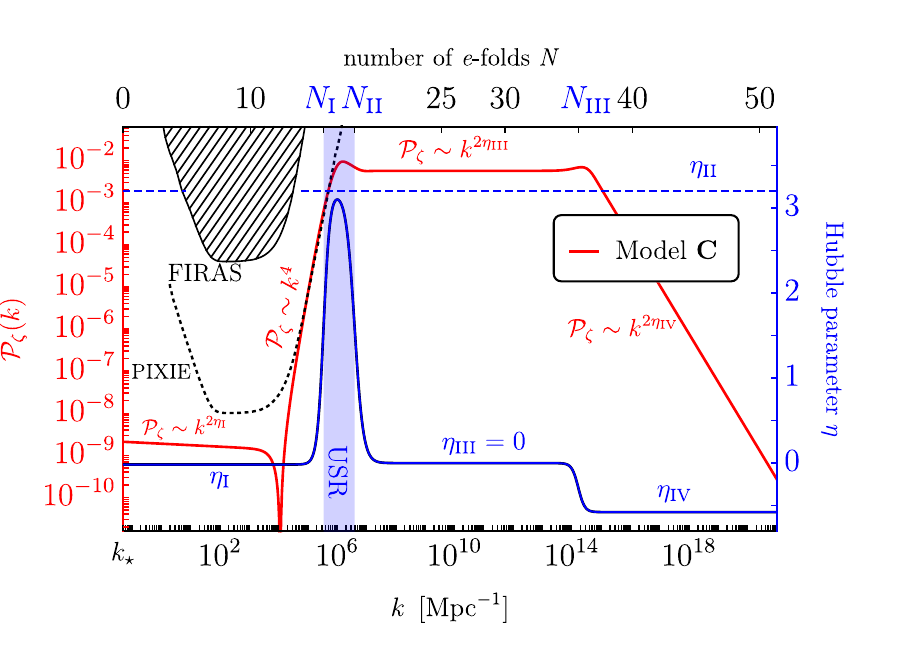}
\quad\includegraphics[width=.495\textwidth]{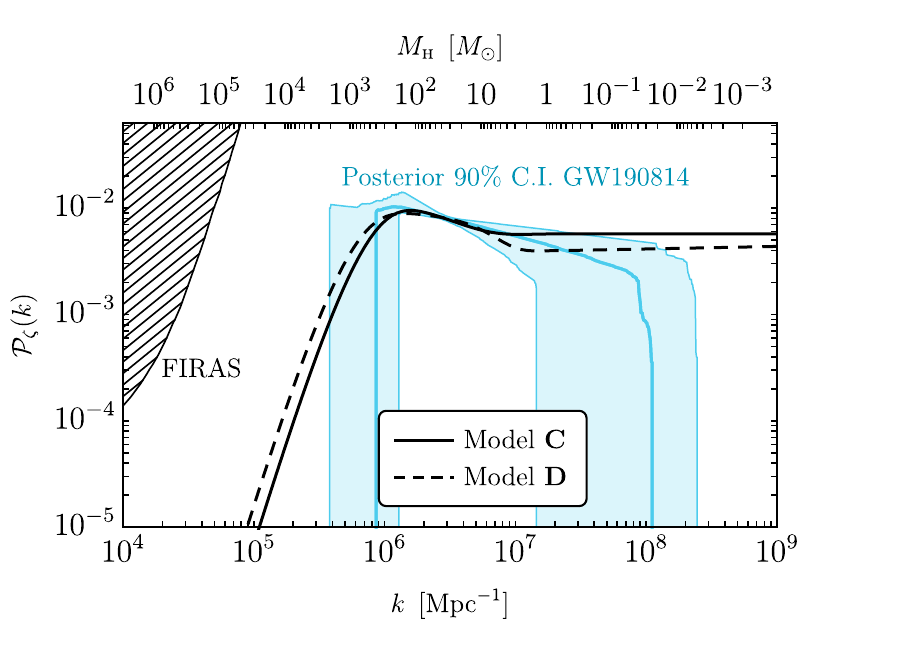}$$
	\caption{
	\auu{
	{\bf Left panel:} Same as in Fig.~\ref{fig:DynSolarPeak} but for Model~\textbf{C} in Table~\ref{Tab:ModelTab}.
	{\bf Right panel:} We zoom in on the 
	left-side edge 
	of the plateau (same as in the left panel of Fig.~\ref{fig:PS_fPBH_SolarPeak}). 
	In addition to Model~\textbf{C} we also show (dashed black line) the curvature power spectrum of Model~\textbf{D} in Table~\ref{Tab:ModelTab}.
	}
	}
\label{fig:PowerSpectrum}
\end{figure*}

At this point of the analysis, we are already in the position to draw a number of relevant conclusions. 
The formation of PBHs is a rare event that requires some finely-tuned underlying dynamics. This statement seems to be true whatever formation mechanism one decides to consider and, in our analysis, we focused on the presence of a phase of USR during inflation. 
Once we are willing to accept the presence of this tuned dynamics, the point that we would like to stress is that the latter naturally comes with a number of features that fully justify the simplified approach taken in our numerical analysis. 
\begin{itemize}[leftmargin=*]
    \item[{\it i)}] First, we note that the cutoff $k_\text{\tiny min}$ (equivalently, $\ml$) arises naturally as a consequence of the sharp enhancement of the power spectrum (with respect to CMB values) that is essential for the generation of a sizable abundance of PBHs; in our explicit realization, such enhancement is provided by the presence of the USR phase. 
    \item[{\it ii)}] Second, curvature modes that cross the horizon during and after the USR phase shape the form of the power spectrum    for $k > k_\text{\tiny min}$. 
    In the context of the parametrization given in Eq.~\eqref{eq:MainEqEta2}, and imposing the Wands duality condition $\eta_{\rm II} = 3-\eta_{\rm III}$, it is possible to get a red tilted power-law functional dependence with $n_s = 1 + 2\eta_{\rm III}$ (cf. Model~\textbf{A} in Table~\ref{Tab:ModelTab} and Fig.~\ref{fig:PS_fPBH_SolarPeak}).
    \item[{\it iii)}] More in general, if we drop the condition $\eta_{\rm II} = 3-\eta_{\rm III}$, the power spectrum is better approximated by a broken power-law with two spectral indices (cf. Model~\textbf{B} in Table~\ref{Tab:ModelTab} and Fig.~\ref{fig:PS_fPBH_SolarPeak}).
    \item[{\it iv)}] Finally, as already noticed, 
    a red tilted spectrum as that suggested by GW data makes the PBH mass distribution practically
    insensitive to the cutoff $k_\text{\tiny max}$ (equivalently, $\ms$). 
    This aspect is well illustrated by our model since the explicit USR dynamics that we consider does not really give any specific value for $k_\text{\tiny max}$; on the contrary, we find that the power spectrum just decreases as $\mathcal{P}_{\zeta}\sim k^{2\eta_{\rm III}}$ following the last part of the dynamics that ends inflation.
\end{itemize}

Overall, our analysis shows that it is possible to devise USR inflationary models that produce the curvature power spectrum in Eq.~\eqref{PS_zeta} assumed as the chief starting ingredient of our GW data-driven population inference. 
In practice, instead of parametrizing the spectrum as in Eq.~\eqref{PS_zeta} one could directly start by parametrizing the evolution of the Hubble parameter $\eta$ (e.g., Eq.~\eqref{eq:MainEqEta2}) or the potential and couplings of the inflaton field(s), and directly run the inference on the values of the inflationary model.

\subsection{Solar-mass PBHs {\it and} dark matter from inflationary dynamics}\label{sec:AlsoDarkMatter}

We now move to consider the second question raised in the introductory part of this section: Is it possible to make the presence of a PBH subpopulation that explains a fraction of GWTC-3 events compatible with the assumption that the entirety of dark matter observed in the universe consists of PBHs?

Answering this question requires devising a realisation of inflationary dynamics tuned in such a way that the logarithmic integral of the mass distribution gives unity, cf. Eqs.~\eqref{eqn:massFunctionlog} and~\eqref{fPBHintegral}. Since in the solar mass range the fraction of dark matter in the form of PBHs is constrained to be at most ${\cal O}(10^{-3})$, the integral must be dominated by the peak in the asteroid mass range (for further details see Appendix~\ref{sec:Tach}).

\subsubsection{Power spectrum with a plateau: how to bridge PBH populations with widely different mass}\label{sec:Plateau}

We consider the full evolution given by Eq.~\eqref{eq:MainEqEta}.  
Compared to the situation discussed in Sec.~\ref{sec:OnlySolar}, we now have $\eta_{\rm III}\neq \eta_{\rm IV}$ and one additional tanh-transition at $e$-fold time $N_{\rm III}$. 
In the left panel of Fig.~\ref{fig:PowerSpectrum} we show (blue, right-side $y$-axis) the evolution of $\eta$ dictated by Eq.~\eqref{eq:MainEqEta} in the explicit realization given by Model~\textbf{C} in Table~\ref{Tab:ModelTab}. We superimpose the (red, left-side $y$-axis) the curvature power spectrum that corresponds to such dynamics. 

As discussed in Ref.~\cite{Franciolini:2022pav}, we impose the condition $\eta_{\rm III} = 0$. 
This condition generates a wide plateau in the power spectrum, raised in amplitude with respect to CMB values because of the preceding USR phase.  
The subsequent transition at $N_{\rm III}$ from $\eta_{\rm III} = 0$ to $\eta_{\rm IV} < 0$ is necessary to end inflation.

The presence of the plateau in the power spectrum provides the concrete possibility to have a mass distribution of PBHs that covers many orders of magnitude. 

What is actually crucial for our analysis is the precise form of the power spectrum at the two edges of the aforementioned plateau. 
The left-side edge is shaped by curvature modes that cross the horizon during and immediately after the USR phase while the right-side edge is shaped by curvature modes that cross the horizon during and immediately after the transition at $N\simeq N_{\rm III}$.

At the left-side edge of the plateau, the 
power spectrum is characterized by a bump-like feature (cf. Ref.~\cite{Franciolini:2022pav} for a detailed discussion about its formation).  
This bump provides the link with our numerical analysis. To make this point more transparent, in the right panel of Fig.~\ref{fig:PowerSpectrum} we zoom in on the 
bump-like feature at the left-side edge of the plateau. 

It is instructive to compare the curvature power spectrum with the 90\% C.I. posterior derived from our numerical analysis assuming GW190814 is a PBH binary.
From this comparison we see that the curvature power spectrum features a cutoff at small wavelengths. 
In full analogy with the previous case (cf. Fig.~\ref{fig:PS_fPBH_SolarPeak}, left), this cutoff is naturally generated by the sharp transition ($\mathcal{P}_{\zeta} \sim k^4$) that, because of the USR phase, brings the power spectrum from CMB values up to the typical amplitudes ${\cal O}(10^{-2})$ that are needed to generate PBHs. 
After the initial $\sim k^4$ growth, the power spectrum decreases before it settles to the constant value of the plateau. 
This decreasing part of the bump plays the role of the red-tilted power spectrum found in our numerical analysis. This is evident from the comparison shown in the right panel of Fig.~\ref{fig:PowerSpectrum} between the posterior distribution and the power spectrum of Model~\textbf{C}. 
It should be noted that the power spectrum of Model~\textbf{C} is constrained to match  the power-law behavior of the posterior distribution, in particular in the interval of comoving wavenumber where the latter is almost precisely nailed down by the numerical analysis (at about $m_\PBH \approx 20 M_\odot$, i.e. the primary mass of GW190814).
On the contrary, away from this $k$-interval deviations are possible. This is consistent with the fact that the numerical analysis is practically
insensitive to $k_\text{\tiny max}$. 
In the explicit realization of our model discussed in Sec.~\ref{sec:OnlySolar}, this freedom was exploited to directly connect the power spectrum to the last part of the dynamics that ends inflation (cf. Fig.~\ref{fig:DynSolarPeak}). In the present scenario, we exploit the same freedom to connect the bump to the subsequent plateau.

\begin{figure*}[t!]
	\centering
$$\includegraphics[width=.495\textwidth]{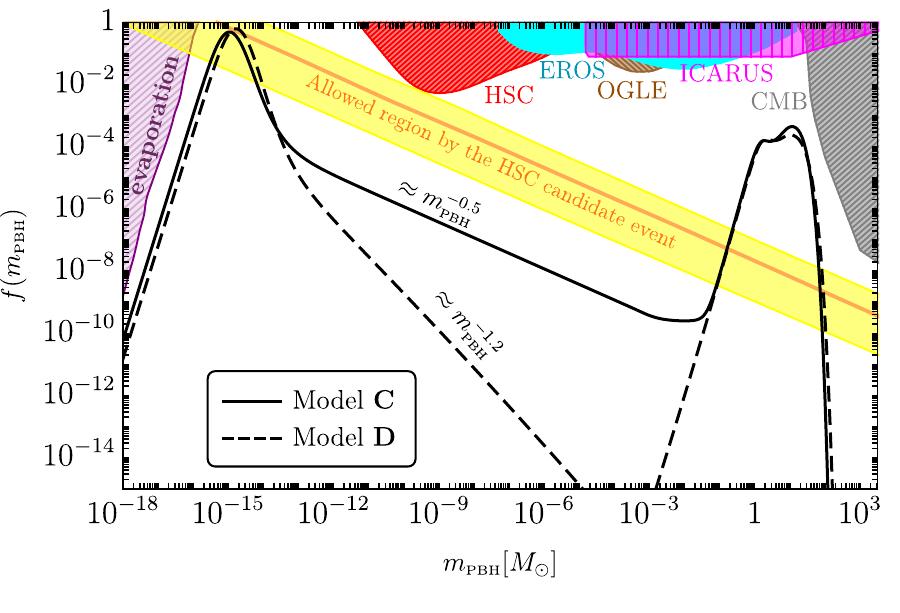}
\quad\includegraphics[width=.495\textwidth]{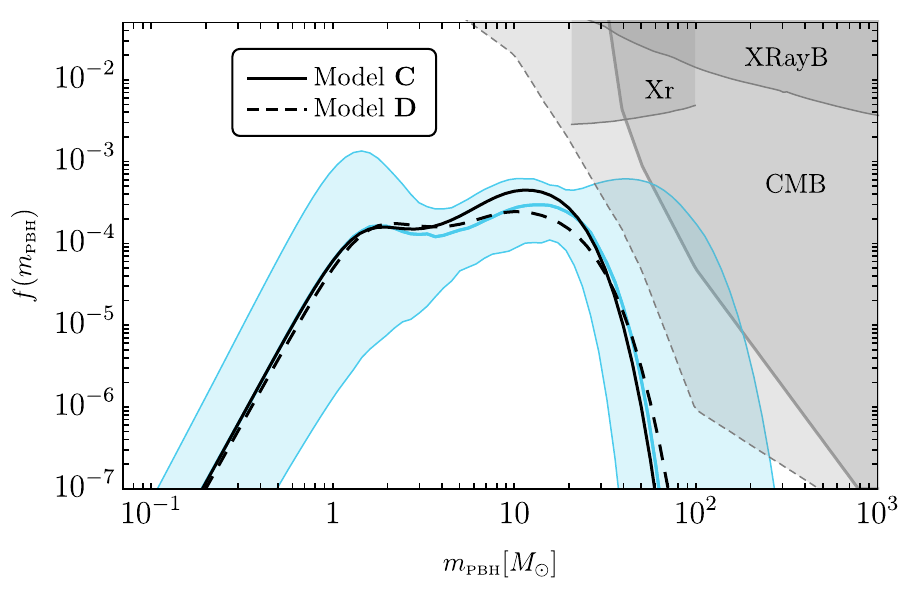}$$
	\caption{
	\auu{
{\bf Left panel:} PBH mass distribution as a function of the PBH mass. 
We show the curves that correspond to Model~\textbf{C} and Model~\textbf{D} in Table~\ref{Tab:ModelTab}.
We show the following constraints (see Ref.~\cite{Green:2020jor} for a review and\,\href{github.com/bradkav/PBHbounds}{\faGithub/bradkav/PBHbounds}).
Envelope of evaporation constraints (see also \cite{Saha:2021pqf,Laha:2019ssq,Ray:2021mxu}): EDGES\,\cite{Mittal:2021egv}, 
CMB\,\cite{Clark:2016nst}, INTEGRAL\,\cite{Laha:2020ivk,Berteaud:2022tws}, 511 keV\,\cite{DeRocco:2019fjq}, Voyager\,\cite{Boudaud:2018hqb}, 
EGRB\,\cite{Carr:2009jm};
microlensing constraints from the Hyper-Supreme Cam (HSC)~\cite{Niikura:2017zjd}; 
microlensing constraints from EROS~\cite{EROS-2:2006ryy}; 
microlensing constraints from OGLE~\cite{Niikura:2019kqi}; 
Icarus microlensing event~\cite{Oguri:2017ock}; constraints from modification of the CMB spectrum due to accreting PBHs~\cite{Serpico:2020ehh}. 
The yellow band 
corresponds to the allowed region for a PBH mass function $\propto m_\PBH^{-0.5}$ consistent with the HSC microlensing candidate event\,\cite{Niikura:2017zjd} (see also \cite{Sugiyama:2020roc}).
{\bf Right panel:} Same as the left panel but we zoom in on the solar-mass region. 
To guide the eye, we add the posterior distribution assuming GW190814 is
interpreted as a PBH binary (cyan band, same as in Fig.~\ref{fig:PBH constraints}).
}
	}
\label{fig:fPBHModel}
\end{figure*}

At the right-side edge of the plateau, the curvature power spectrum is characterized by a second bump-like feature (cf. Ref.~\cite{Franciolini:2022pav} for a detailed discussion about its formation). 
As in Ref.~\cite{Franciolini:2022pav}, we will exploit the bump at the left-side edge of the plateau for the generation of a solar-mass population of PBHs while the bump at the right-side edge of the plateau will be responsible for the generation of much lighter PBHs in the asteroid mass range. 
To this end, Eq.~\eqref{M-k}, together with the approximate horizon crossing condition $N = \log(k/k_{\star})$, gives a good intuition about how to choose the values of $N_{\rm I}$ and $N_{\rm III}$.
We compute the full PBH mass distribution following the formalism introduced in Sec.~\ref{sec PBH MF}.
Intuitively, 
the PBH abundance roughly scales as $\approx \exp \llp -1/\mathcal{P}_{\zeta}(k)\rrp$
and in our model it will be dominated by the two bumps of $\mathcal{P}_{\zeta}(k)$ where the latter takes its largest values. 
In addition, the abundance of heavier PBHs will be further boosted by the effect of the QCD phase transition. 
We expect, therefore, two peaks in the mass distribution of PBHs, one in the solar mass range and the other in the asteroid mass range.
In between the two peaks, we expect the typical redshift-induced scaling $m_\PBH^{-1/2}$ associated to scale invariant power spectra (because of the plateau in between the two bumps).

The above expectations are confirmed by the numerical result shown in Fig.~\ref{fig:fPBHModel}.  
Model \textbf{C} corresponds to the mass distribution given by the solid black line. 
In the same figure, we also plot a fourth realization of our model, dubbed Model~\textbf{D} in Table~\ref{Tab:ModelTab}, that, contrary to the previous case, is characterized by $\eta_{\rm III} \neq 0$; the corresponding mass distribution is given by the dot-dashed black line and features, as expected, a violation of the scaling $m_\PBH^{-1/2}$ in between the two peaks. 
The rationale behind the different choice of $\eta_{\rm III}$ that distinguishes Model~\textbf{C}  from Model~\textbf{D} is discussed in details in Appendix~\ref{sec:Tach}.

Let us summarize here our findings:
\begin{itemize}[leftmargin=*]
    \item[{\it i)}] As discussed in Ref.~\cite{Franciolini:2022pav}, it is possible to tailor an USR dynamics that gives a population of asteroid-mass PBHs consistent with the abundance of dark matter observed in the present-day universe and, at the same time, a subpopulation of solar-mass PBHs. 
    Remarkably, what we have shown with our analysis is that this subpopulation of solar-mass PBHs has the right features to
    explain a fraction of GWTC-3 events.
    \item[{\it ii)}] As in Sec.~\ref{sec:OnlySolar}, the cutoff $k_\text{\tiny min}$ (equivalently, $\ml$) arises naturally as a consequence of the sharp enhancement of the power spectrum (with respect to CMB values) that is essential for the generation of a sizable abundance of PBHs. 
    \item[{\it iii)}] The bulk of the PBH distribution in the solar-mass range is given by the bump at the left-side edge of the plateau in the curvature power spectrum (cf. the right panels of Figs.~\ref{fig:PowerSpectrum} and~\ref{fig:fPBHModel}). The form of this bump is shaped by curvature modes that cross the horizon during and immediately after the USR phase. 
    Despite its simplicity, Eq.~\eqref{PS_zeta} captures well the form of the bump. In particular, the red tilt $n_s<1$ is absolutely crucial since it models the transition between the $k^4$ growth of the power spectrum and the subsequent plateau.
    \item[{\it iv)}] The PBH mass distribution in the solar-mass range is practically
    insensitive to the cutoff $k_\text{\tiny max}$ (equivalently, $\ms$). We exploit such freedom to connect the part of the power spectrum that matches the ansatz in Eq.~\eqref{PS_zeta} with the plateau that in our model bridges solar- to asteroid-mass PBHs.
\end{itemize}

Let us mention that, while in this draft we focused on a model of USR inflation, the key ingredient is the peculiar shape of the power spectrum. 
Thus, we expect similar results would hold for any early universe model that can produce a similar curvature power spectrum.

\subsection{Reconstructed inflaton potential}\label{sec:inflatonpotential}
Once the Hubble parameters $\epsilon$ and $\eta$ are determined, one can derive the inflationary potential by computing~\cite{Franciolini:2022pav}
\begin{align}
V(N) & = V(N_{\rm ref})\exp\left\{
   -2\int_{N_{\rm ref}}^{N}dN^{\prime}\left[\frac{\epsilon(3-\eta)}{3-\epsilon}\right]
   \right\}\,,\label{eq:recPot1}\\
\phi(N) & = \phi(N_{\rm ref}) - \int_{N_{\rm ref}}^N dN^{\prime}\sqrt{2\epsilon}\,.   \label{eq:recPot}
\end{align}
The combination of $V(N)$ and $\phi(N)$ allows reconstructing the 
profile of the inflationary potential $V(\phi)$  in field space. 
Equation~\eqref{eq:recPot1} highlights the advantages of our approach, based on parametrizing the inflationary dynamics in terms of the Hubble parameters, as in Eq.~\eqref{eq:MainEqEta}. 
As both $\epsilon$ and $\eta$ enter in the exponent of Eq.~\eqref{eq:recPot1}, their determination is free from the fine-tuning necessary when working directly on a parametrisation of the potential.

\begin{figure}[!t]
	\centering
	\includegraphics[width=0.495\textwidth]{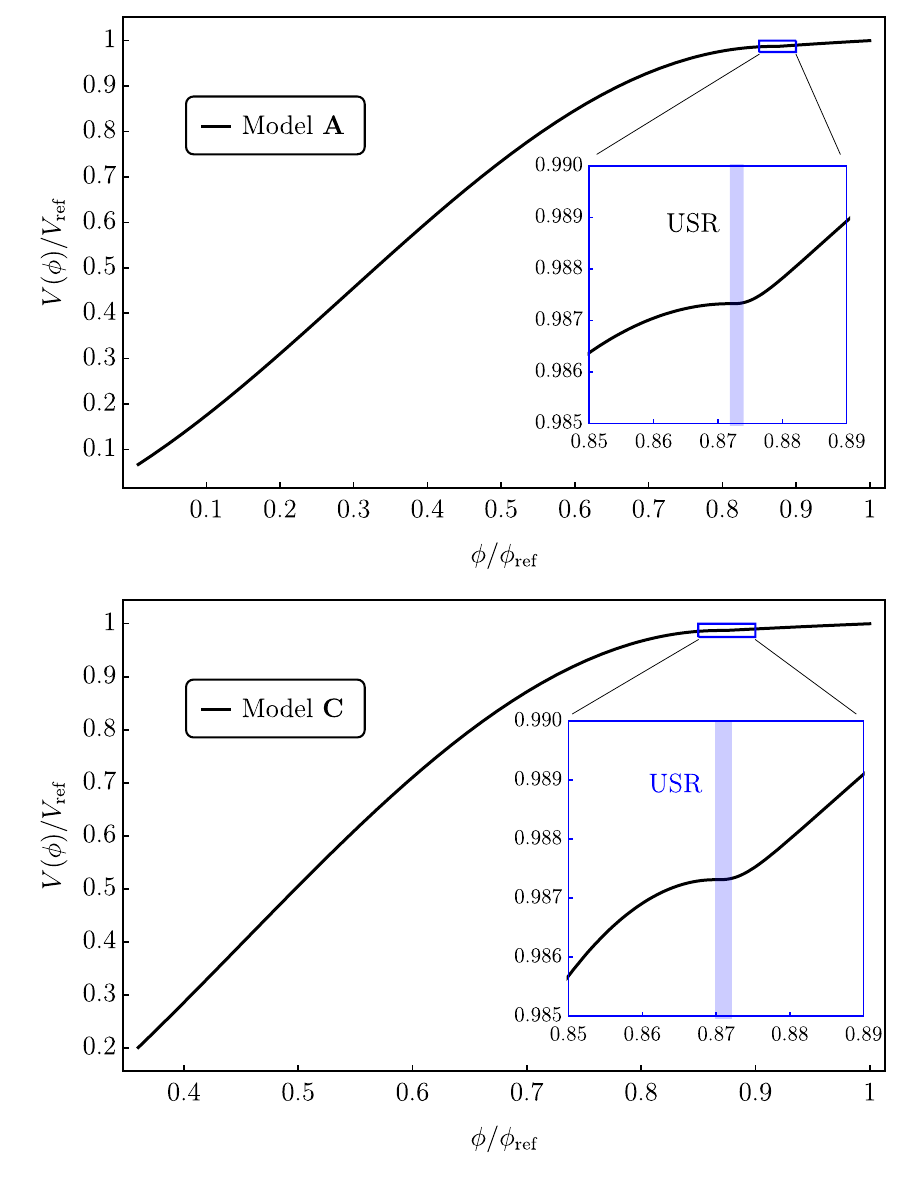}
	\caption{
	Reconstructed inflationary potentials computed by means of the approach discussed in Ref.~\cite{Franciolini:2022pav} in the case of Model~\textbf{A} (top panel) and Model~\textbf{C} (bottom panel). 
	In the inset plots we zoom in on the USR region.
}
\label{fig:ReconstructedPotentials}
\end{figure}

In Fig.~\ref{fig:ReconstructedPotentials} we show the reconstructed inflationary potentials in the case of Model~\textbf{A} and Model~\textbf{C}, cf. Ref.~\cite{Franciolini:2022pav} for details. 
In both cases, we denote as $\phi_{\rm ref}$ the field value at which we fit CMB observables (and define $V_{\rm ref}=V(\phi_{\rm ref})$). 
Both models exhibit the presence of a transition region that corresponds to the USR phase. The blue band limits, in field space, the $e$-fold time interval $N_{\rm II} < N < N_{\rm I}$. 
It is interesting to notice that the inflaton velocity  is drastically reduced by the USR phase enhancing the power spectrum. As a consequence, generating the extended plateau in Fig.~\ref{fig:PowerSpectrum} (which is absent in Models~\textbf{A} and \textbf{B}), only requires the inflaton to remain in the second slow-roll configuration for a very short displacement in field space (still related to numerous e-foldings). 
Therefore, small modifications to the inflaton potential are needed in order to generate the various models, as shown in Fig.~\ref{fig:ReconstructedPotentials}.

We now move to describe the phenomenological consequences of our model as far as stochastic GWs are concerned.

\subsection{Predictions: stochastic GWs from PBHs}\label{Sec:StocaGWs}

Once the free parameters of our model have been fixed by the condition $f_\PBH = {\cal O}(1)$ and the consistency with the posterior spectrum of curvature perturbation compatible with the primordial interpretation of GW190814, 
we are in the position to compute the {\it predicted} signal of scalar-induced stochastic GWs~\cite{Tomita:1975kj,Matarrese:1993zf,Acquaviva:2002ud,Mollerach:2003nq,Ananda:2006af,Baumann:2007zm,Zhou:2021vcw,Zhang:2022dgx,Domenech:2021ztg} and the SGWB produced by PBH mergers \cite{Wang:2016ana,Wang:2019kaf,Bavera:2021wmw}.

\begin{figure}[!t]
	\centering
	\includegraphics[width=0.495\textwidth]{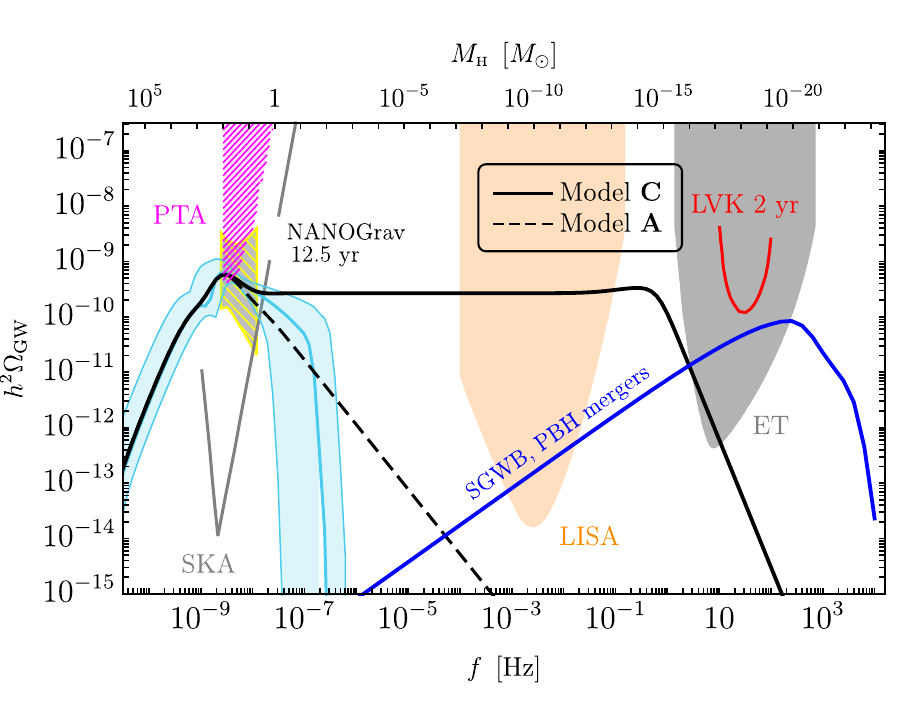}
	\caption{
Fraction of the energy density in GWs relative to the
critical energy density of the Universe as a function of the frequency. We show the power-law integrated sensitivity curves 
of future ground- and space-based GW experiments (two-years observation with LVK at design sensitivity, the Einstein Telescope and LISA, cf. Ref.~\cite{Bavera:2020inc}) as well as previous Parkes Pulsar
Timing Array (PTA) constraint~\cite{Shannon:2015ect}, NANOGrav putative
band~\cite{NANOGrav:2020bcs} and SKA projected sensitivity~\cite{Janssen:2014dka}. We plot the signals predicted by our model in the two realizations \textbf{A} and \textbf{C} proposed in Table~\ref{Tab:ModelTab}. 
}
\label{fig:StochGW}
\end{figure}

As shown 
in Refs.~\cite{DeLuca:2020agl,Franciolini:2022pav}, the 
scalar-induced GW signal in the case of a very broad power spectrum, like the ones we are considering in Model~\textbf{C} and \textbf{D}, covers the wide range of frequencies 
$10^{-9} \lesssim f/{\rm Hz}\lesssim 1$, and its amplitude is both compatible with the putative signal recently reported by the NANOGrav Collaboration~\cite{NANOGrav:2020bcs} at about $f = {\cal O}(10^{-9})$ Hz (also independently supported by other Pulsar Timing Array experiments \cite{Goncharov:2021oub,Chen:2021rqp,Antoniadis:2022pcn}) \footnote{The band compatible with recent NANOGrav observations is in partial tension with previously derived PTA constraints. According to the NANOGrav Collaboration \cite{Arzoumanian:2020vkk}, the improved priors for the intrinsic pulsar red noise used in the most recent analysis relaxes previous bounds.}
and detectable by future space-based GW interferometers like LISA \cite{LISACosmologyWorkingGroup:2022jok} (in the interval $10^{-4} \lesssim f~[{\rm Hz}]\lesssim 10^{-1}$). 
We confirm this expectation in the case of our models in Fig.~\ref{fig:StochGW}. The scalar-induced signal of GWs is proportional to $\mathcal{P}_{\zeta}^2(k)$ (see Eq.~\eqref{eq:Transfer} below) and, therefore, it inherits its shape. The bump at the left-side edge of the plateau falls precisely inside the contour favored, at the 2-$\sigma$ level, by the putative signal reported by NANOGrav (see also \cite{Vaskonen:2020lbd,Kohri:2020qqd,Ashoorioon:2022raz}); this is interesting because it means that our dynamics may predict a peculiar frequency dependence that could be tested by future pulsar timing array measurements.

For completeness, we also show the stochastic signal of scalar-induced GWs computed using the posterior distribution, expressed in terms of the parameters of the power spectrum in Eq.~\eqref{PS_zeta}, assuming GW190814 is a PBH binary (cyan region).  
Interestingly, we note that the part of the signal that is most constrained by the numerical analysis falls precisely in the frequency band of the NANOGrav region.

It should be noted that, as in  Refs.~\cite{DeLuca:2020agl,Franciolini:2022pav}, we compute the scalar-induced GW signal in Fig.~\ref{fig:StochGW} assuming a radiation-dominated universe while a more accurate computation should include the effect of the quark-hadron phase transition through the change of the number of effective degrees of freedom, the EoS parameter $w$, and the sound
speed $c_s$.

To comment more quantitatively on this point, let us write the amplitude of induced GW spectral density measured today in the form~\cite{Domenech:2021ztg}
\begin{align}
 \Omega_{{\rm GW},0}h^2 & = 0.39\, \Omega_{r,0}h^2\underbrace{\left[\frac{g_*(T_H)}{106.75}\right]
 \left[\frac{g_{*,s}(T_H)}{106.75}\right]^{-\frac{4}{3}}
 }_{\equiv\,c_g(T_H)}
 \Omega_{{\rm GW},H},\label{eq:cg}
 \end{align}
 with 
 \begin{align}
 & \Omega_{{\rm GW},H} = \label{eq:Transfer}
 \left(\frac{k}{k_H}\right)^{-2b}
 \int_{0}^{\infty}\d v
 \int_{|1-v|}^{1+v}\d u
 \mathcal{T}(u,v)
 \mathcal{P}_{\zeta}(ku)\mathcal{P}_{\zeta}(kv),
\end{align}
where $b\equiv (1-3w)/(1+3w)$, $\Omega_{r,0}$ is the density fraction of 
radiation, $g_*(T)$ and $g_{*,s}(T)$  the temperature-dependent
effective degrees of freedom for energy density and
entropy density, $\mathcal{T}(u,v)$ the transfer function that fully depends on the universe EoS; the subscript $_H$ stands for the time when
induced GWs of given wavenumber $k$ are sufficiently inside the cosmological horizon to be treated
as a radiation fluid in an expanding universe.

\begin{figure}[!t]
	\centering
	\includegraphics[width=0.495\textwidth]{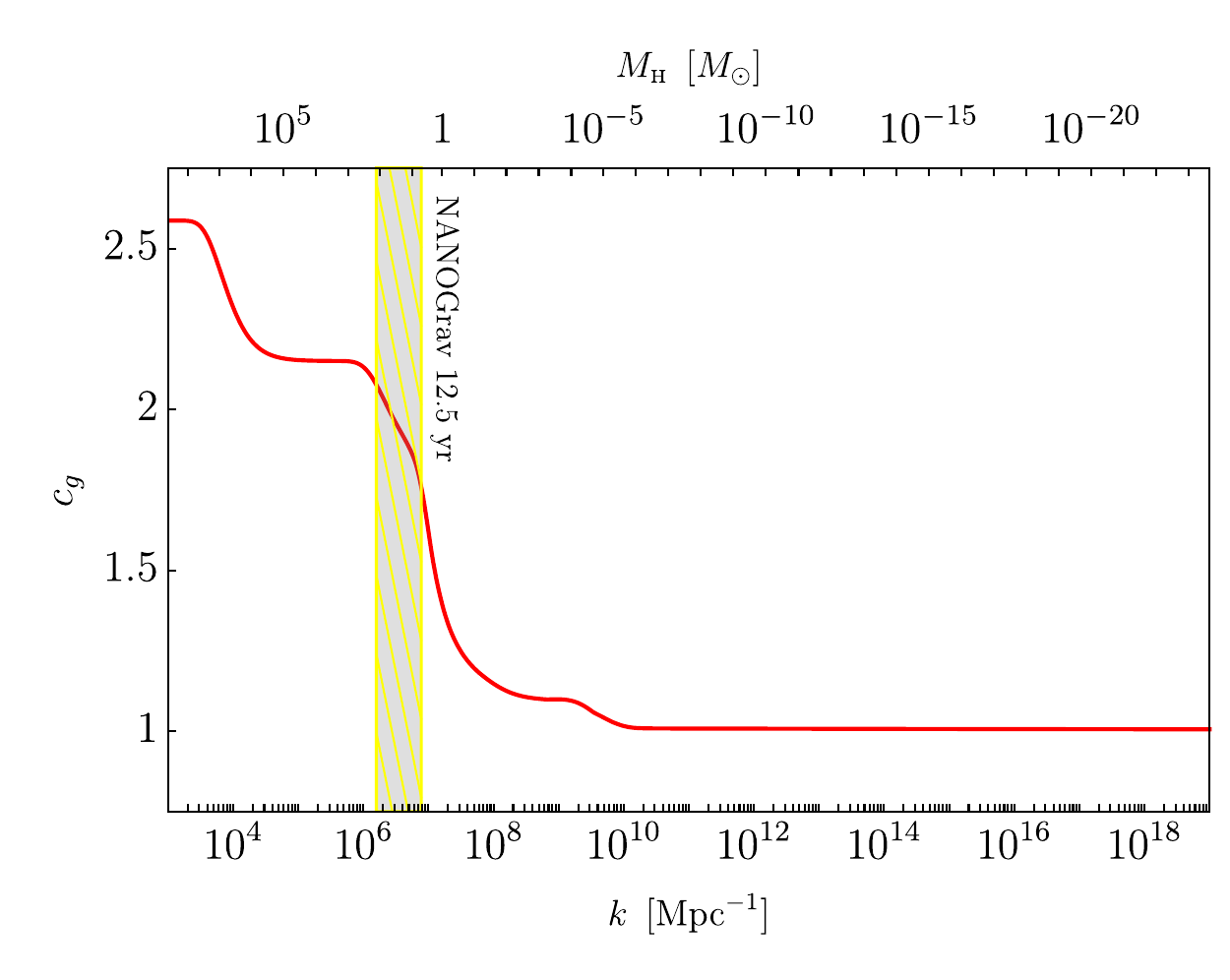}
	\caption{
Evolution of the factor $c_g$ defined in Eq.~\eqref{eq:cg} as a function of the horizon mass $M_H$ (top $x$-axis) and the comoving wavenumber $k$ (bottom $x$-axis). 
The yellow region shaded in gray marks the frequency range (converted into $k = 2\pi f$) 
$2.4\times 10^{-9} \leqslant f\,[{\rm Hz}] \leqslant 
1.2\times 10^{-8}$ favored by the NANOGrav putative signal.
	}
\label{fig:cgPlot}
\end{figure}

There are two effects induced by the thermal history of the universe across the QCD era. 
First, $c_g(T_H)$ is constant and equal to unity only for perturbation modes re-entering the Hubble horizon deep in the radiation epoch;
as the left-side edge of the curvature power spectrum re-enter the Hubble horizon at around the quark-hadron phase transition, the the reduction of $g_*$ and $g_{*,s}$ induce a modulation of the SGWB spectrum.
In Fig.~\ref{fig:cgPlot} we show the evolution of $c_g(T)$ trading its temperature dependence for the dependence on  the horizon mass $M_H$ (top $x$-axis) as \cite{Byrnes:2018clq}
\begin{equation}
M_H = 1.5\times 10^5
\, M_{\odot}
\llp \frac{g_*(T)}{10.75}\rrp^{-1/2}
\lp \frac{T}{{\rm MeV}} \rp^{-2}
\end{equation}
and the comoving wavenumber $k$ (bottom $x$-axis) using Eq.~\eqref{M-k}. 
To guide the eye, we superimpose the frequency range (translated into a wavenumber interval by means of $f = k/2\pi$) favored by the putative NANOGrav signal. 
We conclude that modeling the temperature dependence of the factor $c_g$ enhances the GW signal in the low-frequency part of the spectrum relevant for the comparison with pulsar timing array data (about a factor 2 but with some frequency dependence). 
The second physical effect is induced by the dependence of  ${\cal T}(u,v)$ on both $w$ and $c_s$ in Eq.~\eqref{eq:Transfer}.

Both effects have been discussed in Ref.~\cite{Abe:2020sqb} (see also Ref.~\cite{Saikawa:2018rcs}) specifically addressing the thermal history induced by the QCD phase transition. 
For a scale-invariant power spectrum, it turns out that the evolution of $w$ and $c_s$ only induce a sub-leading modification with respect to the effect of the changing effective degrees of freedom.
However, in view of future tests of the putative NANOGrav signal, it would be certainly important to include, following Ref.~\cite{Abe:2020sqb}, the full effect of the QCD phase transition in the computation of the spectrum of induced GWs, and nail down 
more precisely the frequency dependence of the signal that our inflationary dynamics predicts in the range relevant for pulsar timing array measurements. We leave this investigation for future work.

In Fig.~\ref{fig:StochGW} we also show the SGWB produced by the population of mergers in the solar mass range, again under the assumption of GW190814 being a primordial binary and adopting the best-fit values from Table~\ref{TbPBH:posparameters}. We do not show the astrophysical contribution as it strongly depends on the rate evolution above the peak expected around redshift $z\approx 2$, following the star formation rate~\cite{Madau:2014bja}. This can, therefore, be regarded as a lower bound on such a background from mergers in the stellar mass range. 

We compute the spectrum at frequency $\nu$ 
by integrating the PBH merger rate across the cosmological history as
\begin{equation}
\Omega_\text{\tiny GW}(f)
= 
\frac{f}{\rho_0} \int_0^{{f_3}/{f}-1} 
\d z 
\,\frac{{\cal R}_\PBH }{(1+z)H(z)} 
\frac{\d E_\text{\tiny GW} (f_s)}{\d f_s},
\label{SGWBmerger}
\end{equation}
in terms of the redshifted source frequency $f_s =f (1+z)$, the present energy density $\rho_0 = 3 H_0^2/8\pi$, the Hubble constant $H_0$, and the energy spectrum of GWs denoted $\d E_\text{\tiny GW}/{\d \nu_s}$. 
Notice that Eq.~\eqref{SGWBmerger} implicitly requires an integration over $m_{1,2}$. Finally, $f_3$ controls the maximum redshift beyond which mergers cannot contribute to a given spectral frequency $f$ and it is determined by the effective cut-off of the spectrum (see Appendix A of Ref.~\cite{Bavera:2021wmw} and references therein for more details). 

It may be possible to distinguish the contribution to the SGWB coming from either PBH or ABH/NS mergers thanks to their predicted different merger rate evolution. 
The SGWB results from the integrated contribution of the merger history
\cite{2016PhRvL.116m1102A,2018PhRvL.120i1101A,2020arXiv200109663D,2021PhRvD.103d3002P,2021arXiv211201119P}, and PBHs are characterised by an extended rate growth reaching much before star formation.
Therefore, given the same detection rate of resolved binaries at low redshift, a PBH contribution produces a larger SGWB. Correlating rates of individual detections and the SGWB amplitude may allow to set a lower bound on the primordial contribution at future third-generation experiments \cite{Bavera:2021wmw} (see also \cite{Chen:2019irf,Mukherjee:2021ags}).

The peak frequency of GWs emitted from BH mergers is close to innermost stable circular orbit frequency, 
$f_\text{\tiny ISCO} 
\simeq  4.4 \times 10^3 \, {\rm Hz} 
\lp {M_{\odot}}/{(m_1 + m_2)} \rp$.
As the solar-mass and intermediate-mass PBH population is bounded to be below ${\cal O}(10^2) M_\odot$ by CMB accretion constraints (cf. Fig.~\ref{fig:PBH constraints}) and eventually by FIRAS/PIXIE data, the SGWB cannot get sizeable contributions at frequencies smaller than ${\cal O}(10)$Hz, if not from the $\approx f^{2/3}$ tail produced by the inspiral phase. 
Therefore, the contribution to the SGWB from PBHs with masses smaller than ${\cal O}(10^2) M_\odot$\footnote{
One also expects GWs signals in the LISA band from mergers of supermassive BH binaries \cite{Sesana:2007sh,Banks:2021kbf}, which we do not quantify in Fig.~\ref{fig:StochGW}. As the majority of those mergers will be resolved, and subtracted, they would marginally contaminate SGWB searches. 
} falling in the LISA band cannot overcome the one induced at second order by the formation of an asteroid mass population of PBHs explaining the dark matter~\cite{Bartolo:2018evs,Bartolo:2018rku}.\footnote{These constraints were neglected in Refs.~\cite{Bagui:2021dqi,Braglia:2021wwa}.}
Finally, we neglect the second peak potentially generated by the asteroidal mass PBHs which would fall at much higher frequencies, of interest for UHF-GW experiments \cite{Aggarwal:2020olq} (see in particular Ref.~\cite{ Franciolini:2022htd} and references therein).

\section{Conclusions and outlook}\label{sec:conclusions}

We have performed the first Bayesian PBH population inference on GW data
directly using ab-initio curvature power spectrum parameters and including the effect of the modified threshold due to the QCD EoS~\cite{Muscoinprep}. We critically confronted this state-of-the-art PBH model with LVK phenomenological population models that describe the GWTC-3 catalog both in the NS and in the BH mass range.

We found that the upper bound on the PBH abundance is consistent with previous analyses ($f_\PBH\lesssim 10^{-3}$) and it is stronger than other constraints in this mass range. Nonetheless, we also found marginal evidence for extra information in the data on top of the LVK phenomenological distributions, which may be captured by a primordial subpopulation of binaries. Indeed, a PBH subpopulation can explain a fraction of GWTC-3 events, in particular binaries with light (such as the lower mass-gap event GW190814) or heavy (e.g., GW190521) components. Interestingly, the light events that are assigned the highest PBH likelihood by our inference happen also to be those which are more challenging to accommodate within standard astrophysical scenarios.

Intriguingly, our ab-initio PBH distribution allows us to make some falsifiable predictions: if some of the GWTC-3 events are primordial (in particular the lower mass-gap events GW190814, which is assigned $\approx 29\%$ probability of being primordial by our inference), then the merger rates in the subsolar mass range and in the lower mass gap are high enough to be detectable by future LVK runs.
In particular, the absence of subsolar mergers in O5 would automatically exclude the primordial origin of the light events within GWTC-3. 

Our work is just a first attempt to use ab-initio PBH models in GW population inference, and we hope it will be extended in several ways.
Most importantly, one should perform multi-population Bayesian inference by mixing our PBH model with astrophysical models for BH and/or NS binaries, similarly to what recently done in Ref.~\cite{Franciolini:2021tla}.

It is also possible to improve the PBH modelling, in particular by considering a different parametrised curvature spectrum (e.g. peaked Gaussian bump etc.~\cite{Gow:2020cou,Karam:2022nym}), primordial non-Gaussianities~\cite{Franciolini:2018vbk,Atal:2019cdz,DeLuca:2021hcf,Taoso:2021uvl, Biagetti:2021eep,Ferrante:2022mui}, accretion effects~\cite{DeLuca:2020bjf,DeLuca:2020fpg,DeLuca:2020qqa}, and spin information in the inference~\cite{Franciolini:2022iaa}. 
Eventually, extending the numerical simulations of Ref.~\cite{Muscoinprep} used here for a sufficient set 
of shapes of the collapsing overdensities would allow capturing the threshold and mass dependence on deviations from the nearly scale invariant spectra (i.e. $n_s$), allowing us to include a full dependence of parameters of collapse on each specific spectral mode (or $M_H$ in our formalism) beyond the effect of the QCD EoS.

On the theory side, building on the reverse engineering approach recently devised in Ref.~\cite{Franciolini:2022pav}, we have mapped the GW data-driven curvature power spectrum into an USR inflationary model.
We remark that the reverse engineering approach used in this analysis goes beyond the mere parametrization of the dynamics given in Eq.~\eqref{eq:MainEqEta} since it allows to numerically reconstruct the inflationary potential, 
which we showed in Fig.~\ref{fig:ReconstructedPotentials} in the case of Model~\textbf{A} and \textbf{C}.
It is legitimate to ask whether it would be possible to move directly to the analysis based on some scalar potential and skip the reverse engineering approach of Ref.~\cite{Franciolini:2022pav}. The answer is certainly positive; however, working directly at the level of the potential may not give the same control on the shape of the power spectrum compared to the reverse engineering approach, thus making the analysis much more difficult and way less transparent. 
Ultimately, it may be possible to 
 run the population inference directly on the fundamental coupling constants of a given inflationary model
and investigate the possible quantum field theory origin of the reconstructed potential. We leave these tasks for future work.

We also confirmed a remarkable feature of this approach~\cite{Franciolini:2022pav}, namely that a single USR model can consistently accommodate a double-peaked PBH mass function. The dominant peak occurs in the asteroid-mass range and it is responsible for explaining the totality of the dark matter in small PBHs, while the second (subleading) peak is produced by the enhancement beyond the effect of the QCD phase transition and provides a detectable PBH merger rate in the band of current and future GW detectors.

An important by-product of our analysis is that the inferred value of the (red) tilt of the spectrum makes the above scenario fully compatible with the absence of subsolar mergers in GWTC-3, although it also predicts that subsolar mergers and more lower-mass gap events can be detectable in the future.

Finally, we showed that other falsifiable predictions of the designed curvature power spectrum are: (i)~a detectable scalar-induced SGWB signal compatible with the NANOGrav putative measurement and detectable by future PTA observations and by LISA; and (ii)~a SGWB produced by PBH mergers which will be detectable by the Einstein Telescope. In both cases, an urgent extension of our work is to properly account for the full richness of the ab-initio PBH model (including the effects of the QCD phase transition) in shaping the frequency dependence of these SGWB signals.
These advancements are required in order to fully exploit the constraining power of GW data soon to be available.

\begin{figure*}[t!]
	\centering
	\includegraphics[width=0.39\textwidth]{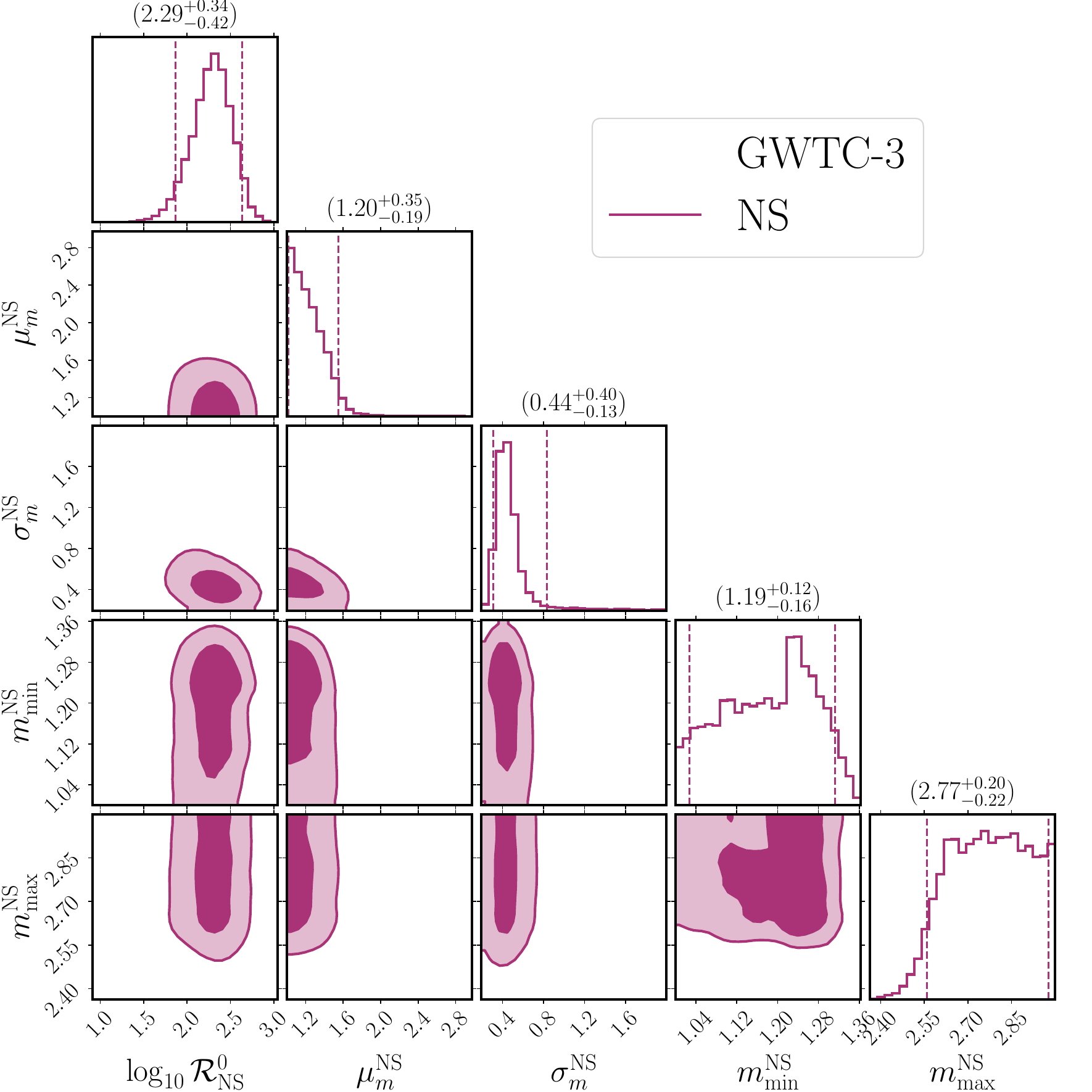}
	\includegraphics[width=0.6\textwidth]{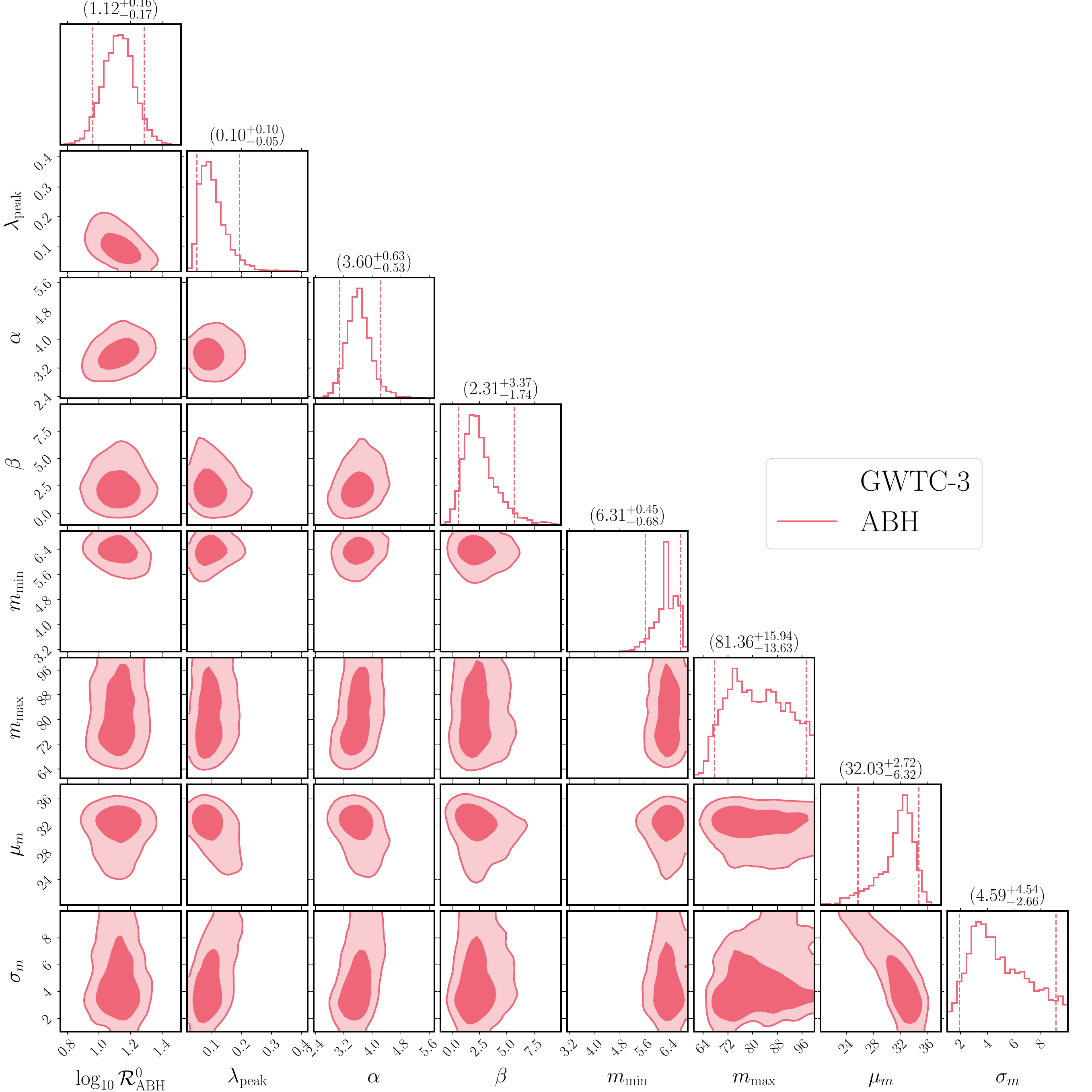}
	\caption{
Posterior distribution for the hyperparameters characterising the LVK NS (left) and ABH (right) phenomenological models, obtained within a single-population inference of the GWTC-3 catalog.}
\label{fig:pos_LVKC}
\end{figure*}

\begin{center}
    {\bf Note added} 
\end{center}

After this work was completed, we became aware of Ref.~\cite{Clesseetalinprep}, which independently explores the role of the QCD phase transition in shaping the PBH mass function and PBH merger rate distribution. Ref.~\cite{Clesseetalinprep} provides a detailed numerical description of the threshold for PBH formation, which looks to be in reasonable agreement with the results of Ref.~\cite{Muscoinprep} used in our analysis. 
However the computation of the mass distribution in~\cite{Clesseetalinprep} does not include the dependence of the critical collapse and density variance on the EoS, leading to an approximate mass function which does not take into account the additional pile up of PBHs around the solar mass produced by the critical collapse.
More importantly, they restrict the discussion to a nearly scale invariant shape of the enhanced spectrum at small scales with tilt $n_s = 0.965 \divisionsymbol  0.975$, 
and do not compare the PBH model to GWTC-3 data through a Bayesian analysis. 

The values of $f_\PBH$ considered in Ref.~\cite{Clesseetalinprep} are larger than ours by $2\divisionsymbol3$ orders of magnitude and compatible with $f_\PBH\approx1$.
This discrepancy can be attributed to various differences, that we list in the following:
{\it i)}~a different suppression factor in the PBH merger rate formula is used in Ref.~\cite{Clesseetalinprep}.
While Ref.~\cite{Clesseetalinprep} used the analytical treatment of Refs.~\cite{Clesse:2020ghq,Bagui:2021dqi}, 
we adopt the results of Refs.~\cite{Raidal:2018bbj,Vaskonen:2019jpv,DeLuca:2020jug} (informed by N-body simulations), which give smaller values of $f_\PBH$, compatible with the analyses of Refs.~\cite{Ali-Haimoud:2017rtz,Hall:2020daa,Wong:2020yig,Hutsi:2020sol,DeLuca:2021wjr,Franciolini:2021tla} and also not excluded by other non-GW constraints.
{\it ii)} Ref.~\cite{Clesseetalinprep} restricts the parameter space to masses above $m_i>M_\odot$ and mass ratio larger than $q\gtrsim 0.1$ when computing\footnote{Also, Ref.~\cite{Clesseetalinprep} neglected the factor $2$ when deriving the merger rate as a function of primary (heavier) mass due to the two possible mass ordering in Eq.~\eqref{eq:diffaccrate}, which allows for both $m_1<m_2$ and $m_2<m_1$ by construction \cite{Raidal:2018bbj}.} 
\begin{equation}
    \frac{\d R_\PBH}{\d m_1}
    \equiv
    \int \d m_2 \times 2\theta(m_1-m_2) \times 
     \frac{\d R_\PBH}{\d m_1\d m_2}.
\end{equation} 
This leads to a drastic reduction of the differential rate, which however would only be justified if LVK were unable to detect such neglected events. 
Using their approximated mass distribution and their choice of merger rate formula with $f_\PBH = 1$, we find LVK would have observed around $N_\text{\tiny det}(m_i<M_\odot) \approx 20$ mergers with at least one subsolar component during GWTC-3. 
With our modelling of the mass distribution, while fixing $n_s = 0.97$ and $f_\PBH=1$, we obtain $N_\text{\tiny det}(m_i<M_\odot) \simeq  36$. 
This means that the absence of subsolar detections
in the various LVK runs is incompatible with such mass function and large values of the abundance, as already pointed out in Ref.~\cite{Juan:2022mir}. This constraint is
automatically included in our MCMC analysis.
{\it iii)} 
Finally, our inclusion of a dominant contribution from ABH mergers to the GWTC-3 catalog only has a minor impact on the constraint on $f_\PBH$ (as showed by our single population analyses).


\begin{acknowledgments}
We are indebted with K.~Jedamzik, P.~Serpico, A.~Sesana, H. Veermäe, and S.~Young for insightful discussions.
We also thank A.~Escrivà, E.~Bagui, S.~Clesse for sharing their draft with us and for useful comments.
Some computations were performed at the Sapienza University of Rome on the Vera cluster of the Amaldi Research Center funded by the MIUR program “Dipartimento di Eccellenza” (CUP: B81I18001170001).
G.F. and P.P. acknowledge financial support provided under the European
Union's H2020 ERC, Starting Grant agreement no.~DarkGRA--757480 and under the MIUR PRIN programme, and support from the Amaldi Research Center funded by the MIUR program ``Dipartimento di Eccellenza" (CUP:~B81I18001170001).
The work of I.M. has received funding from the European Union’s Horizon2020 research and innovation programme under the Marie Skłodowska-Curie grant agreement No 754496. This work was supported by the EU Horizon 2020 Research and Innovation Programme under the Marie Sklodowska-Curie Grant Agreement No. 101007855. The work of A.U. is supported in part by the MIUR under contract 2017FMJFMW (PRIN2017).
This research has made use of data or software obtained from the Gravitational Wave Open Science Center (gw-openscience.org), a service of LIGO Laboratory, the LIGO Scientific Collaboration, the Virgo Collaboration, and KAGRA. LIGO Laboratory and Advanced LIGO are funded by the United States National Science Foundation (NSF) as well as the Science and Technology Facilities Council (STFC) of the United Kingdom, the Max-Planck-Society (MPS), and the State of Niedersachsen/Germany for support of the construction of Advanced LIGO and construction and operation of the GEO600 detector. Additional support for Advanced LIGO was provided by the Australian Research Council. Virgo is funded, through the European Gravitational Observatory (EGO), by the French Centre National de Recherche Scientifique (CNRS), the Italian Istituto Nazionale di Fisica Nucleare (INFN) and the Dutch Nikhef, with contributions by institutions from Belgium, Germany, Greece, Hungary, Ireland, Japan, Monaco, Poland, Portugal, Spain. The construction and operation of KAGRA are funded by Ministry of Education, Culture, Sports, Science and Technology (MEXT), and Japan Society for the Promotion of Science (JSPS), National Research Foundation (NRF) and Ministry of Science and ICT (MSIT) in Korea, Academia Sinica (AS) and the Ministry of Science and Technology (MoST) in Taiwan.
\end{acknowledgments}

\appendix

\section{Posterior distributions}\label{app:posteriors}

Here we present the posterior distributions resulting from the various Bayesian inferences. While the main phenomenological consequences of these posteriors are described in the main text, here we report few interesting insights on these distributions, also highlighting relevant correlations between parameters. 
We note that, in order to simplify the notation, we report mass scales in solar mass units $[M_\odot]$ and rate densities in units of $[{\rm Gpc^{-3} yr^{-1}}]$.

\begin{figure*}[t!]
	\centering
	\includegraphics[width=0.49\textwidth]{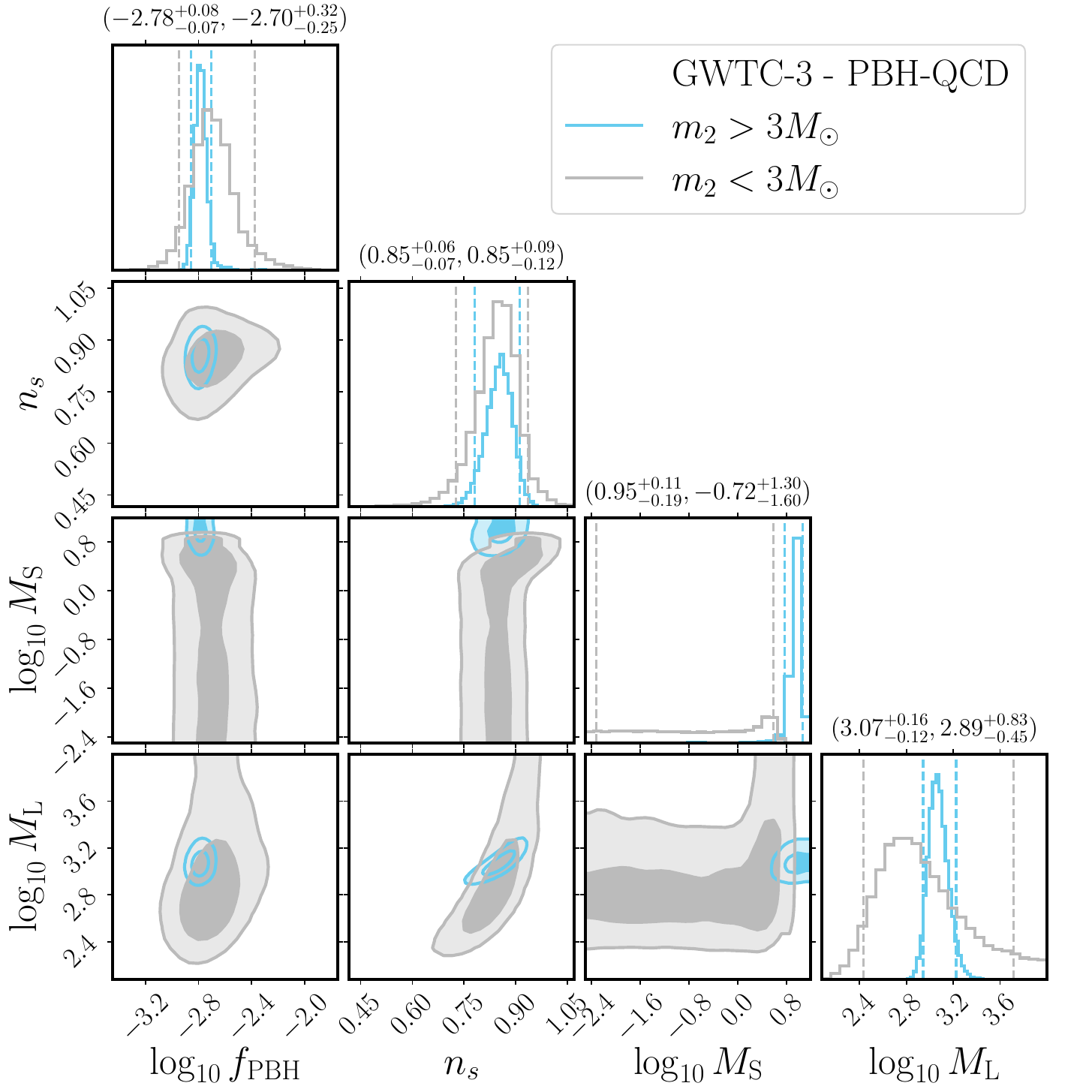}
	\includegraphics[width=0.382\textwidth]{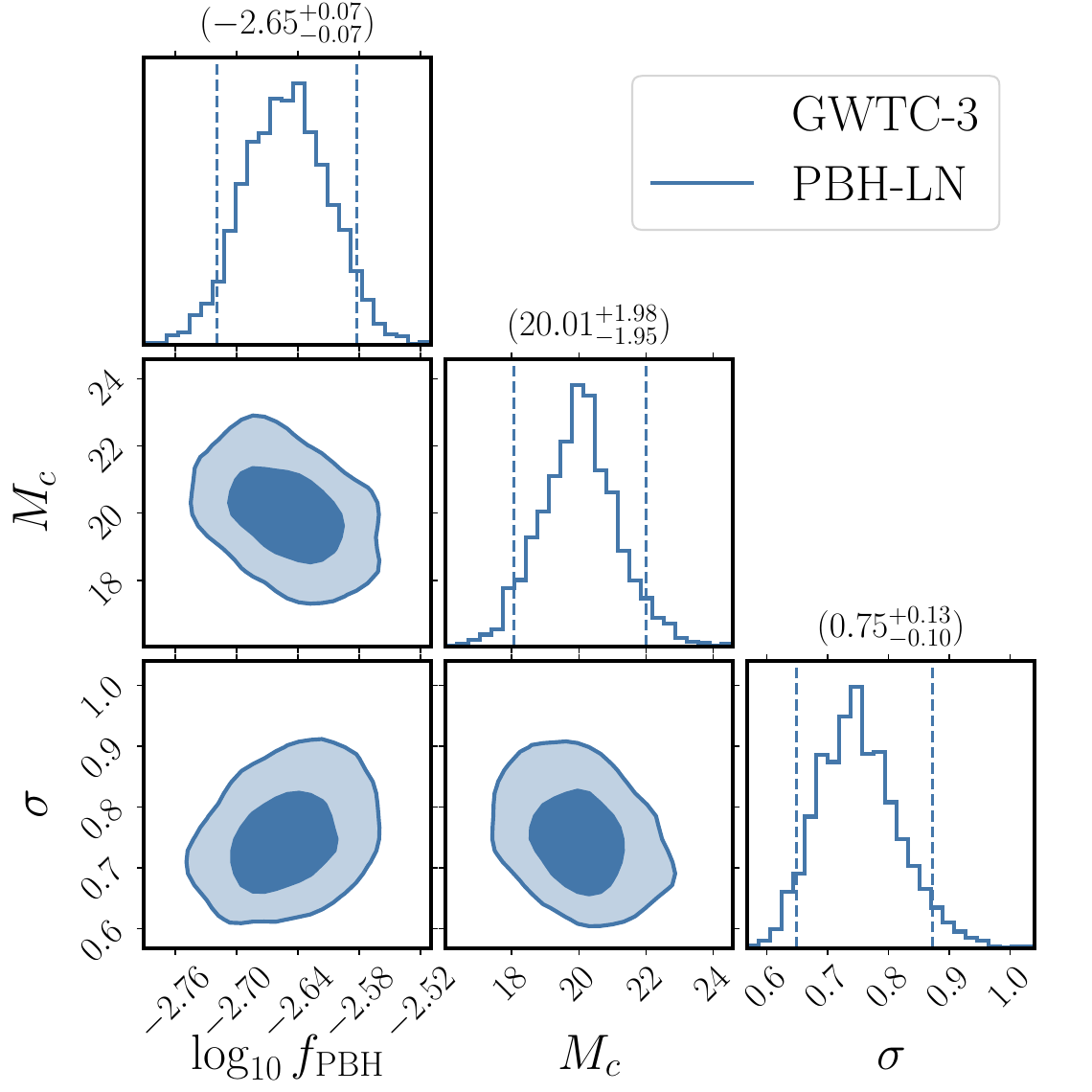}
	\caption{
Same as Fig.~\ref{fig:pos_LVKC} but for the PBH model assuming a mass distribution obtained from first principles with the effect of QCD softening of the EoS (left) or a lognormal mass function (right). Different colors correspond to separate fits of the light (heavy) portion of the catalog, with events characterised by $m_2< 3 M_\odot$ ($m_2> 3 M_\odot$) (canonically referred to events containing NS or not by the LVK analysis).
}
\label{fig:pos_PBH}
\end{figure*}

We start with the phenomenological NS or ABH channels. 
Focusing first on light events with $m_2\leq 3 M_\odot$ (i.e. left panel of Fig.~\ref{fig:pos_LVKC}), which is based on fitting only 7 detections, we observe that the posterior selects narrow mass distributions centered around the solar mass scale, while the mass cut-offs are poorly constrained beyond the basic requirement of encompassing all the events in this mass range. This can be observed by noticing that $m_\text{\tiny min}^\NS$ and $m_\text{\tiny max}^\NS$ flatten reaching the lower, or upper, boundaries of their respective prior range. The latter, in particular, is forced to be above 
$m_\text{\tiny max}^\NS \approx 2.5 M_\odot$ to include the secondary mass of GW190814.

The posterior distribution for the ABH model (i.e. right panel of Fig.~\ref{fig:pos_LVKC}) is instead better constrained, due to the larger number of detections (i.e. 69 in GWTC-3). Similar conclusions as for the NS case can be drawn on the minimum and maximum scales bracketing the ABH population $m_\text{\tiny min}$ and $m_\text{\tiny max}$.
In particular, the latter is bounded to be above $\approx 70 M_\odot$ to capture the mass gap event GW190521, whose primary mass is measured to be $m_1=95.3^{+28.7}_{-18.9} M_\odot$ \cite{2021arXiv211103634T} (see Table~\ref{Tb:parameters}). 
Also, a distinct anti-correlation is observed between the central scale and the width of the Gaussian peak, accounting for a small fraction of the intrinsic population of mergers. This is most probably enforced by the requirement of not overproducing mergers in the heavy tail (i.e. $m_1> \mu_m$) of the Gaussian contribution.

These general features are consistent with the results of the LVK population analysis reported in Ref.~\cite{2021arXiv211103634T}. 
While the overall posterior is fully compatible with LVK findings, slight deviations are observed, most probably introduced by the omission of spin information in our inference,
the absence of subdominant smoothing terms enforced at the tails of the mass distribution of the LVK model, and the adoption of a selection bias solely based on SNR computations, see Sec.~\ref{sec:selectionbias} (instead of the one based on the LVK injection campaign). 
The latter choice, which is customarily adopted in the recent literature (see e.g. \cite{Zevin:2020gbd, Franciolini:2021tla}), is required in our setting, as our analysis necessitate of consistently computing the selection bias also in the subsolar mass range, which is not captured by the LVK injection campaign \cite{ligo_scientific_collaboration_and_virgo_2021_5546676}
and which is crucial to enforce the constraint on the PBH population from the absence of subsolar detections in GWTC-3.
We do not expect these approximations to impact our results.

\begin{figure*}[t!]
	\centering
	\includegraphics[width=1\textwidth]{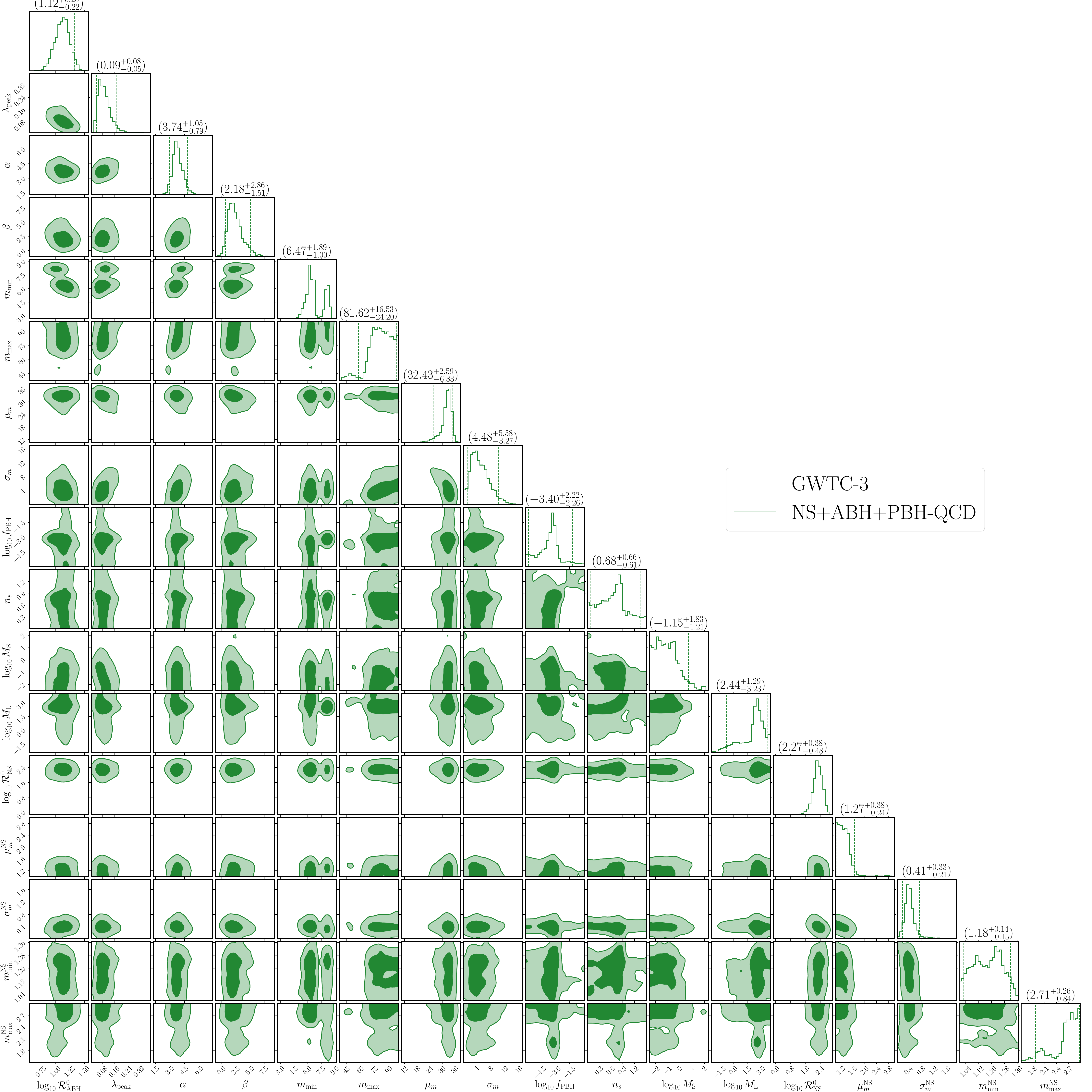}
	\caption{
	Posterior distribution for the hyperparameters in the mixed NS+ABH+PBH analysis where the primordial channels assumes the ab-initio mass distribution shaped by the QCD epoch. 
	}
	\label{fig:posterior_mixedall}
\end{figure*}

In Fig.~\ref{fig:pos_PBH} we report the analogous posteriors obtained assuming the PBH population alone explains the population of mergers. 
In the left panel, we report the result assuming an ab-initio mass distribution of PBHs derived from the curvature spectrum in Eq.~\eqref{PS_zeta} and the effect of the QCD epoch. 
The gray (cyan) color indicates the result of the inference on the light (heavy) events. The much larger uncertainties observed in the gray posterior is due to the aformentioned smaller sample of events with $m_2<3 M_\odot$. Strikingly, both analyses provides similar best-fit values for the hyperparameters ${\bm \lambda}_\PBH$, apart from $\ms$
which is unbounded from below  --~and allows for the presence of the QCD induced bump just above $M_\odot$~-- in the first case,
while it is constrained to be $\log_{10} (\ms/M_\odot) = 0.95_{-0.19}^{+0.11}$ in the second case. 
The correlation between $n_s$ and $\ml$ can be explained by noticing that the smaller values of the tilt, corresponding to \emph{redder} spectra, enhance $\psi(m_\PBH)$ at high masses, and the high mass cut-off $\ml$ needs to adjust to reduce the prominence of heavy mergers. 
Both analyses constrain the abundance to be much smaller than unity in this mass range, namely
$\log_{10}f_\PBH = -2.70_{-0.25}^{+0.32}$
and $\log_{10}f_\PBH = -2.78_{-0.07}^{+0.08}$, respectively.

In order to fully compare the constraint on the PBH abundance 
obtained with this single population analysis of GWTC-3 with previous literature, we also repeat the inference assuming the PBH population is described by a lognormal mass distribution of the form (e.g. \cite{Josan:2009qn})
\begin{equation}
    \psi (m_\PBH) = 
    \frac{1}{m_\PBH \sqrt{2 \pi \sigma^2}} 
    \exp\llp - \frac{\log^2(m_\PBH/M_c)}{2 \sigma^2}\rrp,
\end{equation}
where $M_c$ is the central mass scale and $\sigma$ the width. The right panel of Fig.~\ref{fig:pos_PBH} shows that the best-fit values of such scenario are consistent with results previously derived in the literature \cite{Ali-Haimoud:2017rtz,Raidal:2018bbj,DeLuca:2020qqa,Hall:2020daa,Hutsi:2020sol,Wong:2020yig} under analogous assumptions but with older datasets. 
In particular, the mass distribution is found to be broad and peaked at $\approx 20 M_\odot$. However, as discussed in the main text, such shape would overproduce mergers in the heavy portion of the catalog and give rise to a flat distribution of mass ratio, in sharp contrast with what observed in the data, see the detailed discussion in Sec.~\ref{sec:singlepopinference}. 
Finally, assuming a lognormal mass distribution gives a slightly less stringent, but statistically compatible, bound on the PBH abundance, which is found to be 
$\log_{10}f_\PBH = -2.65_{-0.07}^{+0.07}$.

Let us conclude this appendix by discussing the result of the mixed population inference. The corresponding posterior distribution is shown in Fig.~\ref{fig:posterior_mixedall}. 
In this case, the dataset includes all 76 detections in GWTC-3, and allows for ABH, NS and PBH mergers (with a QCD induced mass distribution) to contribute to the population of mergers. 

First, we notice that the ABH and NS models are mostly uncorrelated with each other, as can be observed by focusing the bottom-left $8\times5$ box. This is because they explain different sets of events, and a cross-talk between them would only be introduced by a dominant contribution from PBH mergers that, instead, can cover both mass ranges. Secondly, bimodal distributions are observed in various mass cut-offs. 
In particular, secondary peaks appear 
in the distributions of 
$m_\text{\tiny min}$,
$m_\text{\tiny max}$,
and 
$m_\text{\tiny max}^\NS$
when extreme events, potentially outliers of the astrophysical populations, 
such as GW190814, GW190924\_021846 and GW190521,
are explained by the PBH channel, respectively. 
In the portions of the posterior where PBHs are \emph{necessary}
to explain the various special events, the PBH abundance $f_\PBH$ is found to be bounded from below and takes values around
$f_\PBH \approx 10^{-3}$ (see also Table~\ref{TbPBH:posparameters}).
We also observe that the posterior shows small support for the simultaneous interpretation of GW190814 and GW190521 as PBH mergers, see the $(m_\text{\tiny max},m_\text{\tiny max}^\NS)$ (or $(6,17)$) panel of the posterior.

Finally, we observe that the hyperparameters of the PBH population are all characterised by a pronounced peak, corresponding to the high likelihood regions where PBHs contribute to the observations and improve the fit (see discussion in Sec.~\ref{sec:multipop}). However, $f_\PBH$ is not bounded from below and have a tail reaching the left boundary $f_\PBH = 10^{-6}$, where PBH contribution is negligible. This also implies that the remaining parameters have posterior distributions with broad tails filling the prior volume, with the small mass scale subject to the condition $\ms<\ml$.
A correlation between $n_s$ and $\ml$ is found also in this case, due to the requirement of not overproducing heavy mergers. 
We conclude by highlighting that the tail at large values of $f_\PBH$ reaching unity is strongly correlated with high values of $n_s>1$ (blue tilts) and small $\ms$. This is because one can evade the constraint from LVK measurements only with light enough populations strongly peaked at light mass scales below ${\cal O}(10^{-1}) M_\odot$, where the LVK sensitivity strongly deteriorates. 
This is reflected in the bound on $f_\PBH$ as a function of the average mass $\langle m_\PBH \rangle$ shown in Fig.~\ref{fig:PBH constraints2}.

\section{PBHs with widely different mass, a technical insight}\label{sec:Tach}

This is a slightly technical appendix providing key details useful to fully understand the rationale behind the numerical values of the parameters chosen in Table~\ref{Tab:ModelTab}. 

The two realizations of our model discussed in Sec.~\ref{sec:Plateau} (Model~\textbf{C} and \textbf{D})  are tuned to give the totality of dark matter in the form of PBHs. Concretely, this means that the logarithmic integral of the mass distribution gives unity, cf. Eqs.~\eqref{eqn:massFunctionlog} and~\eqref{fPBHintegral}. 
Since in the solar mass range the fraction of dark matter in the form of PBHs is constrained to be at most ${\cal O}(10^{-3})$, the integral is dominated by the peak in the asteroid mass range. In turn, 
this implies that the value of the curvature power spectrum in correspondence of the bump at the right-side of the plateau should be high enough to get the desired order-one abundance of PBHs.

\begin{figure*}[!htb!]
\begin{center}
\includegraphics[width=.9\textwidth]{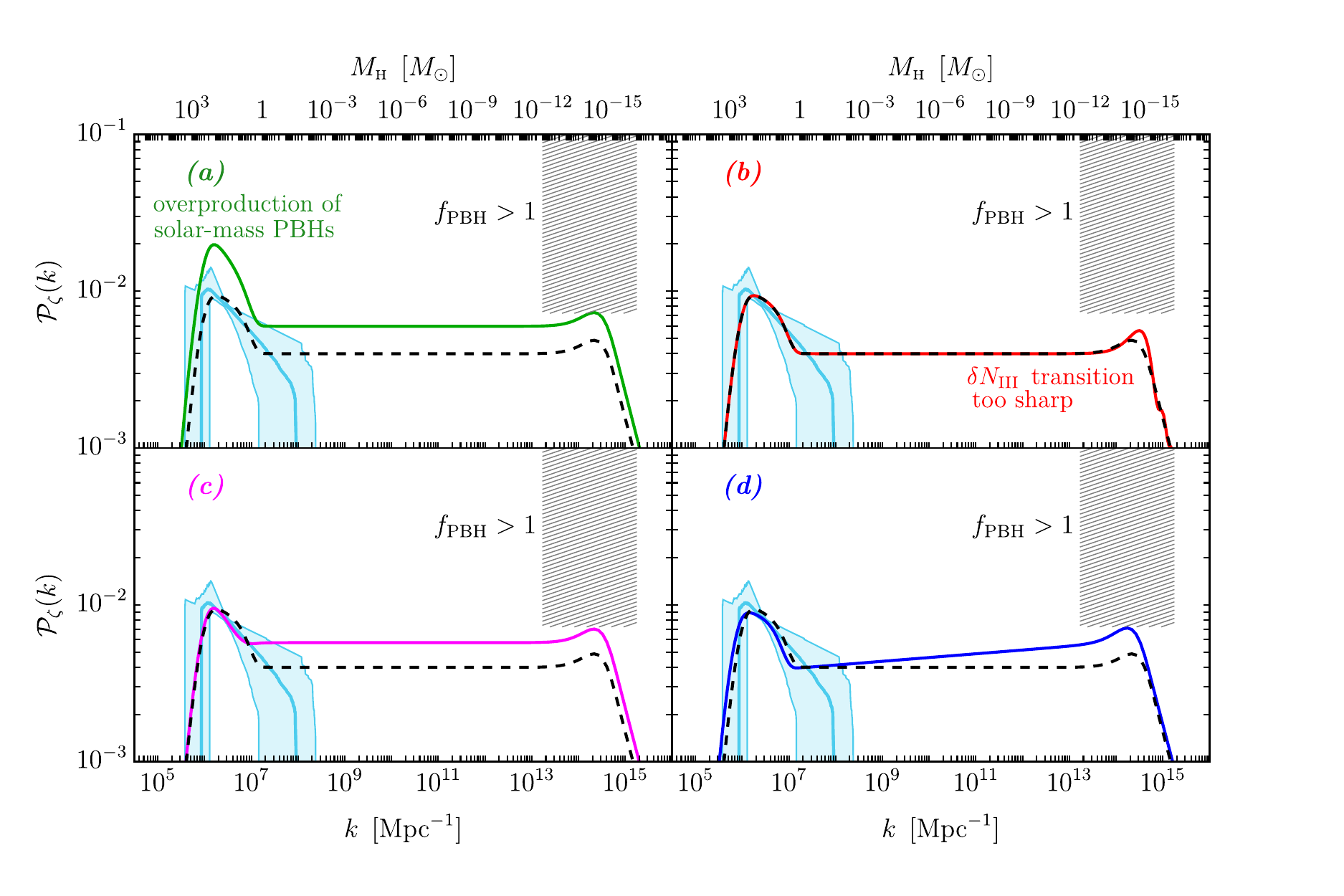}
\caption{
Four examples for the enhancement of the abundance of asteroid mass PBHs while also confronting with the constraints in the solar-mass range.
In all four panels, the dashed black line represents the curvature power spectrum (we zoom in on the very top part) that corresponds to  Model~\textbf{C} in Table~\ref{Tab:ModelTab} but taking $\delta N_{\rm I} = \delta N_{\rm II} = \delta N_{\rm III} = 0.5$ and with $\Delta N_{\rm USR}\eta_{\rm II}$ tuned in order to make the abundance of solar mass PBHs compatible with Fig.~\ref{fig:PBH constraints}. 
In this realization, the model gives $f_\PBH \ll 1$.
The task, therefore, is to enhance the height of the bump at the right-side edge of the plateau (that is, enhance the abundance of asteroid mass PBHs in order to get $f_\PBH = {\cal O}(1)$) without altering the bump at the left-side edge of the plateau. 
{\it (a)}, we take larger $\eta_{\rm II}$; this change shifts the whole power spectrum towards larger values, and overproduces solar mass PBHs. 
{\it (b)}, we take smaller $\delta N_{\rm III}$; the $\tanh$ transition at $N_{\rm III}$ becomes too sharp, and this fact introduces additional non-Gaussianity in the computation of the PBH abundance.
{\it (c)}, we take larger $\eta_{\rm II}$ and larger $\delta N_{\rm II}$; 
as in {\it (a)}, increasing $\eta_{\rm II}$ 
shifts the whole power spectrum towards larger values.
However, $\delta N_{\rm II}$ controls the amplitude of the bump at the left-side edge (without altering the rest of the spectrum). Increasing $\delta N_{\rm II}$ has the consequence of decreasing the amplitude of the left-side bump without altering the amplitude of the right-side one. As shown in the figure, from the combinations of these two effects one gets the desired  enhancement in the amplitude of the right-side bump while the amplitude of the left-side bump is kept at the level of the dashed line. This is Model~\textbf{C}.
{\it (d)}, we take $\eta_{\rm II}$ slightly different from zero and positive.  This introduces a tilt in the plateau that enhances the amplitude of the bump at the right-side edge without changing the left-side one.
This is Model~\textbf{D}. 
 }\label{fig:Solutions}  
\end{center}
\end{figure*}

This is a non-trivial task to accomplish. 
The reason is that one should be careful to enhance the amplitude of the power spectrum at the right-side end of the plateau without also altering too much the amplitude of the left-side edge since, otherwise, the risk is to overproduce solar mass PBHs which are incompatible with LVK merger rates.
We envisage four possible ways to tackle this problem (see also Fig.~\ref{fig:Solutions}):
\begin{itemize}[leftmargin=*]
    \item[{\it (a)}] First, we consider the case in which we
    take $\eta_{\rm III} = 0$ and tune the value of $\eta_{\rm II}\Delta N_{\rm USR}$ appropriately to get $f_\PBH = {\cal O}(1)$. 
    Furthermore, we fix the widths of the three transitions to the benchmark value $\delta N_{\rm I} = \delta N_{\rm II} = \delta N_{\rm III} = 0.5$. 
    
   The above tuning of $\eta_{\rm II}\Delta N_{\rm USR}$ basically corresponds to a rigid shift of the whole power spectrum towards larger values. Consequently, a larger abundance of asteroid mass PBHs will unavoidably enhance also the abundance of solar mass PBHs. Numerically, we find that (in this initial setup with $\delta N_{\rm I} = \delta N_{\rm II} = \delta N_{\rm III} = 0.5$) it is not possible to make the totality of dark matter in the form of PBHs without violating the constraints in the solar mass range (that is, without exceeding the allowed region in Fig.~\ref{fig:PBH constraints}). 
    \item[{\it (b)}] To fix this problem, a possible way out is to keep $\eta_{\rm II}\Delta N_{\rm USR}$ fixed to some value that is compatible with constraints in the solar mass range  and change the shape of the bump at the right-side of the plateau. As discussed in Ref.~\cite{Franciolini:2022pav}, this is possible by tuning the value of $\delta N_{\rm III}$ (smaller values of $\delta N_{\rm III}$ make the bump more pronounced). 
    However, we find that, in order to boost the abundance of asteroid mass PBHs to values $f_\PBH = {\cal O}(1)$, we need $\delta N_{\rm III} \ll 1$. The drawback is that such a sharp transition typically generates sizable non-Gaussianities that may threaten the validity of our computation of the abundance~\cite{Cai:2018dkf,Passaglia:2018ixg,Taoso:2021uvl}. 
    For this reason, we discard this possibility (in addition, it is unclear whether very sharp 
    transitions in the evolution of $\eta$ are realizable in concrete models).  
\item[{\it (c)}] The third possibility is the one we adopted in Model~\textbf{C}.
 As in {\it (a)}, we take $\eta_{\rm III} = 0$ and tune the value of $\eta_{\rm II}\Delta N_{\rm USR}$ appropriately to get $f_\PBH = {\cal O}(1)$; as discussed, we end up with an overabundance of solar mass PBHs. However, as noticed in Ref.~\cite{Franciolini:2022pav}, the value of $\delta N_{\rm II}$ controls the height of the bump at the left-side edge of the plateau. In particular, increasing the value of $\delta N_{\rm II}$ decreases the amplitude of the bump. It is, therefore, sufficient to consider a slightly larger value of $\delta N_{\rm II}$ 
 to smooth out the abundance of solar mass PBHs and get a perfect fit of $f_\PBH = {\cal O}(1)$.
    \item[{\it (d)}] Finally, the fourth possibility 
     is the one we adopted in Model~\textbf{D}. 
    We keep $\eta_{\rm II}\Delta N_{\rm USR}$ fixed to some value that is compatible with the constraints in the solar mass range (in particular, compatible with  the  posterior in Fig.~\ref{fig:PBH constraints}). If we now take  $\eta_{\rm III}$ non-zero and positive, the power spectrum will scale 
    as $\mathcal{P}_{\zeta} \sim k^{2\eta_{\rm III}}$ in the region between the two bumps, and this will enhance the height of the bump at the right-side edge of the  plateau without affecting the one at the left-side edge. Numerically, we find that values of  $\eta_{\rm III}$  as small as few\,$\times 10^{-2}$ are enough to get the desired enhancement that gives $f_\PBH = {\cal O}(1)$.
\end{itemize}    
A bonus possibility is to move the asteroid mass peak towards smaller masses in order to exploit the enhancement of the abundance due to the redshift factor $M_H^{-1/2}$ in Eq.~\eqref{eqn:omega}.
    In our model this means taking larger values of $N_{\rm III}$.
    However, we find that one quickly clashes with the constraint given by Hawking evaporation.
For clarity's sake, we illustrate the four possibilities {\it (a)}-{\it (d)} in Fig.~\ref{fig:Solutions}, see caption for details.

As a technical remark,  we would like to emphasise the power of the parametrization in Eq.~\eqref{eq:MainEqEta}. As clear from the above discussion, the free parameters that enter in the evolution of $\eta$ have a clear connection with the shape of the curvature power spectrum and, therefore, it turns out to be extremely simple to manipulate the dynamics and carve out the desired form of  $\mathcal{P}_{\zeta}(k)$. 

Since in Model~\textbf{D} $\eta_{\rm III}$ is not exactly zero, we expect, as anticipated, a violation of the redshift-induced scaling $m_\PBH^{-1/2}$ associated to scale invariant power spectra. Numerically, we find the power-law scaling  $m_\PBH^{-1.2}$, cf. Fig.~\ref{fig:fPBHModel}. Contrariwise, in Model~\textbf{C} we have $\eta_{\rm III} = 0$. In this model, therefore, the scaling $m_\PBH^{-1/2}$ is recovered,  as confirmed in Fig.~\ref{fig:fPBHModel}.

\bibliography{main}

\end{document}